\documentclass[12pt]{article}
\usepackage{cite}
\usepackage{graphicx}
\usepackage{multicol}
\usepackage{amsfonts}
\usepackage{amssymb}
\usepackage{amsmath}
\usepackage{heck}%
\usepackage{hyperref}
\usepackage{setspace}
\usepackage[all]{xy}
\usepackage{verbatim}
\usepackage{color}
\usepackage{cancel}

\numberwithin{equation}{section}

\newcommand{\cN}{\mathcal{N}}

\newcommand{\Tr}{\, {\rm Tr}}

\begin{document}

\date{February 2011}

\institution{IAS}{\centerline{${}^{1}$School of Natural Sciences, Institute for Advanced Study, Princeton, NJ 08540 USA}}

\institution{SEL}{\centerline{${}^2$School of Physics and Astronomy \& Center for Theoretical Physics}\cr\centerline{Seoul National University, Seoul 151-747 KOREA}}

\title{Baryon and Dark Matter Genesis\\from Strongly Coupled Strings}

\authors{Jonathan J. Heckman\worksat{\IAS}\footnote{e-mail: {\tt jheckman@ias.edu}} and Soo-Jong Rey\worksat{\IAS,\SEL}\footnote{e-mail: {\tt sjrey@snu.ac.kr}}}%

\abstract{D3-brane probes of E-type Yukawa points lead to strongly coupled nearly conformal sectors nearby the Standard Model (visible sector) which are motivated by F-theory GUTs. Realistic visible sector brane configurations induce a seesaw mass hierarchy in the hidden sector with light GUT singlets charged under a strongly coupled hidden sector $U(1)$. Interpreting these GUT singlets as dark matter, this leads to a matter genesis scenario where the freeze out and subsequent decay of heavy mediators between the two sectors simultaneously populates comparable amounts of baryon and dark matter asymmetry. Generating a net matter asymmetry requires a generational structure in the probe sector which is absent at weak string coupling, but is automatically realized at strong string coupling via towers of dyonic bound states corresponding to multi-prong string junctions. The hidden $U(1)$ couples to the visible sector through both electric and magnetic kinetic mixing terms, providing an efficient means to deplete the symmetric component of dark matter. The mass of the dark matter is of order $\sim 10$ GeV.  The dark matter mass and the matter asymmetry are both controlled by the scale of conformal symmetry breaking $\sim 10^{9} - 10^{13}$ GeV, with the precise relation between the two set by details of the visible sector brane configuration.
}

\maketitle

\enlargethispage{\baselineskip}

\setcounter{tocdepth}{2}
\tableofcontents

\newpage

\section{Introduction}

The asymmetry between matter and anti-matter is an outstanding unexplained
feature in the Standard Models of particle physics and cosmology. Though
the ratio $Y_{\Delta B}$ of net baryon number density to entropy density
is exceedingly tiny \cite{Komatsu:2010fb}:
\begin{equation}\label{targetYIELD}
Y_{\Delta B}\equiv{\frac{(n_{\mathrm{B}}-n_{\overline{\mathrm{B}}})}%
{s}}\sim 10^{-10}, %
\end{equation}
it is well known that effects purely within the Standard Model are too small
to explain this value. Perhaps even more curious, baryonic matter forms
only a small component of the total matter density inside the
Universe. Specifically, the ratio of dark matter to visible matter
relic abundances is roughly \cite{Komatsu:2010fb}:
\begin{equation}\label{relicRAT}
{{\Omega_{DM}} \over {\Omega_{\Delta B}}} \ \sim \ 5 \ .
\end{equation}
The origin of dark matter is shrouded in even more mystery. In principle,
it could be that the origin of visible and dark matter have different
origins. However, the relative proximity of these relic abundances naturally suggests a
single origin for both species. This is quite an attractive theoretical
solution, and has been discussed from various viewpoints in for example
\cite{Kaplan:1991ah, Thomas:1995ze, Hooper:2004dc, Kitano:2004sv,
Cosme:2005sb, Farrar:2005zd, Suematsu:2005kp, Banks:2006xr,
Kitano:2008tk, Kohri:2009ka, Kohri:2009yn, An:2009vq,
Cohen:2009fz, Kaplan:2009ag, Cohen:2010kn, Blennow:2010qp, Hall:2010jx}.

Such clues are also of vital importance for potential ultraviolet (UV) completions of the
Standard Model at higher energy scales. Indeed, various proposals for
generating an appropriate baryon asymmetry such as leptogenesis \cite{Fukugita:1986hr}
(see \cite{Davidson:2008bu} for a review) and Grand Unified Theory (GUT)
baryogenesis (see \cite{KT} for a concise overview) provide a potential window into high-energy scale
physics. Given the proximity of the GUT scale to the string scale, this naturally motivates the
search for potential string based mechanisms which can explain such phenomena.

In this work, we propose a stringy mechanism that can explain the
origin of visible and dark matter with correlated relic abundances.
Our setup is as follows. We assume the Standard Model is given by a
configuration of intersecting seven-branes via F-theory \cite{DWI, BHVI, BHVII, DWII} (see \cite{Heckman:2010bq, Weigand:2010wm} for reviews).
The local intersections consist of a stack of seven-branes $7_{vis}$ where the ``visible sector'' e.g.
Standard Model gauge group localizes, as well as an extra flavor seven-brane(s)
which we denote by $7_{hid}$. Matter is localized at the intersections of these seven-branes,
and Yukawa interactions localize at the points of triple intersections between seven-branes.\footnote{Let us note that
for the purposes of this paper, we only require the rather mild condition that in the limit where
all seven-branes are non-compact, such a $7_{hid}$-brane exists.} Our focus will be on the particle physics
content of the theory, and so we shall neglect issues related to gravity, such as moduli stabilization. This approximation is
justified in the context of local model
building \cite{VerlindeWijnholtBottomUp,KiritsisBottomUp,UrangaBottomUp,BerensteinJejjalaLeigh,BHVI,BHVII,DWI,DWII,HeckVerlinde}.

Probe D3-branes constitute another well-motivated sector because they are
locally attracted to E-type Yukawa points \cite{Funparticles} (see also \cite{Martucci, FGUTSNC}).
In many cases of interest, these probe sectors realize a strongly coupled non-Lagrangian
$\mathcal{N} = 1$ superconformal field theory (SCFT) \cite{Funparticles, FCFT, HVW}.
The perturbative string spectrum consists of $3-3$, $3-7_{vis}$ and $3-7_{hid}$ string states.
At strong string coupling, there are also additional $(p,q)$ string
states and multi-prong junctions of comparable mass. In our foregoing considerations, these states and junctions
will be playing an essential and indispensible role.

As in \cite{Funparticles,FCFT}, the probe sector is specified at a UV cutoff scale $M_{\ast} \sim 10^{17}$ GeV by an operator deformation of a
strongly coupled $\mathcal{N} = 2$ theory with $E_8$ flavor symmetry \cite{MNI,MNII}. The value of $M_{\ast}$ defines the scale of
tilting for the intersecting seven-brane configuration and is related to the string length scale $\ell_s$ and string coupling constant
$g_s$ by $M_{\ast}^{8} = \ell_{s}^{-8}/g_{s}$ \cite{BHVII}. At the GUT scale $M_{GUT}$, the visible sector fields become four-dimensional,
and there is a further $\mathcal{N} = 1$ deformation induced through the
coupling of these four-dimensional modes to the probe. Quite remarkably, this further
deformation leads to another IR fixed point in which the scaling dimensions of the Standard Model fields
shift only a \textit{small} amount \cite{HVW}.

\begin{figure}
[ptb]
\begin{center}
\includegraphics[
trim=1.378385in 7.572447in 1.379210in 0.693508in,
height=3.2093in,
width=5.1015in
]%
{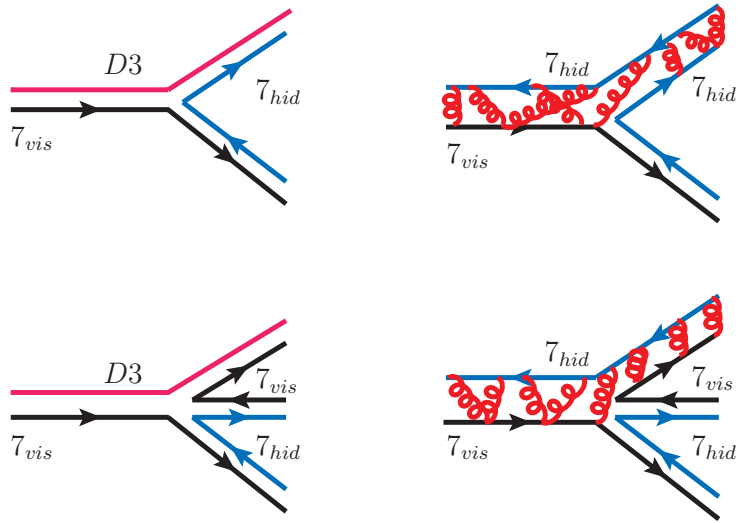}%
\caption{Decay of a heavy $3-7_{vis}$ state connected between the probe D3-brane and the $7_{vis}$-brane. In order for Chan-Paton color flow to be conserved, this state decays into a Standard Model state $7_{vis}-7_{hid}$ string, and a $3-7_{hid}$ string which is a Standard Model singlet. The left column shows a depiction of the decay of a $3-7_{vis}$ string into one (top) and two (bottom) Standard Model states. In the right column the strong coupling analogue of these decays is indicated. At strong coupling there are additional spectator $3-7_{hid}$ states which dress the decay of the $3-7_{vis}$ string.}%
\label{chan-paton}%
\end{center}
\end{figure}

Since the probe sector contains states charged under $SU(5)_{GUT}$, they must have a mass of at least $\sim 500$ GeV
to have evaded detection so far. We find that that there are
vacua where the D3-brane is stabilized close to, but not directly on top of
the $7_{vis}$-brane. This separation from the E-point of the geometry introduces a
SCFT breaking scale $M_{\cancel{CFT}}$ below the scale $M_{GUT}$, which is roughly the characteristic mass for
the $3-7_{vis}$ states of the spectrum:
\begin{equation}
M_{med} \sim M_{\cancel{CFT}}.
\end{equation}
As the notation suggests, the $3-7_{vis}$ strings serve as mediators
between the visible and hidden sector. Below the scale of CFT breaking,
the probe sector appears as a strongly coupled $U(1)_{D3}$ gauge theory.

By analyzing the holomorphic geometry of the brane configuration,
we show that the characteristic mass scale of the $3-7_{hid}$ GUT
singlets is set by the monodromic seesaw relation:
\begin{equation}
M_{hid}\sim M_{\cancel{CFT}}\cdot\left(  \frac{M_{\cancel{CFT}}}{M_{\ast}}\right)
^{\alpha}, \label{37primemass}%
\end{equation}
where $\alpha \sim O(1)$ is a parameter set by the details of seven-brane monodromy \cite{Hayashi:2009ge, BHSV, Marsano:2009gv, TBRANES}.
In weakly coupled models, we have $\alpha = 1 ,2 ,3 $, though there can be D-term
contributions as well. Typically, there is only a small shift in the scaling dimensions of the operators
associated with such states, so we expect such shifts from D-term effects to be small. This leads
to highly suppressed masses for the light $3-7_{hid}$ states.

This hierarchy of scales between $3-7_{vis}$ and $3-7_{hid}$ strings
sets up a natural scenario for matter genesis based on the dynamics of the probe D3-brane.
The basic idea is that when a heavy $3-7_{vis}$ string decays, its
end products will involve a $7_{vis}-7_{hid}$ string and a $3-7_{hid}$ string. The
former corresponds to a Standard Model particle, while
the latter corresponds to a much lighter GUT singlet, which constitutes the dark
matter of the model.\footnote{Let us note that in earlier work on F-theory GUTs such as \cite{FGUTSCosmo},
it was found that for an appropriate range of parameters, leptogenesis could potentially
generate a viable amount of matter asymmetry, and that $10-100$ MeV mass gravitinos
could provide a dominant component of the dark matter. This is certainly a
possibility, but it requires two separate physical mechanisms to generate the
visible and hidden sector relic abundance.}
At a qualitative level, these considerations apply equally well whether we are at strong coupling, as
in F-theory GUTs, or simply in perturbative IIB string theory. See figure \ref{chan-paton}
for a depiction of such decays at weak and strong coupling of the probe D3-brane.

To generate an appropriate baryon asymmetry, a matter genesis scenario must satisfy the
Sakharov conditions\cite{Sakharov:1967dj}:
\begin{itemize}
\item Departure from thermal equilibrium

\item Baryon number violation

\item C and CP violation
\end{itemize}

In the probe sector, these conditions are satisfied as follows. After the end
of inflation at a temperature $T_{RH}$, the Universe begins to cool.
Eventually, it cools below the mass of the $3-7_{vis}$ strings and their anti-particle counterparts,
at which point these states freeze out from the thermal bath. This leads to a departure from
thermal equilibrium.

At strong coupling, there are multiple decay channels of the $3-7_{vis}$ states to visible sector states.
Conservation of the $SU(3)_{C} \times SU(2)_L \times U(1)_Y$ and $U(1)_{D3}$ quantum numbers implies that the $3-7_{vis}$ strings
always decay simultaneously to Standard Model states ($7_{vis} -7_{hid}$ strings) and hidden sector states ($3-7_{hid}$
strings). Moreover, conservation of the $7_{hid}$-brane quantum number correlates the effective baryon charge violation
created in the visible and hidden sectors. Because the probe is at strong coupling, such decays will be dressed by additional spectator $3-7_{hid}$ strings.
Decay processes which violate baryon number correspond to those channels where two or more Standard Model
states are present in the final state. Examples of such baryon number violating interactions include $SU(5)_{GUT}$ invariant
operators $5 \times 10 \times 10$ and $\overline{5} \times \overline{5} \times 10$
as well as many higher order processes which are generically present at strong coupling. Here, at least
two of the states correspond to Standard Model states, and at least one involves the $3-7_{vis}$ states.

By appealing to various toy models as well as a
gauge invariant operator analysis, we find that the decay rate for
$3-7_{vis}$ strings is then given by:
\begin{equation}
\Gamma_{D} \sim \vert \epsilon \vert^{2} \cdot M_{\cancel{CFT}}
\end{equation}
with $\epsilon$ a number less than one, so that the decay is slow compared to the time scale $1/M_{\cancel{CFT}}$.
It turns out
\begin{equation}
\epsilon \sim \left(\frac{M_{\cancel{CFT}}}{M_{GUT}} \right)^{\nu},
\end{equation}
where $\nu$ is an order one number typically less than one, which is determined
by the scaling dimensions of operators in the approximately conformal sector.

Generating a net C and CP violation requires two or more independent decay processes, leading to nontrivial interference among them. In weakly coupled IIB string theory, we find that
the tree level and loop level interaction terms carry an overall common complex phase,
which at first loop order in perturbation theory leads to identical decay rates for a
particle and the CP conjugate decay process. In field theory language, this is due to the
fact that there is only a single generation of modes in the hidden sector which participate
in such processes. The existence of a single generation is basically due
to the fact that, at weak string coupling, only a single type of string (the fundamental string)
participates in the field theory.

At strong coupling, however, the decay of heavy $3-7_{vis}$ strings generically violates C and CP.
This is basically due to the existence of an entire tower of states, all of which now contribute in the strongly coupled
analogue of ``loop order processes''. We motivate from purely field theoretic as well as from string based descriptions
the existence of the required C and CP violating decays, which in turn generate a net baryon asymmetry.

Kinetic mixing between the $U(1)$ gauge bosons of the visible and hidden sectors allows
the two sectors to remain in thermal contact. As the Universe
cools below the mass scale of the dark matter, kinetic mixing between the visible and hidden sectors
depletes the symmetric component of the dark matter, leaving only the net matter asymmetry. \footnote{See for
example \cite{Davoudiasl:2010am} for related discussions of kinetic
mixing as a mechanism to deplete the symmetric component of dark matter
in an asymmetric dark matter scenario.} This leaves us with a correlated
matter asymmetry in the visible and hidden sectors.

Generating an appropriate relic abundance ratio as in \eqref{relicRAT} then fixes the mass of the dark matter to be on the order of $m_{DM} \sim 10$ GeV,
similar to most asymmetric dark matter models. Rather than converting the relic abundance from the visible to hidden sector, or the other way around, here,
the relic abundances are generated simultaneously. It therefore has some (mild) qualitative overlap with recent discussions in \cite{Davoudiasl:2010am,Hall:2010jx,Falkowski:2011xh}.

In order to match to observation, it is also necessary for the proposed mechanism
to generate the correct amount of baryon asymmetry. Computing the
yield of dark matter and visible matter, we find:
\begin{equation}
Y_{DM} \sim Y_{\Delta B} \sim \mathcal{D}_{net} \cdot \frac{g_{eff,med}}{g_{\ast}(M_{\cancel{CFT}})} \cdot \left( {M_{\cancel{CFT}} \over M_{GUT}} \right)^{2 \nu}.
\end{equation}
Here, $g_{eff,med} \sim 1 - 10$ is the effective number of mediator states which freeze out and decay (counted as GUT multiplets), $g_{\ast}(M_{\cancel{CFT}}) \sim 10^{3}$ is the number of ``free field'' degrees of freedom of the probe/MSSM system, and $0 < \mathcal{D}_{net} \leq 1$ takes into account possible dilution from late entropy production, as well as thermal washout effects. Typically, we find there is little washout due to the slow decay rate of the $3-7_{vis}$ strings. By adjusting the various parameters of the scenario, the model can also overproduce a matter asymmetry, which can then be diluted by a non-zero value of $\mathcal{D}_{net}$.

A highly non-trivial feature of this scenario is that the mass of the dark matter, the required relic abundance, and the absence of thermal washout effects can all be satisfied for an appropriate $n$ and $M_{\cancel{CFT}} \sim 10^{9} - 10^{13}$ GeV with dynamically determined $\nu \sim O(1)$ typically somewhat less than one. The precise value of these parameters depends on various non-holomorphic data, which in turn will affect the exact spectrum and mass scale. The basic point, however, is that various a priori \textit{independent} considerations are simultaneously satisfied for a natural range of parameters for the probe D3-brane.

The rest of this paper is organized as follows. In section \ref{sec:SETUP}, we
briefly review the main elements of F-theory GUTs and D3-brane probe theories.
In section \ref{sec:MASS}, we estimate the mass scales
of the probe sector when the D3-brane is displaced from the Yukawa point.
We next study in section \ref{sec:DECAY} the decay of heavy $3-7_{vis}$ states
as a mechanism for generating a baryon asymmetry. Putting these elements
together, in section \ref{sec:COSMO} we discuss the cosmological timeline for
this scenario, and show that for parameter ranges natural for the probe, we obtain
the correct baryon asymmetry, dark matter asymmetry and dark matter mass.
Section \ref{sec:CONCLUDE} contains our conclusions and potential
directions for future investigation. Related material
is discussed in the Appendices.

\section{Basic Setup\label{sec:SETUP}}

In this section, we review some basic features of F-theory GUTs, and in particular
the motivation for considering probe D3-branes in this setup. After this, we discuss
some additional aspects of such probe sectors studied in \cite{Funparticles, FCFT, HVW}.

Our starting point is compactifications of F-theory to four dimensions.
F-theory can be viewed as a strongly coupled variant of IIB string theory in
which the axio-dilaton $\tau \equiv (C_0 + i / g_{s})$ is allowed to have non-trivial profile on the
internal directions, and moreover, can be order one. In order to preserve
$\cN=1$ supersymmetry in four dimensions, we consider F-theory compactified on
an elliptically fibered Calabi-Yau fourfold $X_{4}$ fibered over a threefold
base $B_{3}$. Working in a local patch of $B_{3}$, we can introduce three
local coordinates $z_{1}$, $z_{2}$ and $z_{3}$. In Weierstrass form, the
elliptic curve is locally given by
\begin{equation}
y^{2}=x^{3}+f(z_{1},z_{2},z_{3})x+g(z_{1},z_{2},z_{3}).
\end{equation}
The modulus $\tau(z_{1},z_{2},z_{3})$ is set by the coefficient functions $f,g$ according
to the relation:%
\begin{equation}
j(\tau)=\frac{4\left(  24f\right)  ^{3}}{4f^{3}+27g^{2}}, %
\end{equation}
where $j(\tau)$ is the modular invariant $j$-function. The locations of
seven-branes in the base $B_3$ are given by the zeroes of the discriminant of
the cubic equation for $x$:%
\begin{equation}
4f^{3}+27g^{2}=0.
\end{equation}
Physics near such singularities is dictated by the local behavior
of eight-dimensional super Yang-Mills theory, with the specifics of the
intersection dictated by the choice of symmetry breaking
pattern of this theory \cite{BHVI, DWI, DWIII, TBRANES}.

In F-theory GUTs, the basic idea is to realize an eight-dimensional GUT on the
worldvolume of a seven-brane. We choose the local
coordinates $(z_{1},z_{2})$ to locally parametrize the worldvolume of the
GUT\ brane, and $z\equiv z_{3}$ to parametrize the normal direction to this
hypersurface. For concreteness, we shall take the GUT group to be
$SU(5)_{GUT}$. Matter fields such as the $\overline{5}$ and $10$ of an
F-theory GUT are then realized at the intersection of the GUT\ seven-brane
with additional branes, which we generically refer to as $7_{hid}$-branes.
Yukawa couplings are localized at points of the geometry, and are computed in
terms of the triple overlap of wavefunctions in the geometry.

The local intersections and interactions can be modelled in terms of a
parent eight-dimensional $E_{8}$ gauge theory \cite{BHVI, DWI, DWIII, TBRANES}. The basic idea is that the
$z_{1}$ and $z_{2}$ dependent vacuum expectation value (vev) of a complex scalar field $\Phi$ taking
values in the adjoint representation of $E_{8}$ dictates the local tilting of
seven-branes in the geometry. In order to realize an $SU(5)$ F-theory GUT, we
consider breaking patterns where $\Phi$ takes values in the $SU(5)_{\bot}$
subalgebra of $SU(5)_{GUT}\times SU(5)_{\bot}\subset E_{8}$. All of the matter
fields then descend from the adjoint representation of $E_{8}$ via the
breaking pattern:%
\begin{align}
E_{8}  &  \supset SU(5)_{GUT}\times SU(5)_{\bot}\nonumber\\
248  & \rightarrow(1,24)\oplus(24,1)\oplus(5,\overline{10})\oplus(\overline
{5},10)\oplus(10,5)\oplus(\overline{10},\overline{5}).
\end{align}
As a point of terminology, we refer to states charged under just $SU(5)_{GUT}$
as $7_{vis}-7_{vis}$ strings, those charged under just $SU(5)_{\bot}$ as $7_{hid}-7_{hid}$
strings, and states charged under both factors as $7_{vis}-7_{hid}$ strings.
To a certain extent, this terminology is imprecise because the
actual states also involve non-perturbative $(p,q)$ strings and their junctions. These extra states
are important for our considerations, so we shall later make this more precise.

One of the important characteristic features of an F-theory GUT is that the
geometric data of the visible sector is determined by the local behavior of
the parent eight-dimensional $E_{8}$ gauge theory. A remarkable feature of this sort of scenario
is that it is flexible enough to accommodate all of the interaction terms for
a viable visible sector \cite{BHSV, EPOINT}. However, embedding in $E_{8}$ also
imposes some rigid constraints. In particular, the condition that all matter and interactions
descends from the adjoint of $E_{8}$ is quite stringent.

\subsection{Probing the Standard Model with a D3-Brane}

In \cite{Funparticles} it was noted that there is another natural source of
interactions originating from D3-branes, rather than seven-branes. Indeed, in
the context of the flavor physics scenario considered in \cite{HVCP, BHSV}, the
requisite fluxes which induce flavor hierarchies in the seven-brane
superpotential also induce a superpotential for D3-branes which tends to
attract D3-branes to the Yukawa points of an F-theory GUT. Furthermore, such
D3-branes are naturally a part of the compactification due to the tadpole
constraint in global models:%
\begin{equation}
N_{\mathrm{D3}} = \int_{B}H_{\mathrm{NS}%
}\wedge H_{\mathrm{RR}} + {\frac{1}{24}}\chi_{E}(\mbox{CY}_{4}), %
\end{equation}
where $\chi_{E}(\mbox{CY}_{4})$ is the Euler character of the Calabi-Yau
fourfold and $H_{\mathrm{NS}},H_{\mathrm{RR}}$ are the NS and RR 3-form
fluxes, and $N_{\mathrm{D3}}$ is the number of D3-branes.

The dynamics of such probe D3-branes have been studied in \cite{Funparticles, FCFT, TBRANES}.
There, it was found that the D3-brane coupling to the configuration
of intersecting seven-brane provides a novel way to extend the Standard Model
at higher energy scales by coupling it to a strongly coupled superconformal field theory.

\begin{figure}
[t!]
\begin{center}
\includegraphics[
trim=0.000000in 7.109330in 0.000000in 0.712219in,
height=2.8573in,
width=6.0329in
]%
{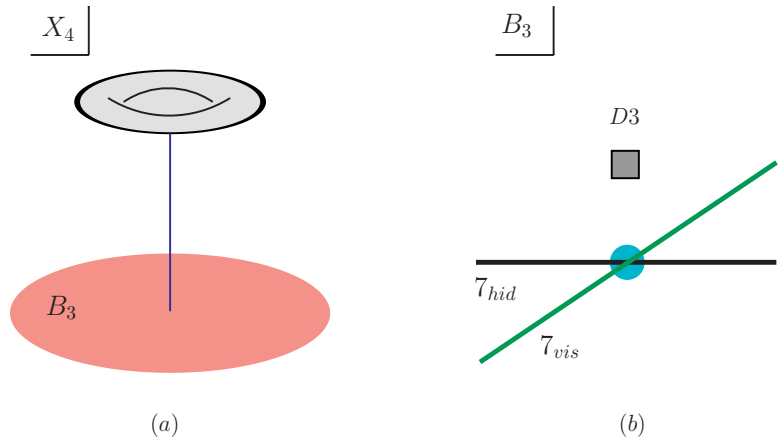}%
\caption{Setup of F-theory GUTs. (a) In F-theory compactified to four dimensions with $\mathcal{N} = 1$ supersymmetry, the six internal directions and the profile of the axio-dilaton are combined into an elliptically fibered Calabi-Yau fourfold $X_4$ with threefold base $B_3$. (b) Various seven-branes are located at hypersurfaces where the torus degenerates. One stack gives the Standard Model gauge group, while the other stack corresponds to a flavor seven-brane. Intersections of matter seven-branes define Yukawa couplings and E-points. Also depicted is a D3-brane on the Coulomb branch of its moduli space.}%
\label{FGUTS}%
\end{center}
\end{figure}

Taking the location of the Yukawa point to be $z_{i}=0$, when the D3-brane is
located at $|z_{i}|\gg M_{\ast}$, the worldvolume dynamics is given by a $U(1)$ gauge theory
with holomorphic coupling $\tau_{D3}=\tau_{IIB}$. As the D3-brane moves close
to a seven-brane, additional strings and their junctions stretched between the
seven-branes and the D3-brane become light, and can potentially give rise
to a non-trivial interacting SCFT. For example, the D3-brane probe
of an $E_8$ seven-brane realizes the $\mathcal{N} = 2$ Minahan-Nemeschansky theory with $E_8$ flavor symmetry \cite{MNI, MNII}.

The D3-brane probes of $\mathcal{N} = 1$ seven-branes are all defined in terms of $\mathcal{N} = 1$ deformations of
$\mathcal{N} = 2$ probe theories. In turn, the $\mathcal{N} = 2$ theories are associated with a stack of parallel
seven-branes with gauge group $G$. This gauge group becomes the flavor symmetry of the D3-brane theory. The operator
content of the $\mathcal{N} = 2$ theory includes operators $Z$, $Z_1$ and $Z_2$ parameterizing the motion of the D3-brane
away from the seven-brane, with $Z_1 \oplus Z_2$ a decoupled hypermultiplet. In addition, there are dimension-two operators
$\mathcal{O}$ in the adjoint representation of $G$ which parameterize the Higgs branch of the theory. A
remarkable feature of such probe theories is that the $\cN=2$ Seiberg-Witten curve is identical to the
F-theory geometry of the compactification. The $\mathcal{N} = 2$ curve also serves to characterize
the possible holomorphic mass deformations of the probe theory.

The $\mathcal{N} = 1$ theories are then defined at the scale of seven-brane tilting $M_{\ast}$ via the deformations:
\begin{equation}\label{defdef}
\delta W_{tilt} = Tr_{E_{8}} (\Phi( Z_1 , Z_2) \cdot \mathcal{O}),
\end{equation}
where $\Phi(Z_1 , Z_2)$ is a field-dependent mass deformation which dictates the tilting of the intersecting seven-brane
configuration \cite{Funparticles}. In phenomenological applications, $\Phi$ takes values in the $SU(5)_{\bot}$ factor
of $SU(5)_{GUT} \times SU(5)_{\bot} \subset E_{8}$. Realistic tilting configurations require
$\Phi(0,0)$ nilpotent but non-zero, realizing a T-brane configuration \cite{TBRANES}. Other breaking
patterns involving $E_6$ and $E_7$ are possible, but for simplicity we shall focus on the case of $E_8$ breaking,
as this choice also leads to attractive phenomenology for the visible sector \cite{BHSV, EPOINT}.

An important feature of the tilting of the seven-brane configuration is the notion of ``seven-brane monodromy''
\cite{Tatar:2009jk, BHSV, TBRANES}. This is partially dictated by how $\Phi(0,0)$ splits up into a direct
sum of nilpotent Jordan blocks. The infrared (IR) behavior of the probe theory
is mainly determined by the choice of $\Phi(0,0)$ \cite{FCFT}.

Evidence for additional strongly coupled $\mathcal{N} = 1$ fixed points was given in \cite{FCFT}. An interesting feature
of such theories is that the holomorphic mass deformations of the $\mathcal{N} = 1$ theory are still captured by
the F-theory geometry. In particular, there is an overall homogeneity constraint which must be satisfied by such
deformations.

In a physically realistic setting, we must compactify the GUT seven-brane.
This can be characterized in terms of an $\mathcal{N} = 1$ deformation which is added at the GUT scale:
\begin{equation}
\delta W_{SM \otimes D3} = \Psi^{SM}_{R} \cdot \mathcal{O}_{\overline{R}}.
\label{sm-coupling}
\end{equation}
The state $\Psi^{SM}_{R}$ is associated with a dynamical
field of the Standard Model in a representation $R$ of
$SU(3)_C \times SU(2)_L \times U(1)_{Y} \subset SU(5)_{GUT}$.\footnote{Note that
R-parity of the Minimal Supersymmetric Standard Model (MSSM)\ will automatically be preserved by the interaction terms
if we assume it descends from a $\mathbb{Z}_{2}$ subgroup of an abelian factor of $SU(5)_{\bot}$. In some scenarios such
as the Majorana neutrino scenarios of \cite{BHSV} and \cite{EPOINT}, R-parity is instead
assumed to originate from an additional approximate geometric symmetry. In
the present case, we assume that a similar $\mathbb{Z}_{2}$
action also extends to the probe SCFT sector.} Here, the operator
$\mathcal{O}_{\overline{R}}$ is given by decomposing the original dimension-two operators in the adjoint of
$E_8$ of the $\mathcal{N} = 2$ probe theory into representations of $SU(3)_{C} \times SU(2)_L \times U(1)_Y$ which can pair with Standard Model states.
A non-trivial consequence of correlating the choice of seven-brane tilting in
the visible sector with the particular operator deformation is that the
scaling dimensions of the Standard Model fields typically remain quite
close to their free field values in most examples \cite{HVW}.

In other words, the probe defines a \textit{small} perturbation of the Standard Model. An interesting feature of the scaling dimension of such operators is that it depends on the choice of the monodromy in $\Phi$ and the associated $SU(5)_{\bot}$ quantum numbers \cite{FCFT}. In this paper, our main interest will be in the scaling dimensions of the operators $\mathcal{O}_{R}$ and $\Psi^{SM}_{R}$ for $R$ a representation of the Standard Model gauge group. The scaling dimension of the $\mathcal{O}_{R}$'s is typically around $2$, but it can be lower or higher depending on the $SU(5)_{\bot}$ quantum numbers.
See Appendix A for further discussion of operator scaling dimensions in the $\Phi$-deformed probe theories.

In principle, there is an additional deformation of the $\mathcal{N} = 1$ fixed point, which is the flux superpotential $W_{flux}$ which attracts the D3-brane to the Yukawa point. Generically, the form of this superpotential deformation will be a power series in the $Z_i$ subject to the
condition that $W_{flux}$ has a critical point at $Z_i = 0$ \cite{FGUTSNC}.
The detailed form of $W_{flux}$ is actually fixed by considerations in the visible sector, and in particular, the requirement that we realize
an appropriate flavor hierarchy along the lines of \cite{HVCKM, FGUTSNC}. One complication is that the required form of this superpotential has not been determined in the case of the most interest with seven-brane monodromy. In this paper we treat the form of $W_{flux}$ as a small perturbation to the $\mathcal{N} = 1$ probe theory. This is technically natural since $W_{flux}$ breaks additional symmetries of the probe theory \cite{FCFT}. See Appendix B for further discussion.

Let us now turn to the interactions between the probe sector and the Standard Model.
There are two basic ways that the probe sector interacts with the states of
the Standard Model. First, there are couplings between matter fields such as the F-term deformation (\ref{sm-coupling}).
There are additional couplings between the probe and Standard Model which we shall estimate
later in this paper. For appropriate T-brane configurations and
in the absence of SCFT breaking effects, some of such couplings are marginal in the infrared \cite{HVW}. The dominant coupling between the probe sector
and the Standard Model is via those matter fields which do not vanish at the Yukawa point. In the context of $E_8$-point unification scenarios
such as \cite{EPOINT}, such matter fields primarily correspond to the Higgs fields and the
third generation of chiral superfields \cite{Funparticles, HVW}. Let us note that there
will be additional but small mixing with the first and second generations because the
profiles of these wave functions are not exactly zero at the Yukawa point.

The second main that the probe interacts with the Standard Model is via gauge interactions. These include the obvious couplings based on the presence of operators of the probe sector which are charged under the $SU(5)_{GUT}$ group. Such contributions will affect the running of the Standard Model gauge couplings, though in a way which preserves one-loop gauge coupling unification \cite{Funparticles,HVW}. Additionally, there are kinetic mixing terms which couple the abelian gauge field strength of the visible sector (e.g. $U(1)_{Y}$) with the $U(1)_{D3}$ of the hidden sector \cite{Holdom:1985ag,Babu:1996vt,Dienes:1996zr,Babu:1997st,Lust:2003ky,Abel:2003ue,Abel:2004rp,
Abel:2006qt,Feldman:2006wd,Feldman:2007wj,Abel:2008ai,Brummer:2009cs,Bruemmer:2009ky,Benakli:2009mk}.
Given the gauge field strengths of the visible and hidden sectors $F_{vis}$ and $F_{hid}$, we have \cite{Funparticles}:
\begin{equation}
L_{mix} \supset -\frac{1}{4} F_{vis}^{2} -\frac{1}{4} F_{hid}^{2} + \frac{\theta_{hid} \alpha_{hid}}{8 \pi} F_{hid} \widetilde{F}_{hid} +   {\kappa_{elec} \over 2} F_{vis} F_{hid} + {\kappa_{mag} \over 2} F_{vis} \widetilde{F}_{hid}
\end{equation}
This effective Lagrangian includes both electric \textit{and} magnetic mixing terms. Let us note that although some of these terms appear as total derivatives, the presence of $F_{hid} \widetilde{F}_{hid}$ and related dualized terms cannot be neglected, because the probe D3-brane sector typically contains electric and magnetic states and hence the gauge group is compact $U(1)$ (see \cite{Witten:1979ey} for an early discussion of the importance of such terms). The precise value of the mixing at strong coupling depends on many details which we do not at present know how to compute. However, a reasonable first estimate which we shall sometimes
use is that the $\kappa_{mix}$ couplings are of order
\begin{equation}
\kappa_{mag} \sim \frac{\sqrt{\alpha_{vis} \alpha_{hid}}}{4 \pi} \sim 10^{-3} - 10^{-2}.
\end{equation}
See Appendix C for further discussion of kinetic mixing at strong coupling.

\subsection{Toy Model: Weakly Interacting $D_{4}$ SCFT}

Though our ultimate interest is in strongly interacting D3-branes probing an E-type Yukawa point,
the absence of a weakly coupled Lagrangian description significantly limits
our ability to study detailed properties of such theories. As a way to gain
additional intuition into such probe theories, it is useful to study a case
which admits a weakly coupled Lagrangian description.

The main example of such a weakly coupled probe theory is
a D3-brane probe of a seven-brane with gauge group
$SO(8)$. This corresponds to a D3-brane probe of four D7-branes and an
O7-plane. The D3-brane probe theory is then an $\cN=2$ superconformal
gauge theory with gauge group $SU(2)_{D3}$ and four quark flavors. In terms of $\cN=1$ supermultiplets,
the field content consists of a vector multiplet and a chiral multiplet $\varphi$ in
the adjoint representation of the gauge group, and four chiral multiplets $Q_{I}\oplus\widetilde{Q}_{I}$
for $I=1,2,3,4$ in doublets of $SU(2)_{D3}$. In addition, there is a
decoupled hypermultiplet $Z_1\oplus Z_2$. $\cN=2$ supersymmetry uniquely fixes the superpotential:%
\begin{equation}
W=\sqrt{2}\sum_{I=1}^{4} Q_{I} \varphi \widetilde{Q}_{\overline{I}}.
\end{equation}

Restricting for the moment to the case with $\cN=2$ supersymmetry, the moduli space consists of two branches. The Coulomb branch is parameterized by
\begin{equation}
Z={\frac{1}{2}}\mbox{Tr}_{SU(2)_{D3}} \left(\varphi^{2} \right),
\end{equation}
which encodes motion of the D3-brane transverse to the seven-branes.
It is manifestly a singlet of $SO(8)$. The origin $Z=0$ defines the $\mathcal{N}=2$ SCFT with $SO(8)$
symmetry. The Higgs branch emanates from $Z=0$, and is parametrized by quark
bilinear operators:
\begin{equation}
\mathcal{O}_{[IJ]}=Q_{[I} Q_{J]},\qquad \mathcal{O}_{[\overline{IJ}]}=\widetilde{Q}_{[\overline{I}}\widetilde{Q}_{\overline{J}]},\qquad \mathcal{O}_{I\overline{J}}=Q_{I} \widetilde{Q}_{\overline{J}}.
\end{equation}
We see that they transform in the adjoint representation of the $SO(8)$ flavor symmetry.

From the perspective of the probe theory, tilting the configuration of
seven-branes corresponds to a field dependent mass deformation of the form:%
\begin{equation}
\delta W_{tilt} = \Tr_{SO(8)}\left(  \Phi(Z_1,Z_2)\cdot \mathcal{O}\right),
\end{equation}
where $\Phi(Z_1,Z_2)$ is in the adjoint representation of $SO(8)$.

Although the group $SO(8)$ is too small to accommodate the actual Standard Model gauge
group or $SU(5)_{GUT}$, we can consider decomposing
\begin{equation}
SO(8) \supset (SU(2)\times U(1))_{vis} \times
(SU(2) \times U(1))_{hid}
\end{equation}
and view $(SU(2) \times U(1))_{vis}$ as a toy version of the \textquotedblleft Standard Model\textquotedblright\ gauge group. With respect to this decomposition, the four chiral multiplets $Q_I, \widetilde{Q}_I$ are decomposed to
\begin{eqnarray}
&& (2_\pm, 1_0) \oplus (2_\mp, 1_0) = Q^{med}_{a} \oplus \widetilde{Q}^{med}_{a}  \qquad (a = 1, 2), \nonumber \\
&& (1_0, 2_\pm) \oplus (1_0, 2_\mp) = Q^{hid}_{b} \oplus \widetilde{Q}^{hid}_{b}  \qquad (b = 1, 2).
\end{eqnarray}
These modes define a toy version of the ``$3-7_{vis}$'' mediator strings and the ``$3-7_{hid}$'' hidden sector strings. They are indexed with $a = 1,2$ under $(SU(2) \times U(1))_{vis}$ and $b = 1,2$ under $(SU(2) \times U(1))_{hid}$, respectively. In all cases, they transform as doublets under the $SU(2)_{D3}$ gauge group present at the origin of the Coulomb branch.

We also have various \textquotedblleft Standard Model fields\textquotedblright\ which
descend from the adjoint of $SO(8)$:%
\begin{align}
SO(8)  &  \supset (SU(2) \times U(1))_{vis}\times
(SU(2) \times U(1))_{hid}\nonumber \\
28  &  \rightarrow(3_{0},1_{0})\oplus(1_{0},3_{0}) \oplus (1_{0},1_{0}) \oplus (1_{0},1_{0}) \nonumber \\
& \oplus (1_{-2},1_{0}) \oplus (1_{0},1_{2})\oplus (1_{2},1_{0}) \oplus (1_{0},1_{-2}) \nonumber \\
& \oplus (2_{+1},2_{+1}) \oplus (2_{+1},2_{-1}) \oplus (2_{-1},2_{+1}) \oplus (2_{-1},2_{-1}) .
\end{align}
To mimic the Standard Model, we gauge the $(SU(2) \times U(1))_{vis}$ factor, and assign our toy Standard Model fields to the representations:
\begin{eqnarray}
&& h_u = (2_+, 2_+), \qquad h_d = (2_-, 2_-) \nonumber \\
&& q_L = (2_+, 2_-), \qquad u_L^c = (1_{-2}, 1_0) \nonumber \\
&& \ell_L = (2_-, 2_+), \qquad e_L^c = (1_{+2}, 1_0)
\end{eqnarray}
and also replicate for multiple generations.
In addition, the gauge fields of the toy Standard Model transform in the representations $(3_0, 1_0)$ plus $(1_0, 1_0)$. Note that
all mixed gauge and gauge/gravity anomalies cancel for appropriate numbers of ``generations''.
We couple our probe sector to the toy Standard Model sector via the fields $h_u$ and
$\ell$ which transform in the representations $(2_{+},2_{+})$ and $(2_{-},2_{+})$. Suppressing
the gauge theoretic indices with respect to both $(SU(2) \times U(1))_{vis}$ and $(SU(2) \times U(1))_{hid}$,
we can then write the interaction between the probe and visible sector as:
\begin{equation}
W_{SM \otimes D3} = \sum_{i} \lambda^{h}_{(i)} h_{(i)} \widetilde{Q}^{med}
\widetilde{Q}^{hid} + \sum_{g} \lambda^{\ell}_{(g)} \ell_{(g)} Q^{med} \widetilde{Q}^{hid},
\label{toy-interaction}
\end{equation}
where the sums are over the ``generations'' of Higgs and lepton
fields.

\subsection{Toy Models at Strong Coupling}

In much of this paper our main focus will be on strongly coupled probes of E-points. In contrast to the case of the $D_4$ probe theory, here, we do not
have a Lagrangian description. Even listing the microscopic degrees of freedom is more challenging for these $\mathcal{N} = 1$ theories, because
they are defined in terms of $\mathcal{N} = 1$ deformations of strongly coupled $\mathcal{N} = 2$ theories. To make further progress, we shall therefore appeal to various toy models which appear to capture the qualitative features of these theories.

As a starting point, we begin by discussing some additional qualitative features of the $\mathcal{N} = 2$
$E_8$ Minahan-Nemeschansky theory. In this case, we can appeal to a picture of the physical states in terms of
$(p,q)$ strings and their junctions, as well as the associated Seiberg-Witten curve, which is the F-theory geometry
probed by the D3-brane. First consider the string junction picture. So long as we work at $\vert z \vert \gg M_{\ast}$, we can reliably characterize
the spectrum of strings stretching from the D3-brane to the E-type seven-branes in terms of networks of string junctions. The basic idea is that the various operators of the $E_8$ $\mathcal{N}=2$ Minahan-Nemeschansky theory fill out representations of $E_8$. This includes the operators $\mathcal{O}$ in the adjoint of $E_8$, as well as many additional operators in higher dimension representations. For example, the $248$ and the $3875$ of $E_8$ decompose into irreducible representations of $SU(5)_{GUT} \times SU(5)_{\bot}$ as \cite{Slansky}:
\begin{align}
E_{8}  & \supset SU(5)_{GUT}\times SU(5)_{\bot}\\
248  & \rightarrow(1,24)\oplus(24,1)\oplus(5,\overline{10})\oplus(\overline
{5},10)\oplus(10,5)\oplus(\overline{10},\overline{5})\\
3875  & \rightarrow(1,1)\oplus(24,24)\oplus(1,24)\oplus(24,1)\oplus(1,75)\oplus(75,1)\\
& \oplus(5,\overline{10})\oplus(\overline{5},10)\oplus(10,5)\oplus
(\overline{10},\overline{5})\\
& \oplus(5,\overline{15})\oplus(\overline{5},15) \oplus(\overline{15},\overline{5}) \oplus (15,5) \\
& \oplus(5,\overline{40})\oplus(\overline{5},40)\oplus(40,5)\oplus
(\overline{40},\overline{5})\\
& \oplus(10,45)\oplus(\overline{10},\overline{45})\oplus(45,\overline
{10})\oplus(\overline{45},10).
\end{align}
These additional representations are realized in the probe sector as additional multi-prong string junctions between the D3-brane and the $E_8$ seven-brane. The precise correspondence between string junctions and weights of $E_8$ representations has been worked out in \cite{DeWolfe:1998zf, DeWolfe:1998bi}.
The important point for us is that the decomposition of these higher dimensional $E_8$ representations contains additional states with the same $SU(5)_{GUT} \times SU(5)_\bot$ representation content as descendants from the $248$.

As we pass to smaller values of $z$, one can expect this qualitative picture to receive various corrections, due to the breakdown of the classical string picture. Nevertheless, we still expect some qualitative link with the string junction picture. Indeed, in the case of $\mathcal{N} = 2$ supersymmetric theories, we can use the Seiberg Witten curve to read off the BPS spectrum of particles on the Coulomb branch of the theory. Moreover,
deformations of this curve can be interpreted as deformations of the $\mathcal{N} = 2$ theory of the form (in $\mathcal{N} = 1$ language):
\begin{equation}
\delta L = \int d^{2} \theta Tr_{E_8}(\delta m \cdot \mathcal{O})
\end{equation}
for $\delta m$ an element in the adjoint of $E_8$. Periods of the Seiberg-Witten differential then translate to masses of states which are
electric, magnetic, or dyonic under the $U(1)$ gauge group factor. We can roughly view these as the analogues of the $Q$'s appearing in the weakly coupled $D_4$ probe theory. In contrast to the case of the $D_4$ probe theory, however, the stringy realization will typically contain many $Q$'s with similar mass. Indeed, this follows from the fact that at strong coupling, both F1-strings, D1-strings, and string junctions will all have comparable mass, and can stretch from the D3-brane to the seven-brane. The main complication here is that as opposed to one set of $Q \oplus \widetilde{Q}$'s associated with a given operator $\mathcal{O}$, we can now expect more complicated relations of the form:
\begin{equation}
\mathcal{O} \sim \sum_{i,j} Q_{i} \widetilde{Q}_{j} + \cdots
\end{equation}
where $i$ and $j$ are generic labels which could potentially run over a large list of electric and magnetic states. In what follows, we shall adopt a ``meson approximation'' where we associate the $\mathcal{O}$'s with quadratic terms. In particular, the mass of corresponding meson will then be roughly the same as that of the electric and magnetic constituent particles. This is quite analogous to what would happen in QCD if we consider mesons composed of constituent quarks with mass comparable to the scale of chiral symmetry breaking.

Similar qualitative considerations also hold for the $\mathcal{N} = 1$ $\Phi$-deformed theories. Here, we must exercise additional caution, because even if we had managed to give a characterization of the original UV degrees of freedom, in the IR, this description may be cumbersome. Nevertheless, we still expect there to be light electric and magnetic states which will affect the running of $\tau$ for the D3-brane. Indeed, the existence of the $\mathcal{N} = 1$ curve, and the fact that it is typically a non-trivial function of the Coulomb branch parameter and deformations of the curve provides evidence for the existence of light charged states, as well as their composites.

To take account of these different types of states, we shall refer to all states which transform non-trivially under $SU(5)_{GUT}$ as ``$3-7_{vis}$'' strings. Those which transform as singlets we shall refer to as ``$3-7_{hid}$'' strings. Finally, we shall also refer to the Standard Model chiral matter as ``$7_{vis}-7_{hid}$'' strings. In what follows, we shall sometimes use the notation:
\begin{align}
\text{SM states}  & : \Psi_{SM} \label{SMdef}\\
3-7_{vis}\text{ states}  & : Q_{med} \\
3-7_{hid}\text{ states}  & : Q_{hid}
\end{align}
to indicate respectively Standard Model states, mediators between the visible and hidden sector states, and the hidden sector.
As noted above, this is a slight abuse of notation, because a given state will involve multi-prong string junctions.

\section{Mass Hierarchies and Monodromy\label{sec:MASS}}

In realistic model building applications, we must break conformal symmetry in the probe sector. This is because the probe contains fields charged under the Standard Model gauge group, which we do not wish to keep at low-energy. Conformal symmetry breaking introduces a characteristic mass scale $M_{\cancel{CFT}}$ for the probe sector. Below this scale, the probe sector consists of a collection of strongly coupled particles of various masses.

A geometrically natural way to break conformal symmetry is to displace the D3-brane away from the Yukawa point \cite{Funparticles} which we accomplish through a non-zero vev for the operator $Z$ of the probe theory. We show using the holomorphic geometry of the compactification
that in vacua where $\langle Z \rangle \neq 0$ and $\langle Z_1 \rangle = \langle Z_2 \rangle = 0$, states charged under the Standard Model gauge group ($3-7_{vis}$ strings) have characteristic mass of order $M_{\cancel{CFT}}$ while states charged under other seven-branes nearby the Yukawa point ($3-7_{hid}$ strings) have masses which are hierarchically lower. We refer to these two mass scales as:
\begin{align}\label{monoseesaw}
M_{med} & \sim M_{\cancel{CFT}} \times \text{K\"ahler}\\
M_{hid} & \sim M_{\cancel{CFT}} \cdot \left( \frac{M_{\cancel{CFT}}}{M_{\ast}}\right)^{n} \times \text{K\"ahler}
\end{align}
where $n$ is an integer specified by the Jordan block type of $\Phi(0,0)$. Here, we have separated out two contributions. The first factor of each line
corresponds to what can be computed based on F-term considerations. In addition, there is a correction due to the K\"ahler potential of the theory. At strong coupling, we do not know how to compute this factor, but we can still perform a crude estimate, and we shall argue that it does not affect the main estimates we shall perform. This hierarchical separation in mass scales arises from a monodromic seesaw mechanism, and naturally suggests the $3-7_{hid}$ states as a potential dark matter candidate which are generated through the decay of much heavier $3-7_{vis}$ states.

The rest of this section is organized as follows. In order to gain more intuition into the monodromic seesaw mechanism of \eqref{monoseesaw}, we first discuss a weakly coupled analogue of this effect for the probe $D_{4}$ theory. Next, we show that even without a Lagrangian description, holomorphy considerations and homogeneity of the $\mathcal{N} = 1$ geometry leads to the \textit{same} relation. For completeness, we also discuss how a non-zero vev of $Z$ far below $M_{\ast}$ is compatible with the general form of the probe sector coupled to the flux induced superpotential $W_{flux}$. This also leads to mass scales for the position modes $Z_{i}$. We next discuss how the estimates we have
given are compatible with supersymmetry. Finally, we discuss some
additional properties of the light GUT singlets of such theories.

\subsection{Monodromic Seesaw: Weakly Coupled Case}

Though our ultimate interest is in the mass spectrum of the strongly coupled probe theories, we first discuss the representative mass scales for $\Phi$-deformations of the weakly coupled $\mathcal{N} = 2$ $SU(2)_{D3}$ theory with four flavors.

As a simple example, we consider $\Phi(0,0)$ a nilpotent $2 \times 2$ matrix taking values
in the $(SU(2) \times U(1))_{hid}$ factor of our toy model. This produces a mass term:
\begin{equation}
\delta W_{tile}=\left[
\begin{array}
[c]{cc}%
Q_{1}^{hid} & Q_{2}^{hid}%
\end{array}
\right]  \left[
\begin{array}
[c]{cc}%
0 & M_{\ast}\\
0 & 0
\end{array}
\right]  \left[
\begin{array}
[c]{c}%
\widetilde{Q}_{1}^{hid}\\
\widetilde{Q}_{2}^{hid}%
\end{array}
\right]  =M_{\ast}Q_{1}^{hid}\widetilde{Q}_{2}^{hid}
\end{equation}
Integrating out the heavy quarks, the effective superpotential for the $3-7_{vis}$ and $3-7_{hid}$ strings is:
\begin{equation}
W_{eff}=\underset{a}{\sum}\sqrt{2}Q^{med}_{a}\varphi\widetilde{Q}^{med}_{a}
-2 Q^{hid}_{2}\frac{\varphi^{2}}{M_{\ast}} \widetilde{Q}^{hid}_{1}.
\end{equation}
We observe that when $\varphi$ develops a non-zero vev and conformal symmetry is broken,
the characteristic mass for the $3-7_{vis}$ strings is far higher than that of the $3-7_{hid}$
strings. Indeed, we have:%
\begin{align}
M_{med}  &  \sim M_{\cancel{CFT}}\\
M_{{hid}}  & \sim M_{\cancel{CFT}}\cdot\left(  \frac{M_{\cancel{CFT}}%
}{M_\ast}\right)  . \label{M37}%
\end{align}
In other words, a seesaw hierarchy is generated for the $3-7_{hid}$ string masses.
We shall sometimes write these mass terms as:
\begin{equation}
W_{masses} = M_{med} Q_{med} \widetilde{Q}_{med} + M_{{hid}} Q_{hid}
\widetilde{Q}_{hid}
\end{equation}

One can also consider the more realistic situation where $\Phi$ also contains some $Z_1$ and $Z_2$ dependence as well. So long as these $Z_i$
do not develop large non-zero vevs, the same estimates will apply in this situation as well. When $Z_i$ does develop a vev, there is another source of
mass terms for the quarks of the toy model. If these vevs are the dominant source of SCFT breaking, this will lift some of the quark masses to at least the
SCFT breaking scale, and some up to the higher scale $M_{\ast}$. In what follows we shall therefore assume that the $Z_i$ have zero vev. As we explain
later, this assumption is well-justified because when treated as canonically normalized modes, the $Z_i$ have higher masses than $Z$.

\subsection{Monodromic Seesaw: Strongly Coupled Case}

Let us now turn to the monodromic seesaw of the strongly coupled case. Strong coupling can occur in various ways. For example, in the $D_4$ probe theory, we can increase the value of the $SU(2)_{D3}$ gauge coupling. Even so, by a judicious choice in the $\mathcal{N} = 2$ theory, we can appeal to a weakly coupled microscopic description in the UV. In the case of the strongly coupled Minahan-Nemeschansky theories, even this is unavailable.

This is an important subtlety, because if we are to interpret the various relevant deformations of the theory in terms of a characteristic mass scale,
we must have some notion of the mass scales for the underlying degrees of freedom. What allows us to make some progress is the existence of the $\mathcal{N} = 1$ curve for such systems. As noted for example in \cite{Intriligator:1994sm, FCFT},
even in the case of $\mathcal{N} = 1$ deformations, we still retain a characterization of the holomorphic gauge coupling $\tau$ associated with the strongly coupled $U(1)$ gauge theory defined on the Coulomb branch of the theory. This characterization allows us to track the dependence of $\tau$ as a function of the Coulomb branch parameter $z$, as well as various deformations $m$ of the $\mathcal{N} = 1$ curve, which we can loosely think of as ``mass deformations'' of the theory. The reason these further deformations are strictly speaking not quite mass deformations is that in general, they will not have scaling dimension one. For example, in the $\Phi$-deformed theories, the dimension of some of the $\mathcal{O}$'s will be less than two, and some will be greater than two. This means that in deforming the theory by a term such as:
\begin{equation}
\delta L = \int d^{2} \theta Tr_{E_{8}}(\delta m \cdot \mathcal{O})
\end{equation}
the parameter $\delta m$ will in general not have scaling dimension one in the UV.

However, in the original $\mathcal{N} = 2$ probe theories, all of the $\mathcal{O}$'s parameterizing the Higgs branch have scaling dimension exactly two, and so these deformations can still be thought of as mass deformations. What we are going to do in this section is to estimate the characteristic size of this holomorphic F-term deformation of dimension one, and interpret it as a characteristic mass scale for the $\mathcal{N} = 1$ theory. In other words, we are going to extract the characteristic scaling of the F-term deformation. This will neglect various scaling dependence from D-terms, which we discuss afterwards.

Our starting point is the original $\Phi$-deformed theory deformed onto the Coulomb branch. We assume that $z \neq 0$ and $\langle Z_1 \rangle = \langle Z_2 \rangle = 0$. It is therefore enough to consider the mass scales associated with $\Phi$ a direct sum of constant nilpotent Jordan blocks. We assume the only non-zero entries of $\Phi$ are in the $SU(5)_{\bot}$ factor of $SU(5)_{GUT} \times SU(5)_{\bot} \subset E_8$. By separating the D3-brane from the intersecting seven-branes, all of the string states will pick up some mass. To determine this mass, we shall consider \textit{another} theory obtained by adding to $\Phi$ a small perturbation by a term $\delta M$:
\begin{equation}
\Phi(0,0) \rightarrow \Phi(0,0) + \delta M
\end{equation}
such that $\Phi(0,0) + \delta M$ is no longer nilpotent. For $\delta M$ sufficiently small, the two theories deformed onto the Coulomb branch will have very similar mass spectra. However, as we increase $\delta M$, this perturbation will exercise more of an effect on the low energy spectrum of the theory, inducing a flow to another theory.

Our task is to determine how large the entries of $\delta M$ can become while keeping the spectrum of the theory close to the original $\Phi$-deformed theory.
To accomplish this, we exploit the connection between the holomorphic $\mathcal{N} = 1$ curve, and its dependence on $z$ and the mass deformations of the original $\mathcal{N} = 2$ theory. Characteristic mass scales of the supersymmetric probe theory correspond to holomorphic expressions in the mass deformations which are also invariant under the flavor symmetries of the system. The crucial point for our present considerations is that even with only $\cN=1$
supersymmetry, with underlying ${\cal N}=2$ supersymmetry, we can characterize the form of the holomorphic mass
deformations. As explained for example in \cite{Intriligator:1994sm, FCFT}, the
value of the holomorphic gauge coupling $\tau$ defined on the Coulomb branch
depends holomorphically on the mass deformations, as well as the Coulomb
branch parameter $z = \langle Z \rangle$. These mass deformations must
in turn form singlets under the associated flavor symmetry in order to appear
in $\tau$. Geometrically, this can be stated as the condition that the
homogeneity of the mass parameters is related to the scaling of the Coulomb
branch parameter $z$.

Denote by $f_{(n)}(m_{i \overline{j}})$ the degree $n$ polynomial homogeneous in the matrix of mass parameters $m_{i \overline{j}}$ taking values in $SU(5)_{GUT} \times SU(5)_{\bot}$. In the case of a nilpotent $\Phi$-deformation, there is no non-trivial holomorphic invariant we can form. This means
that the only dependence we can consider is a function of $z$ alone. Once we break the SCFT, we expect that in the infrared, all degrees of freedom will develop a mass. This is what we have seen in the weakly coupled case, where we have also observed a hierarchy of scales. The mass of the
states charged under $SU(5)_{GUT}$ is then clearly up near the SCFT breaking scale.

The case of the states neutral under $SU(5)_{GUT}$ is more subtle. For concreteness,
we phrase our discussion in terms of $\Phi(0,0)$ given by a $2 \times 2$ nilpotent block,
with $\Phi(0,0) + \delta M$ given by:
\begin{equation}
\Phi(0,0) + \delta M =
\left[
\begin{array}
[c]{cc}%
0 & M_{\ast}\\
\delta m & 0
\end{array}
\right]
\end{equation}
We interpret $\delta m$ as the characteristic mass scale of potentially light states.
We will discuss the extension to other nilpotent Jordan block structures later. The
crucial point for us is that in order for $\delta M \neq 0$ to not significantly distort
the mass spectrum from the case $\delta M = 0$, the flavor invariant formed from
$\Phi(0,0) + \delta M$ must scale as a power of $M_{\cancel{CFT}}$. In other words, we obtain the estimate:
\begin{equation}
\delta m \cdot M_{\ast} \sim M_{\cancel{CFT}}^{2}
\end{equation}
or:
\begin{equation}
\delta m \sim M_{\cancel{CFT}} \cdot \left( \frac{M_{\cancel{CFT}}}{M_{\ast}} \right),
\end{equation}
which recovers the monodromic seesaw relation obtained earlier!

We now generalize this discussion to other choices of $\Phi(0,0)$.
Given a degree $n$ homogeneous polynomial
$f_{(n)}(m_{i\overline{j}})$ in the mass parameters, these parameters will
satisfy the relation:%
\begin{equation}
f_{(n)}(m_{i\overline{j}})\sim M_{\cancel{CFT}}^{n}.
\end{equation}
From this, we conclude that when some of these mass parameters enter into the
definition of the $\cN=2$ to $\cN=1$ deformation, the other mass parameters are
then specified as corresponding ratios. This recovers the basic
relation of (\ref{M37}). We also conclude that since no mass
deformations will typically be turned on in the $SU(5)_{GUT}$ sub-block, the
characteristic mass of the $3-7_{vis}$ strings will instead be set by $M_{\cancel{CFT}}$.
Again, this produces a hierarchy of scales.

The precise hierarchy depends on the choice of $\Phi$-deformation we consider.
For example, when $\Phi(0,0)$ consists of one nilpotent $2\times2$ Jordan block and
three $1\times1$ nilpotent blocks, the characteristic mass scales are:%
\begin{equation}
\Phi(0,0)=\left[
\begin{array}
[c]{ccccc}%
0 & 1 &  &  & \\
0 & 0 &  &  & \\
&  & 0 &  & \\
&  &  & 0 & \\
&  &  &  & 0
\end{array}
\right]  \Longrightarrow\text{Mass Scales}=\left\{  M_{\cancel{CFT}}\text{, }%
M_{\cancel{CFT}}\cdot\left(\frac{M_{\cancel{CFT}}}{M_{\ast}}\right)\right\}
\end{equation}
where we assume $M_{\ast}$ is the scale of the original seven-brane
tilting. As another example, we can consider $\Phi$ given by a $3\times3$
block and two $1\times1$ nilpotent blocks. The characteristic mass scales in
this case are:%
\begin{equation}
\Phi(0,0)=\left[
\begin{array}
[c]{ccccc}%
0 & 1 & 0 &  & \\
0 & 0 & 1 &  & \\
0 & 0 & 0 &  & \\
&  &  & 0 & \\
&  &  &  & 0
\end{array}
\right]  \Longrightarrow\text{Mass Scales}=\left\{  M_{\cancel{CFT}}\text{, }%
M_{\cancel{CFT}}\cdot\left(  \frac{M_{\cancel{CFT}}}{M_{\ast}}\right)  ^{2}\right\}  .
\end{equation}
Thus, depending on the details of the seven-brane tilting, we can realize
different seesaw mass hierarchies.

The precise details of the mass spectrum will certainly depend on the
K\"ahler potential. The basic point, however, is that the mass
of the $3-7_{vis}$ strings is up at the SCFT breaking scale,
while the D3-brane theory still contains far lighter
states such as the $3-7_{hid}$ strings.

What sorts of corrections to the mass spectrum can we expect from the K\"ahler potential?
This involves knowing far more about the strongly coupled $\mathcal{N} = 1$ theory, but a reasonable
first expectation is that there is a notion of canonical normalization for the various fields. For example,
if we work with respect to the ``meson approximation'' mentioned earlier, it is appropriate to treat the $\mathcal{O}$'s
as composites of two elementary fields. In this approximation, we can canonically normalize each of these modes. This leads to a shift in
the scaling dimension of the form:
\begin{equation}
\mathcal{O} \sim M_{\cancel{CFT}}^{\Delta_{IR}(\mathcal{O}) - \Delta_{UV}(\mathcal{O})}\mathcal{O}_{can}
\end{equation}
where we have canonically normalized according to the ``UV scaling dimension'' of the operator. Note that this is expected
to be a small effect when the IR and UV dimensions are sufficiently close. Obtaining more exact results is not particularly clear either,
and so in what follows we shall ignore such subtleties, treating the holomorphic mass deformations as an approximate guide to the relative
mass spectra.

In this approximation, we have that the hidden sector $3-7_{hid}$ strings have characteristic mass:
\begin{equation}
M_{hid} \sim M_{\cancel{CFT}} \cdot \left(\frac{M_{\cancel{CFT}}}{M_{GUT}} \right)^{n}.
\end{equation}
The parameter $n$ is set by the size of the nilpotent Jordan blocks
in $\Phi(0,0)$. In most realistic examples $n = 1,2,3$ \cite{TBRANES}.

The most important feature of this spectrum is that the lightest states of the
probe theory are neutral under $SU(5)_{GUT}$. In particular, this means that
the heavy colored states will eventually decay to much lighter singlets.

\subsection{SCFT Breaking Vacua}

Our discussion so far has assumed that $Z$ has attained an appropriate vev to break the conformal dynamics. The main
thing we now wish to motivate is that the mass scale associated with this vev can naturally be below the scales $M_{\ast}$ and $M_{GUT}$.

Though our ultimate interest is in vacua associated with $\cN=1$ supersymmetry,
this is also the case where we have much less control over the details of the
K\"ahler potential. To illustrate the main points of our discussion, we first
discuss a simplified version with $\cN=2$ supersymmetry. After this, we discuss
the extension to $\cN=1$ systems.

Metastable vacua associated with $\cN=1$ deformations of $\cN=2$ systems have been
studied, for example, in \cite{Ooguri:2007iu, Auzzi:2010kv}. There, the main idea
is to consider $\mathcal{N}=2$ supersymmetric mass deformations, as well as $\mathcal{N}=1$
deformations by the Coulomb branch parameter. A general feature of this analysis is that metastable vacua
with SCFT breaking scale below $M_{\ast}$ and $M_{GUT}$ can indeed be arranged.

In this section, we show that SCFT breaking involving $\mathcal{N}=1$ supersymmetric vacua are also present even in
the limit where all mass deformations are switched off. We begin our discussion by returning to an $\mathcal{N}=2$
superconformal theory with Coulomb branch parameter $Z$. To this system, we imagine adding a small $\mathcal{N}=1$
deformation of the form:%
\begin{equation}
W=M^{3}\left(  \frac{a}{\Delta}\frac{Z}{M^{\Delta}}-\frac{b}{2\Delta}\left(
\frac{Z}{M^{\Delta}}\right)  ^{2}+...\right)  \label{Wz}%
\end{equation}
at a mass scale $M$. Physically, we identify $M$ with the scale associated
with the localized fluxes, which is naturally on the order of the GUT scale \cite{FGUTSNC}.
Here, $a$ and $b$ are dimensionless coefficients and $\Delta$ is the scaling dimension of $Z$.
We assume that quadratic terms involving the other $Z_{i}$ fields can be ignored.

Assuming both $a$ and $b$ are not large, it is legitimate to use the K\"ahler
potential for $Z$ obtained in the supersymmetric case. The metric on moduli
space is conical at the origin, with corresponding K\"ahler potential:
\begin{equation}
K_{\cN=2}=\left(  Z^{\dag}Z\right)^{1 \over \Delta} + \cdots, %
\end{equation}
where, for example, $\Delta=6$ for the $E_{8}$ Minahan-Nemeschansky theory. The
effective Lagrangian for the $Z$-field therefore contains the terms:%
\begin{equation}
L_{\rm eff}(Z)=g_{Z\overline{Z}}\left\vert \partial_{\mu}Z\right\vert
^{2}+g^{Z\overline{Z}}\left\vert \partial_{Z}W\right\vert ^{2}.
\end{equation}
To analyze the vacua of this system, it is convenient to introduce a
canonically normalized field $\widetilde{Z}$. Since we are working at the
level of the classical Lagrangian, we can perform the field redefinition
$\widetilde{Z}^{\Delta}=Z$. In terms of this variable, we have:%
\begin{align}
L_{eff}(\widetilde{Z})  &  =\left\vert \partial_{\mu}\widetilde{Z}\right\vert
^{2}+M^{4}\Big\vert a\left(  \frac{\widetilde{Z}}{M}\right)  ^{\Delta
-1}-b\left(  \frac{\widetilde{Z}}{M}\right)  ^{2\Delta-1}\Big\vert ^{2}\\
&  =\left\vert \partial_{\mu}\widetilde{Z}\right\vert ^{2}+M^{4}\Big\vert
\frac{\widetilde{Z}}{M}\Big\vert ^{2\Delta-2}\Big\vert a-b\left(
\frac{\widetilde{Z}}{M}\right)  ^{\Delta}\Big\vert ^{2}.
\end{align}
From this, we observe that there are two branches of supersymmetric vacua. One
is located at the origin of the field space, $\widetilde{Z}=0$. On the other hand,
we also observe that there is another branch given by:%
\begin{equation}
\widetilde{Z}=M\left(  \frac{a}{b}\right)  ^{1/\Delta}.
\end{equation}
Typically, it is appropriate to ignore this latter branch of vacua. The reason
is that when $a$ and $b$ are comparable, allowing field ranges of order $M$ means
it is then also appropriate to include an infinite set of additional terms of the form:%
\begin{equation}
\left(  \frac{\widetilde{Z}}{M}\right)  ^{n\Delta}\sim\left(  \frac{a}%
{b}\right)  ^{n}.
\end{equation}
Note however that, when $a/b$ is small, it is appropriate to
restrict attention to the leading order terms given by $n=1$ and $n=2$. In
particular, we see that the vev of the normalized field $\widetilde{Z}$ is accurately approximated
by including just the first two terms.

In order to work in this controlled approximation, it is therefore necessary
for $a$ to be hierarchically smaller than $b$. Indeed, in the
limit where $a=0$, we observe that the $W$ of equation (\ref{Wz}) enjoys the
$\mathbb{Z}_2$ symmetry, $Z\rightarrow-Z$. Small $a/b$ is therefore
technically natural.

This sort of condition is also geometrically natural. The
local configuration of seven-branes is controlled by the
spectral equation \cite{Hayashi:2009ge,DWIII,TBRANES}:
\begin{equation}
P_{\Phi}(z)=\det\left(  z-\Phi(z_{1},z_{2})\right)  =0
\end{equation}
where $\Phi(z_{1},z_{2})$ is a $5\times5$ matrix valued in $SU(5)_{\bot}$. For
generic $\Phi$, the polynomial $P_{\Phi}(z)$ is a quintic in $z$ with Galois
group $S_{5}$, the permutation group on five letters. Detailed model building
considerations, however, lead to significantly different structures for
$\Phi(z_{1},z_{2})$. For example, in a $Dih_{4}$ monodromy scenario, $P_{\Phi
}(z)$ can factorize into a linear and a quartic polynomial of the form:%
\begin{equation}
P_{\Phi}(z)=z(z^{4}+\alpha z^{2}+\beta)=0.
\end{equation}
From this, we observe that the roots are invariant under $z\mapsto-z$. This
provides a geometric realization of the $\mathbb{Z}_2$ symmetry $Z\rightarrow-Z$. Similar
considerations hold for other choices of monodromy groups. In
this more general situation, it is natural to expect both $a$ and $b$ to be
small parameters. Note that in such situations, the ratio $a/b$ can also
be small, because $a$ and $b$ multiply different powers of the $Z$ field.

In fact, this symmetry is only expected to be a feature of the
geometry near the E-point. In a global compactification, we expect that the branes will have a
more generic profile, which will destroy the symmetry which is present near
the Yukawa point. This in turn means that we can expect violations of the
symmetry, determined by the ratio $M/M_{\ast}$ since $M_{\ast}$ sets the
characteristic mass scale specified by the global completion of model. We
therefore conclude that $a$ and $b$ will be given by powers of $M/M_{\ast}$.
The precise power depends on details of the compactification. We shall
therefore treat the ratio $a/b$ as a tunable small parameter:%
\begin{equation}
\frac{a}{b}\equiv r .
\end{equation}

Let us now turn to the $\Phi$-deformed $\cN=1$ theories. Here we have far less
information available as to the exact behavior of K\"ahler potential. This in
turn limits our ability to determine the exact properties of such vacua.
However, we can still use holomorphy as a general guide. Indeed, for
supersymmetric vacua away from the origin of moduli space, it is appropriate
to impose the F-term equation of motion:%
\begin{equation}
\partial_{Z}W=0
\end{equation}
from which we deduce:%
\begin{equation}
\left\langle Z\right\rangle =M^{\Delta} \ r.
\end{equation}
Let us comment that in the limit where $r $ is very small, the
corresponding vev $\left\langle Z\right\rangle $ is far below $M^{\Delta}$. In
particular, this means that it is appropriate to evaluate the dimension of $Z$
as given by its IR value:%
\begin{equation}
\left\langle Z\right\rangle =M^{\Delta_{IR}}\  r .
\end{equation}
When $r$ is not infinitesimal, however, the value of $\Delta$ will
be somewhere in between its IR and UV values. In what follows we shall
typically assume that $r$ is sufficiently small to avoid such subtleties.

In the more realistic case of a full compactification, we should also include terms quadratic in the other $Z_i$ fields. Indeed, the form of the flux induced superpotential consistent with the flavor physics considerations of \cite{FGUTSNC} is of the form:
\begin{equation}
W_{flux}=\sum_{a,b,c=1}^3 ( m_{ab}Z_{a}Z_{b}+\lambda_{abc}Z_{a}Z_{b}Z_{c} ),
\label{WFLUX}%
\end{equation}
where $a$, $b$ and $c$ are indices running from $1$ to $3$. In typical monodromic
scenarios, the scaling dimension of $Z_1$ and $Z_2$ is somewhat larger than one,
but less than $3/2$ \cite{FCFT}. Hence, such terms correspond to an additional
potentially relevant $\mathcal{N} = 1$ deformation. In what follows we will
treat such terms as a small perturbation. This is technically natural, because adding such terms breaks additional flavor asymmetries of the probe sector \cite{FCFT}. Though not the case of interest in this paper, it is of independent
interest to study what happens to the probe theory when we do not treat $W_{flux}$ as
a small perturbation. See Appendix B for additional discussion of the
fixed point obtained from including such a deformation.

For our present purposes, the main point is that $Z_1$ and $Z_2$ have lower scaling dimension than $Z$. We interpret these modes as
characterizing motion of the D3-brane. The higher the scaling dimension of the fields, the lighter the corresponding mode.
To see this, let us suppose that at some scale $m$, we add by hand mass terms involving the $Z_i$ fields:
\begin{equation}
\delta W = m \cdot (Z_{i})^{2}.
\end{equation}
We canonically normalize by introducing a mode $\widetilde{Z}_{i}$ defined via
$Z_{i} = M_{\cancel{CFT}}^{\Delta(Z_{i}) - 1} \widetilde{Z}_{i}$. Substituting back, we
learn that the mass of the canonically normalized mode is:
\begin{equation}
M_{\widetilde{Z}_{i}} = m \cdot \left( \frac{M_{\cancel{CFT}}}{M_{\ast}} \right)^{2 \Delta (Z_{i}) - 2}
\end{equation}
for the field of dimension $\Delta(Z_{i})$. In other words, the higher the scaling dimension, the lower the mass. Let us note that this analysis
is certainly \textit{not} justified for a generic field of a strongly coupled SCFT. Here we are exploiting the additional stringy
insight that these modes are associated with motion of the D3-brane.

\subsection{Supersymmetry Breaking Effects}\label{ssec:SUSYbreak}

The main reason we have been able to obtain an estimate of the various mass scales of the probe sector
is due to $\mathcal{N} = 1$ supersymmetry. Since supersymmetry must also be broken,
it is important to check that such effects do not eliminate our candidate light states.
The effects of supersymmetry breaking can be quite model-dependent. Our
aim in this section will therefore be to argue that there exist scenarios where
our estimate of the hierarchical $3-7_{hid}$ masses due to the monodromic seesaw mechanism persist.

Supersymmetry breaking effects can be expected to generate corrections to the scalar masses of the hidden sector. For example,
if the $3-7_{vis}$ strings develop a non-supersymmetric mass spectrum they will induce gauge mediation via the dark $U(1)_{D3}$
to the hidden sector \cite{HVW}. There are additional contributions to the scalar masses from various higher dimension operators, such as the Peccei-Quinn (PQ) deformation of the gauge mediated supersymmetry breaking scenario considered in
\cite{HVGMSB, HKSV, Heckman:2010xz, EPOINT, Pawelczyk:2010xh}. By comparison, there
are fewer ways for fermions to develop a mass. Indeed, the mass scale $M_{{hid}}$ can be thought of as a supersymmetric $\mu$-term. It is therefore reasonable to expect that even in the strongly coupled setting, our previous estimate $M_{{hid}}$ is accurate for the fermionic states of the theory. Interpreting the dark matter as $3-7_{hid}$ strings, this also means that we predict these modes to be fermionic.\footnote{Let us note that in
the specific context of the models considered in \cite{EPOINT} where the $\mu$-term is also generated at the E-point, generating a weak scale $\mu$-term in
the visible sector might also induce a similar size $\mu$-term in the hidden sector, though this depends on non-chiral details of the probe. One way to sidestep this issue is to consider scenarios where $\mu$ of the visible sector is generated at a different point of the geometry. Even if we still choose to generate the $\mu$-term as in \cite{EPOINT}, it is not clear that at a practical level this is much of a concern. The reason is that in the scenarios we consider, the mass scale of the dark matter will be on the order of $10$ GeV, while the value of the $\mu$-term is perhaps one to two orders of magnitude removed from this. Given some of the ambiguities in extracting the K\"ahler potential from these strongly coupled theories, it is not clear this is an issue.}

The non-supersymmetric correction to the scalar masses of the probe sector is actually a welcome feature, because such contributions will sometimes induce tachyonic modes which cause the $3-7_{hid}$ modes to roll away from the origin. When this occurs, the theory will develop a mass gap for the hidden $U(1)_{D3}$, which we refer to as $M_{U(1)_{D3}}$. In the weakly coupled regime, this is simply achieved by the usual Higgs mechanism. At strong coupling, however, the phases of such strongly coupled systems can be quite complex. A full study would go beyond the scope of the present paper, and we will relegate the analysis to future work.

The exact mass of the hidden $U(1)_{D3}$ depends on details of this non-Lagrangian theory, so giving an exact value is not possible with present methods. A reasonable expectation is that the masses of the $3-7_{hid}$ scalars will typically set the mass of this $U(1)_{D3}$. Since there is already a supersymmetric mass term for the $3-7_{hid}$ strings, this means that the characteristic vev for such modes, and hence the mass of the hidden $U(1)_{D3}$ is expected to at least be on the order of $M_{hid}$, though this could be significantly higher, depending on the details of the supersymmetry breaking scenario.

It is also possible to consider scenarios where the mass of the $U(1)_{D3}$ is lighter than the $3-7_{hid}$ strings. This can occur in scenarios where the mass of the $U(1)_{D3}$ is primarily driven by kinetic mixing with the visible sector, which induces a small effective Fayet-Iliopoulos (FI) parameter in the hidden sector (see for example \cite{Morrissey:2009ur}). For kinetic mixing of order $\kappa_{mix} \sim 10^{-3}$, this can induce masses for the hidden sector $U(1)_{D3}$ on the order of a few GeV or less, so this is also a possibility to consider. Since we do not know that much about the strongly coupled $U(1)_{D3}$ sector anyhow, for now we shall treat the mass $M_{U(1)_{D3}}$ as a parameter which can in principle range within a few orders of magnitude of the dark matter mass. Our expectation is that phenomenological constraints from either cosmology or other model building considerations will further constrain this value.

\section{The Decay of Heavy States\label{sec:DECAY}}

In the previous section we studied the spectrum of $3-7_{vis}$ and $3-7_{hid}$ strings
once the D3-brane is displaced away from the Yukawa point. The important
point from this analysis is that the $3-7_{vis}$ states are far heavier than their
singlet counterparts. It follows from this that the $3-7_{vis}$ strings will decay to lighter
states.

The $3-7_{vis}$ strings are charged under both the $7_{vis}$-brane and the D3-brane, and so their decay products
must conserve these charges. Note, however, that the only light modes involving $7_{vis}$ strings
are Standard Model states. From this we conclude that the $3-7_{vis}$ strings will always decay to
Standard Model states and hidden sector $3-7_{hid}$ GUT singlets. Though we cannot give a microscopic description of the various
decay processes, it is convenient to work in terms of this heuristic picture. The actual decays will typically involve
states which are $U(1)_{D3}$ neutral, and so will also include ``spectator quarks'' such
as the light $3-7_{hid}$ strings.  A helpful analogy is to QCD, where we can imagine
that the decay of a $3-7_{vis}$ state ``hadronizes'' to many
soft $3-7_{hid}$ states in addition to a few Standard Model states. See figure \ref{chan-paton} for a depiction of such decay processes.
This is not much of a complication for cosmology, because most of these soft states will annihilate away to radiation.

Our aim in this section will be to estimate the net amount of baryon number violation expected from the decay
of the heavy $3-7_{vis}$ states. Since the D3-brane probe theory is vector-like, it will also contain $7_{vis}-3$ strings with opposite orientation to the $3-7_{vis}$ states. These modes carry conjugate quantum numbers to the $3-7_{vis}$ strings. Because the matter content of the Standard Model is chiral, the coupling to Standard Model fields will be quite different for the $3-7_{vis}$ states and $7_{vis}-3$ states. It therefore suffices to focus on the decay of the $3-7_{vis}$ states. The net CP violation in decays of these heavy modes is given by
comparing the decay rate of $3-7_{vis}$ states and the decay of their anti-matter counterparts to
CP conjugate final states. Given a decay channel for $3-7_{vis}$ states, the amount of CP violation in the decay of a $3-7_{vis}$ string
to $J_1,...,J_n$ soft $3-7_{hid}$ states and $K_{1},...,K_{m}$ Standard Model states is:
\begin{equation}
\varepsilon^{IJK}_{CP} \equiv \frac{\Gamma( Q^{I}_{med} \rightarrow  Q^{\overline{J}_1 \dag}_{hid} \cdots Q^{\overline{J}_n \dag}_{hid} \otimes \Psi^{\overline{K}_1 \dag}_{SM} \cdots \Psi^{\overline{K}_{m} \dag}_{SM} )- \Gamma( Q^{\overline{I} \dag}_{med} \rightarrow Q^{J_1}_{hid} \cdots Q^{J_n}_{hid} \otimes \Psi^{K_1}_{SM} \cdots \Psi^{K_m }_{SM} )}{\Gamma( Q^{I}_{med}  \rightarrow any^{\dag}  ) + \Gamma( Q^{\overline{I} \dag}_{med} \rightarrow any )}.
\end{equation}
where the denominator is a sum of the two inclusive decay rates. Assigning baryon number $b_{K}$ to the Standard Model state $\Psi^{K_1}_{SM} \cdots \Psi^{K_{m}}_{SM}$, the net baryon asymmetry created in the visible sector from the decay of the $I^{th}$ $3-7_{vis}$ string is then given by summing over all final states, weighted by the overall baryon number:
\begin{equation}
\Delta B_{vis}^{I} = \sum_{J,K} b_{K} \cdot \varepsilon^{IJK}_{CP}.
\end{equation}
As a point of notation, in writing decay amplitudes we shall not distinguish in
our discussion between the components of a chiral superfield, and the chiral
superfield itself, since the statistics of the various states should be clear.

Our aim in this section will be to analyze in more detail the decay of these heavy states.
The decay of $3-7_{vis}$ strings and the estimate of C and CP violation shares some
similarities with both thermal leptogenesis and ordinary GUT baryogenesis, since it
involves the decay of a heavy state to lighter states. Indeed, in these scenarios as
well as ours, the resulting C and CP asymmetry is generated through interference terms.\footnote{Though
there are other heavy states of the probe theory besides the $3-7_{vis}$ strings, it is sufficient to focus on
the decay of probe theory states charged under $SU(5)_{GUT}$. The reason is that the decay
of heavy modes uncharged under the Standard Model gauge group does not generate a
sizable baryon asymmetry. Indeed, since they are singlets under the Standard Model,
these heavy modes dominantly couple to other states of the probe theory.
Likewise, seven-brane states which are Standard Model singlets but are charged under a $7_{hid}$-brane
will have dominant decay modes to the probe theory.\ These decay processes are governed by a vector-like spectrum and interactions because the D3-brane probe is defined as an $\mathcal{N} = 1$ deformation of an $\mathcal{N} = 2$ theory. This makes it difficult to generate a matter asymmetry,
though one might speculate that the $\theta$-angle of the hidden sector $U(1)$ gauge theory may
help with generating an asymmetry.}

A novel feature of our matter-genesis scenario is that the mechanism requires strong coupling of $\tau \sim O(1)$
for the probe D3-brane sector. We show that at weak coupling, the decay of heavy $3-7_{vis}$ states
fails to generate enough baryon asymmetry. At strong coupling, however, fundamental string states as
well as $(p,q)$-string states and their string junctions all participate in the decay process, producing an infinite tower of ``dyon resonance states''. These resonance states and their interactions are then the source of interference terms in the decay amplitudes.

The rest of this section is organized as follows. To frame the discussion to follow, we first illustrate in the case of the weakly coupled $D_4$ theory that the decay of heavy $3-7_{vis}$ states made from \textit{perturbative} string states fails to generate enough CP violation. We show that this obstacle can be overcome once we include a generational structure for these modes. After this, we illustrate
how this generational structure is automatically present at strong coupling, due to the similar masses of electric, magnetic and dyonic states. We then estimate the decay rate of heavy states at strong coupling.

\subsection{The Limitations of Weak Coupling}

Before proceeding to the realistic situation, we first consider the probe sector at weak coupling and show that there
are simply not enough degrees of freedom to achieve the required C and CP violation in this case.

Our example is the toy Standard Model discussed in section \ref{sec:SETUP}. The details of monodromy are not so important for this discussion, and so we shall be schematic in indicating the quantum numbers under $(SU(2) \times U(1))_{hid}$.
What is quite important is the structure of the Standard Model to probe sector couplings, as well as the general form of the mass terms. Reproducing the discussion given there, we have:
\begin{equation}
W_{SM \otimes D3} = \sum_{i} \lambda^{h}_{(i)} h_{(i)} \widetilde{Q}_{med} \widetilde{Q}_{hid} + \sum_{g} \lambda^{\ell}_{(g)} \ell_{(g)} Q_{med} \widetilde{Q}_{hid} + M_{med} Q_{med} \widetilde{Q}_{med} + M_{hid} Q_{hid} \widetilde{Q}_{hid},
\end{equation}
where the sum is over the ``generations'' of Higgs fields and lepton fields.

At first glance, this appears quite promising for generating a baryon asymmetry because of the appearance of so many complex couplings. Let us count the C and CP violating phases in this toy model. Fixing the phases of the Standard Model fields based on the visible sector Yukawas, we see that there are additional CP violating phases from the $\lambda$'s. We are still free to rotate the fields $\widetilde{Q}_{med}$, $\widetilde{Q}_{hid}$, so two of the complex phases in $\lambda^{h}_{(i)}$ can be set to zero. The remaining complex phases are C and CP violating. Similar considerations hold for $\lambda^{l}_{(g)}$ and the mass terms $M_{med}$ and $M_{hid}$.

\begin{figure}
[ptb]
\begin{center}
\includegraphics[
trim=0.799793in 8.531429in 0.800619in 0.714558in,
height=2.3531in,
width=6.2742in
]%
{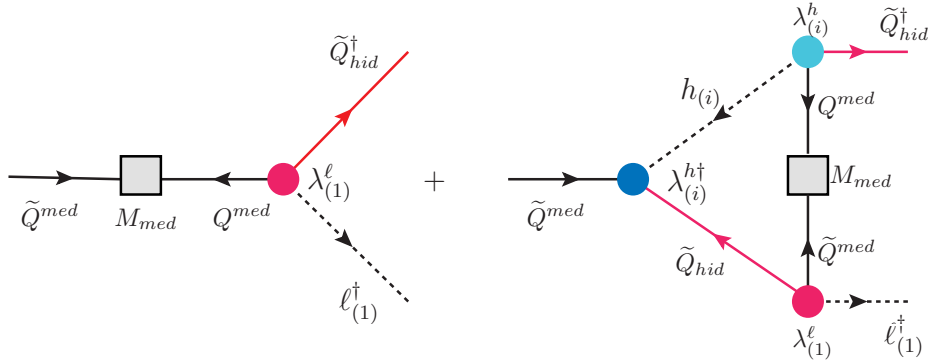}%
\caption{Decay of $\widetilde{Q}^{med}$ to $\widetilde{Q}^{hid}$ and $\ell_{(1)}$. The tree amplitude (left) is mediated from $\widetilde{Q}^{med}$ to $\widetilde{Q}^\dagger_{hid}$ by mass insertion for helicity flip and by Yukawa interaction $\lambda^\ell_{(1)}$. The one-loop amplitude (right) is mediated from $\widetilde{Q}^{med}$ to $\widetilde{Q}^\dagger_{hid}$ by virtual $\ell_{(i)}$ exchange and mass insertion for helicity flip. There are additional one-loop diagrams involving self-energy corrections which we have not drawn here. The Yukawa interactions $|\lambda^h_{(i)}|^2\lambda^\ell_{(1)}$ has the same phase as the tree amplitude.  Consequently, there is no asymmetry between CP conjugate processes.}%
\label{pert-asymmetry}%
\end{center}
\end{figure}

Let us now consider the decays of the $3-7_{vis}$ strings. Taking account of $\vert M_{med} \vert \gg \vert M_{hid} \vert$, there will be a kinematically allowed decay of the form:
\begin{equation}
\widetilde{Q}_{med}\rightarrow  \widetilde{Q}^\dag_{hid} \ + \ l_{(1)}^{\dag} %
\label{weakdecay}
\end{equation}
which populates both the $3-7_{vis}$ and $7_{vis}-7_{hid}$ states. Let us see if the decay of the $Q^{SM}$ states
also violates C and CP. Computing the decay process $\widetilde{Q}_{med}\rightarrow \widetilde{Q}^\dag_{hid}+l_{(1)}^{\dag}$, we have, to one-loop order in the Yukawa coupling:
\begin{equation}
\mathcal{A}(\widetilde{Q}_{med} \rightarrow \widetilde{Q}^\dag_{hid} + l_{(1)}^{\dag}) = \lambda^{l}_{(1)} + \lambda^{l}_{(1)}\cdot \sum_{i} \lambda^{h \dag}_{i} \lambda^{h}_{i} \cdot \mathcal{I}_{toy} (M_{med})
\end{equation}
where in the above, $\mathcal{I}_{toy}(M_{med})$ is the one-loop integral, and we have suppressed all order one constants. For matters of CP violation, the crucial feature of $\mathcal{I}_{toy}$ is that it must contain a complex phase. This is just a consequence of the optical theorem and unitarity thereof. Its dependence on $M_{med}$ also drops out since the probe sector has only one generation. This amplitude is to be compared with the decay of the anti-particle $\widetilde{Q}^\dag_{med}$:
\begin{equation}
\mathcal{A}(\widetilde{Q}^\dag_{med} \rightarrow \widetilde{Q}_{hid} + l_{(1)}) = \lambda^{l \dag}_{(1)} + \lambda^{l \dag}_{(1)}\cdot \sum_{i} \lambda^{h}_{i} \lambda^{h \dag}_{i} \cdot \mathcal{I}_{toy}(M_{med}^\dagger).
\end{equation}
Here, the loop integral is identical to that for the other terms. This is required in order for the theory to be unitary. Now, if the sum over $\lambda^{h \dag}_{i} \lambda^{h}_{i}$ had been imaginary, we would have obtained a difference between the two decay rates, and hence obtained a net CP violation. However, since this quantity is the norm squared of a vector and hence real, we instead find:
\begin{equation}
\Gamma(\widetilde{Q}_{med} \rightarrow \widetilde{Q}^\dag_{hid} + l_{(1)}^{\dag})
- \Gamma(\widetilde{Q}^\dag_{med} \rightarrow \widetilde{Q}_{hid}+ l_{(1)}) = 0.
\end{equation}
In the absence of additional mixing effects from the Standard Model Yukawa interactions, this would be an exact result. The reason is that because there is only one decay mode available for the $\widetilde{Q}_{med}$ fields, the decay rate for a $\widetilde{Q}_{med}$ state and its anti-particle counterpart will be the same. To one-loop order, $\varepsilon_{CP} = 0$. This is already problematic, and makes it difficult to generate enough baryon asymmetry.

This is a basic but important point, so let us reiterate it. To get a C and CP violating decay, it is necessary to have two sources of complex numbers. One is from a relative complex phase among multiple independent decay process, and the second is from the imaginary part of the loop integral. The latter is automatically present, but as we have seen, the existence of a relative phases between the couplings of loop and tree level effects is less apparent.

In our case this can be traced to the presence of only one family of string states present on a D3-brane. To have a net baryon asymmetry, we must have more than one family of $3-7_{vis}$ or $3-7_{hid}$ states. This is a feature built in for both conventional GUT baryogenesis and leptogenesis, but is missing from our weakly coupled toy model. Let us note that in contrast to other situations involving intersecting seven-branes, switching on a gauge field flux would fail to produce a generational structure for probe D3-branes since there is no index theorem in zero dimensions.\footnote{Generating chiral matter from probe D3-branes is possible, however, it requires probing a singular geometry in the threefold base, which is not the present situation.} The absence of a generational structure at weak coupling is thus a significant obstruction to generating C and CP violating decays.

In principle, we can make our model more realistic by including additional Yukawa couplings to the Standard Model fields. This would lead to some additional suppression based on the ratio of masses in the Standard Model. Moreover, such processes would only arise in our example at higher loop order. This requires significant fine tuning in the model to generate enough baryon asymmetry.

Though it is not present in our weakly coupled model, let us now track down the types of interaction terms we would need to generate larger C and CP violating decay processes. Later, we shall argue that such interaction terms are automatically present at strong coupling. To achieve C and CP violating decays in our toy model, we generalize the $\lambda$'s to matrices of couplings:
\begin{equation}\label{wishful}
W_{SM \otimes D3} \supset \lambda^{h}_{(i)\overline{I}\overline{J}} h_{(i)} \widetilde{Q}^{med}_{\overline{I}} \widetilde{Q}^{hid}_{\overline{J}} + \lambda^{l}_{(g) K \overline{L}} l_{(g)} Q^{med}_{K} \widetilde{Q}^{hid}_{\overline{L}} + M_{med} Q^{med} \widetilde{Q}^{med} + M_{hid} Q^{hid} \widetilde{Q}^{hid}
\end{equation}
where the indices $I,J,K,L$ are the needed ``generational indices'' of the probe sector. In fact, even with only one generation of Standard Model fields, the amplitude:
\begin{equation}
\mathcal{A}(\widetilde{Q}^{med}_{I} \rightarrow \widetilde{Q}^{hid \dag}_{\overline{J}} + l_{(1)}) =  \lambda^{l}_{(1)I\overline{J}} + \lambda^{l}_{(1) I \overline{B}} \cdot \lambda^{h \dag}_{(i) B \overline{C}} \cdot \lambda^{h}_{(i) C \overline{J}} \cdot \mathcal{I}_{loop} (M_{med})
\end{equation}
for generational indices $A,B,C$, and a loop integral $\mathcal{I}_{loop}(M_{med})$ contains the necessary interference terms to generate CP violating effects and also depends nontrivially on the mass spectrum. This is because the matrix product of the couplings at one-loop will now contain relative complex phases to the tree-level contribution. Again, we stress that this is a combination of two effects, the relative phase between products of Yukawas in tree level and loop level amplitudes, and the common phase present in the loop integral $\mathcal{I}_{loop}$. In the following subsections we show that this type of generational structure is realized at strong coupling.

\subsection{Effective Generations From String Junctions}

In this subsection we show that one of the limitations of working at weak coupling, namely the absence of a generational structure is eliminated at strong coupling. We begin by stating in more precise terms the form of the probe sector in the limit where all Standard Model fields are non-dynamical. In this limit, the probe sector is not perturbed by the Standard Model. Later, we shall add back in the coupling to the Standard Model, treating it as a small perturbative correction.

Prior to breaking conformal invariance, we have a strongly coupled $\mathcal{N} = 1$ superconformal field theory. We do not have a complete characterization of the operators, let alone their scaling dimensions. However, we do know that such operators must fill out representations in the unbroken part of $SU(5)_{GUT} \times SU(5)_{\bot} \subset E_8$ specified by the choice of $\Phi$-deformation. We label these operators as $\mathcal{O}_{R}$ for an operator in a representation $R$ in the unbroken part of $SU(5)_{GUT} \times SU(5)_{\bot}$.

Again, it is helpful to draw an analogy to the case of QCD. Although the microscopic description of QCD is in terms of $N_f$ quark flavors charged under a strongly coupled $SU(N_{c})$ gauge theory, at low energies we can still deduce some basic information in terms of the $SU(N_f)_L \times SU(N_f)_R$ flavor symmetry of the gauge-invariant hadrons of the chiral Lagrangian. In the spirit of low-energy effective field theory, we therefore expect interaction vertices in the effective Hamiltonian take the form:
\begin{equation}
H^{(0)}_{eff} \supset \sum_{R_{i} , p_{i}} \kappa_{R_{1} , R_{2}, R_{3}} \mathcal{O}_{R_{1}} \mathcal{O}_{R_{2}} \mathcal{O}_{R_{3}} + \kappa_{R_{1}, R_{2}, R_{3}, R_{4}} \mathcal{O}_{R_1} \mathcal{O}_{R_2} \mathcal{O}_{R_3} \mathcal{O}_{R_4} + \cdots
\end{equation}
where the ellipses denote additional multi-state interactions which at strong coupling \textit{cannot} be ignored when the scale of momentum transfer is comparable to the CFT breaking scale.

As we now explain, at strong coupling there is an effective generational index for the $\mathcal{O}_{R}$ operators which is the main ingredient necessary to realize CP violating decays of the heavy states. This is most easily seen by appealing to the stringy realization of the strongly coupled field theory. The basic idea is that the various operators of the $E_8$ $\mathcal{N}=2$ Minahan-Nemeschansky theory fill out representations of $E_8$. This includes the operators $\mathcal{O}$ in the adjoint of $E_8$, as well as many additional operators in higher dimension representations. Taking this feature into account, we see that it is actually more appropriate to write the effective interactions of the probe sector as:
\begin{equation}\label{intHam}
H^{(0)}_{eff} \supset \sum_{R_{i} , d_{i}} \kappa^{d_{1},d_{2},d_{3}}_{R_{1} , R_{2}, R_{3}} \mathcal{O}^{d_{1}}_{R_{1}} \mathcal{O}^{d_{2}}_{R_{2}} \mathcal{O}^{d_{3}}_{R_{3}} + \kappa^{d_1, d_2, d_3, d_4}_{R_1, R_2, R_3, R_4}
\mathcal{O}_{R_1}^{d_1} \mathcal{O}_{R_2}^{d_2} \mathcal{O}_{R_3}^{d_3} \mathcal{O}_{R_4}^{d_4} + \cdots
\end{equation}
where in the above, the $d_i$ is a generational index we identify with a dyonic charge vector and the ellipses denote additional multi-point interactions which at strong coupling cannot be ignored. Here, the products of the $\mathcal{O}$'s in different representations must form gauge invariant operators.\footnote{Let us note that fixing the representation content $R_{i}$ only partially fixes the junction content once we consider breaking patterns of $E_8$ down to $SU(5)_{GUT}$.} The existence of these additional string junctions leads to an effective generational structure. Indeed, these additional generations are non-perturbative, and decouple when $g_{s} \rightarrow 0$.

\begin{figure}
[ptb]
\begin{center}
\includegraphics[
trim=0.000000in 6.544466in 0.000000in 0.714558in,
height=3.1592in,
width=5.8349in
]%
{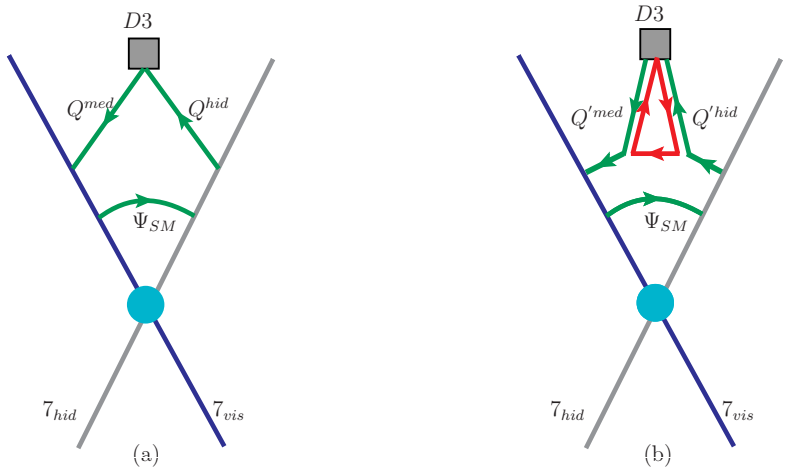}%
\caption{The interaction of $(3-7_{vis})$ and $(3-7_{hid})$ states with the Standard Model particle $(7_{vis}-7_{hid})$. For a given $(7_{vis}-7_{hid})$ particle, there is an interaction with D3-brane states created by fundamental strings (diagram (a)) and also an interaction with states created by $(p,q)$ string junction (diagram (b)).}%
\label{p-q-junction}%
\end{center}
\end{figure}

The $\kappa$'s correspond to the wave function overlaps between these junction states. Due to strong coupling, we expect that, if allowed by selection rules, these couplings are all order one. Let us now present some additional evidence in favor of this statement. With present techniques it is not known how to compute from first principles such overlap terms in a precise manner.\footnote{The computation of higher stringy modes is in principle possible through a generalization of string field theory to the non-perturbative setting. Since we are discussing F-term couplings, a more practical route may involve a computation in terms of wave function overlaps for massive modes in holomorphic Cherns-Simon theory, perhaps along the lines of \cite{BHSV}.} Nevertheless, we can estimate such contributions geometrically, at least when the classical string picture is valid. Though this limits our discussion to cases where the D3-brane is separated from the seven-branes by more than a string length, we can at least see the presence of additional interaction terms becoming prominent as we move the D3-brane close to the E-point. See figure \ref{p-q-junction} for a depiction of such interaction terms in terms of classical string junctions.

In string frame, the tension of a $(p,q)$ string with $p$ units of F1 charge and $q$ units of D1 charge is:
\begin{equation}
T_{p,q} = \frac{1}{l^{2}_{s}} \vert p + \tau q \vert \ .
\end{equation}
Recall that $l_{s} \sim M_{\ast}^{-1}$ for Im$\tau \sim 1$. We can form even more bound states based on various junctions. A $(p,q)$ string junction with $n$ end-points as a junction will be composed of $n$ two-component vectors $({p}_{1},q_{1}),...,(p_{n},q_{n})$. For a junction $j$ we refer to this entire collection of vectors by $d_{j}$ for ``dyonic charge''. A junction involving three endpoints will satisfy a balancing condition:
\begin{equation}
(p_{1} , q_{1}) + (p_{2},q_{2}) + (p_{3},q_{3}) = 0
\end{equation}
with similar considerations for additional endpoints. From this perspective, the generational indices $I,J,K,L$ now correspond to the entire tower of dyonic $(p,q)$ junction resonances. This will involve both $(p,q)$ strings, as well as junctions involving multi-prong strings.

F-term couplings between string zero modes are computed in the topological B-model by disk diagrams which localize at a point of the geometry. For massive modes, this localization will be distorted, and we can expect for a string junction $j$ some dependence of the form ${\rm{exp}}(- A_{j} \cdot T_{j})$ for an appropriate measure of the area $A_{j}$ subtended by a dyonic loop, and $T_{j}$ the tension of a string junction $j$ with endpoint charges $d_{j}$. Note that when $g_{s} \rightarrow 0$, $T_{j} \rightarrow \infty$ and the Yukawa coupling is significantly suppressed.

\subsection{Coupling to the Standard Model}

We now consider coupling our probe sector to dynamical Standard Model fields. We achieve this by compactifying some of the curves of the geometry. Note that we can still work in terms of non-compact seven-branes, but with some compact curves. This compactification occurs at the GUT scale, and from the perspective of the probe sector it corresponds to adding the deformation at the GUT scale \cite{Funparticles}:
\begin{equation}
\delta W_{SM \otimes D3} = \Psi^{SM}_{\overline{R}} \cdot \mathcal{O}_{R}.
\label{deformation}
\end{equation}
At weak coupling this deformation would be identified with a cubic interaction of the form $\Psi_{SM} Q_{med} \widetilde{Q}_{hid}$. At strong coupling, the dynamics of the strings attached to the D3-brane is most conveniently described in terms of gauge-invariant operators $\mathcal{O}_{R}$, carrying representation $R$ under the $SU(5)_{GUT}$. We view such operators as composite operators made from ``quarks'' of the strongly coupled probe sector. For an operator with quantum numbers under both $SU(5)_{GUT}$ and the unbroken part of $SU(5)_{\bot}$, this should be viewed as a composite involving both the $3-7_{vis}$ and $3-7_{hid}$ strings.

Proceeding to the IR, the deformation (\ref{deformation}) will induce mixing between operators of the SCFT and the Standard Model fields. Diagonalizing the mixing, we also see that the canonical basis of Standard Model fields is a linear combination of $\Psi^{SM}_R$ and $\mathcal{O}_R$. Here, we claim that, quite generally, such mixing gives rise to baryon and lepton number violating interactions, while correlating visible and hidden sector energy densities.\footnote{Since the notion of particle number is somewhat ill-defined at strong coupling, the more precise notion is in terms of the overall charge under the $7_{hid}$-brane, and the expected energy density. To avoid being overly pedantic, however, we shall often use the intuition of correlated ``number densities''.}

Using some basic features of the approximately conformal sector, we can estimate the decay rate for $3-7_{vis}$ strings. To motivate our answer, let us return to the case of a weakly coupled theory, but one in which the fields have experienced some amount of wave-function renormalization, and so in a holomorphic basis of fields, do not have canonical kinetic terms. In a supersymmetric theory, the F-terms will not be renormalized. The Lagrangian we consider is then:
\begin{equation}
L_{toy} = \int d^{4} \theta (\mathcal{Z}_{Q} Q_{med}^{\dag} Q_{med} + \mathcal{Z}_{\widetilde{Q}} \widetilde{Q}_{hid}^{\dag} \widetilde{Q}_{hid} + \mathcal{Z}_{\Psi} \Psi_{SM}^{\dag} \Psi_{SM}) + \int d^{2} \theta \lambda Q_{med} \widetilde{Q}_{hid} \Psi_{SM} + ... +  h.c.
\end{equation}
where the additional terms include various supersymmetric mass terms. To compute a decay rate, we first
canonically normalize the various fields. This in general cannot be done in a way which maintains manifest holomorphy of the
F-terms. Doing this, we find that the resulting decay rate is:
\begin{equation}
\Gamma_{toy}(Q_{med} \rightarrow Q^{\dag}_{hid} \Psi^{\dag}_{SM}) \sim M_{med} \cdot \frac{\vert \lambda \vert^{2}}{\mathcal{Z}_{Q} \mathcal{Z}_{\widetilde{Q}} \mathcal{Z}_{\Psi}}.
\end{equation}
In other words, we see that there will be some effect from wave function renormalization.

Similar considerations are expected to hold for decays in the strongly coupled regime. Here, we must exercise caution because we do not know the exact form of the kinetic terms after allowing the $\mathcal{N} = 1$ deformation $\delta W_{SM \otimes D3}$ to proceed. The other issue is that defining a notion of ``canonical normalization'' for the $\mathcal{O}$'s is less clear. However, we can still appeal to the correlators of the CFT. Treating $\mathcal{O}_{R}$ as a composite of $3-7_{vis}$ strings and $3-7_{hid}$ strings, we view this type of coupling as inducing a decay of the form $Q_{med} \rightarrow Q^{\dag}_{hid} \Psi^{\dag}_{SM}$. In practice, this process will be dressed by additional spectator $3-7_{hid}$ states. In this case, we expect that the analogue of the wave-function renormalization factors $\mathcal{Z}_{Q} \mathcal{Z}_{\widetilde{Q}} \mathcal{Z}_{\Psi}$ are given by appropriate ratios of $M_{\cancel{CFT}}/M_{GUT}$. More precisely we expect:
\begin{equation}
\mathcal{Z}_{Q}\mathcal{Z}_{\widetilde{Q}} \sim \left(\frac{M_{\cancel{CFT}}}{M_{GUT}} \right)^{ \Delta_{UV}(\mathcal{O}^{\dag} \mathcal{O}) -  \Delta_{IR}(\mathcal{O}^{\dag} \mathcal{O})} \,\,\,\,\,\, \mathcal{Z}_{\Psi} \sim \left(\frac{M_{\cancel{CFT}}}{M_{GUT}} \right)^{ \Delta_{UV}(\Psi^{\dag} \Psi) - \Delta_{IR}(\Psi^{\dag} \Psi)}.
\end{equation}
For the case of $\Psi_{SM}$ states this is clear, though for $\mathcal{O}$ and the associated ``quark states'' this notion of canonical normalization is somewhat more ambiguous, because it requires us to fix the particle number for the external states. Here, we are fixing this ambiguity by appealing to the normalizations of fields defined at the GUT scale. Finally, this approximation is only expected to be valid if there is a small amount of mixing between the SM and probe states in passing from the UV to the IR. Fortuitously, this is indeed the case \cite{HVW}.

In terms of the scaling dimensions of the operators in the IR, we find that the decay rate is:
\begin{equation}
\Gamma(Q_{med} \rightarrow Q_{hid}^{\dag} \Psi_{SM}^{\dag}) \sim M_{\cancel{CFT}} \cdot \left(\frac{M_{\cancel{CFT}}}{M_{GUT}} \right)^{2 \nu}
\end{equation}
where to this level of approximation, $\nu$ is given by an effect associated with wave function renormalization:
\begin{equation}
\nu = \frac{1}{2}\Delta_{IR}(\mathcal{O}_{\overline{R}}^{\dag} \mathcal{O}_{\overline{R}}) + \frac{1}{2}\Delta_{IR}(\Psi_{R}^{\dag} \Psi_{R})- \frac{1}{2}\Delta_{UV}(\mathcal{O}_{\overline{R}}^{\dag} \mathcal{O}_{\overline{R}}) - \frac{1}{2}\Delta_{UV}(\Psi_{R}^{\dag} \Psi_{R}).
\end{equation}
with the UV scale identified as the GUT scale, e.g., the scale at which the probe to Standard Model coupling is first added.

The exact value of $\nu$ depends on the specific monodromy scenario, but numerically, the difference between the UV and IR scaling dimensions is expected to be an order one number somewhat less than one. In the examples of \cite{HVW} it has been found that when one ignores the effects of $W_{flux}$ on the scaling dimensions, the difference in IR and UV scaling dimension for $\Psi$ is typically on the order of $0.1 - 0.2$ (though in some scenarios it can be slightly more) while the change to $\mathcal{O}$ is typically smaller. This provides a first estimate of $\nu \sim 0.2$. In practice, this number may receive additional corrections due to the scaling of non-chiral operators. Contributions from $W_{flux}$ provide another source of corrections to scaling dimensions. In Appendix B we show that if not treated as a small perturbation, such corrections can lead to somewhat larger shifts in the scaling dimensions of operators. Treated as a perturbation to the system, adding a small $W_{flux}$ and allowing some RG evolution provides a way to effectively tune the value of $\nu$ to match observation. For all of these reasons, in what follows we view $\nu$ as an order one number somewhat less than one, to be fit by various phenomenological constraints. In many of the cosmological scenarios, the required value of $\nu$ is on the order of $0.5$, which is remarkably close to these crude considerations.

There are two points which the above computation highlights. Firstly, in passing from the UV to the IR, the presence of wave-function renormalization can influence the decay rate. This is basically the same as the fact that the decay rate depends on the differences of the IR and UV scaling dimensions. Secondly, we see that in general, there is some suppression of the decay rate due to this difference.

Let us now generalize this result to other types of decays. Basically, the idea is that for each additional insertion of a Standard Model external state, there is a further suppression by a factor of $\epsilon$:
\begin{equation}
\epsilon \equiv \left(\frac{M_{\cancel{CFT}}}{M_{GUT}} \right)^{\nu}
\end{equation}
To provide additional motivation for this prescription, we now consider an approximation scheme in which we view $\mathcal{O}$ as an operator
associated with a single particle meson state. In this case, such meson states will have similar quantum numbers to those of a Standard Model field. Canonically normalizing $\Psi_{SM}$ and $\mathcal{O}$, the mixing between the visible and probe sector can
be characterized as a shift in the interaction terms of $H_{eff}$ by operators $\mathcal{O}_{R}$ of the form:
\begin{equation}\label{firstmix}
\mathcal{O}_R \rightarrow \mathcal{O}_R + \sum_{g=1}^3 \epsilon_{g} \Psi_{R,g}^{SM}
\end{equation}
where $\epsilon_g$ is the mixing coefficient with the $g^{th}$ generation Standard Model field. The dominant source of mixing is with the third generation of Standard Model fields, with additional suppression by the profile of the lighter generations near the Yukawa point. For simplicity, let us focus on the coupling to the third generation only, as this is the largest source of matter/probe mixing.

Having the mixing term \eqref{firstmix} switched on, we obtain from the effective interaction Hamiltonian \eqref{intHam} couplings between the strong coupling states and the visible sector fields:
\begin{eqnarray}
\delta H_{eff} &\supset& \sum_{R} \epsilon^{SM}_{R} \mathcal{O}_{\overline{R}} \Psi^{SM}_{R} + \sum_{R_{1} , R_{2} , R; d_{1} , d_{2}} \epsilon^{d_{1},d_{2},SM}_{R_{1} , R_{2}, R} \mathcal{O}^{d_{1}}_{R_{1}} \mathcal{O}^{d_{2}}_{R_{2}} \Psi^{SM}_{R} + \cdots
\nonumber \\
&+& \sum_{R_1, R_2, R, R'; d_1, d_2}
\epsilon^{d_1, d_2,SM,SM}_{R_1, R_2, R, R'} \mathcal{O}^{d_1}_{R_1} \mathcal{O}^{d_2}_{R_2}
\Psi^{SM}_{R} \Psi^{SM}_{R'} + \cdots
\label{SMcoupling}
\end{eqnarray}
for a Standard Model field $\Psi^{SM}_{R}$ of representation $R$. Here we are assuming an appropriate notion of canonical normalization for the various modes. The first line denotes terms involving one Standard Model field, where the ellipses denote terms having more than two $\mathcal{O}$'s. Likewise, the second line denotes terms involving two Standard Model fields, where the ellipses denote terms having more than two $\mathcal{O}$'s. For now, it is important to recall that the representations $R_1, R_2, \cdots$ are constrained by invariance under $SU(5)_{GUT}$ and $SU(5)_\bot$. Here, we have inserted an explicit factor of $\epsilon^{SM} \sim \epsilon$ into the term linear in $\mathcal{O}$ and $\Psi$, based on our discussion of the suppressed decay rates. In this sense, we are working in terms of a basis of ``canonically normalized fields'', though as we have already noted, the exact definition of this normalization is somewhat ambiguous at strong coupling. Here, we have fixed this ambiguity by working in terms of the scaling dimensions of fields at the GUT scale.

In (\ref{SMcoupling}), the couplings $\epsilon^{d_{1},d_{2},SM}_{R_{1} , R_{2}, R}$ are of order $\epsilon$ while
the couplings $\epsilon^{d_1, d_2,SM,SM}_{R_1, R_2, R, R'}$ are of order $\epsilon^2$, as it involves two Standard Model insertions. Let us note that in general we do not expect a simple relation such as $\epsilon^{d_{1},d_{2},SM}_{R_{1} , R_{2}, R} = \epsilon \cdot \kappa^{d_{1},d_{2},d}_{R_{1} , R_{2}, R}$ to hold, since the actual coupling involves a complicated overlap integral between the dyonic modes and the Standard Model field. We also note that in the context of F-theory, the Standard Model states will be composed of multi-prong string junctions, though of fixed type dictated by the embedding into the adjoint of $E_8$. Finally, the ellipses now contains additional terms of higher order in the number of Standard Model fields and also in the number of the gauge invariant dyonic operators. The interaction term involving $\Psi^{n}$ is suppressed by a factor of $\epsilon^{n}$.

\subsection{Baryon Number Violating Processes}\label{BVIOL}

For the purposes of generating a baryon asymmetry, we are especially interested in decays which violate baryon number, and can therefore generate an asymmetry. From the perspective of the $SU(5)_{GUT}$ group, such interactions will descend from interactions such as the $5 \times 10 \times 10$ and $\overline{5} \times \overline{5} \times 10$ interactions as well as higher order analogues. It is well-known in the context of conventional four-dimensional GUTs that such terms induce baryon number violating processes, basically because they involve couplings of the Higgs triplets to the Standard Model fields. The states of the probe sector can function as Higgs triplets, inducing similar baryon number violating processes. Let us note that generating a baryon number violating process requires at least two Standard Model external states. Such amplitudes are therefore of order $\epsilon^2$.

As an example of this type, consider the coupling of the $3-7_{vis}$ string ``Higgs triplet'' of the probe sector (denoted by $\mathcal{T}^{med}_{u}$) to the Standard Model states. Disregarding the $U(1)_{D3}$ charge which we can always dress by appropriate $3-7_{hid}$ spectator quarks, there are interaction terms descending from the $5 \times 10 \times 10$ of the form:
\begin{equation}
\delta L_{int} \supset \mathcal{T}^{med}_{u} Q^{SM} Q^{SM} \prod_{Q_{hid}} Q_{hid} + \mathcal{T}^{med}_{u} U^{SM} E^{SM} \prod_{Q_{hid}} Q_{hid}
\end{equation}
where $Q^{SM}$, $L^{SM}$ and $E^{SM}$ are the quark doublet, lepton doublet and right-handed lepton superfields of the MSSM, and each interaction has been dressed by some additional spectator $Q_{hid}$ states.

An important distinction between usual GUT baryogenesis and the present context is that the $3-7_{vis}$ strings participate in all decay processes. This means the $(B-L)$ charge assigned to the Higgs triplets of a usual four-dimensional GUT model will differ from that of the $3-7_{vis}$ string. In particular, some of the $3-7_{hid}$ strings will also carry non-zero $(B-L)$ charge. Similar mechanisms with ``dark baryons'' have been discussed for example in \cite{Davoudiasl:2010am,Buckley:2010ui}.

In order to track the amount of baryon number violation in a given decay channel, we appeal to the symmetries of our theory. In most realistic examples of F-theory GUT scenarios, there is at least one, and sometimes two approximate global $U(1)$ symmetries realized on the $7_{hid}$-branes which persist to low energies. These are typically referred to as $U(1)_{PQ}$, and in the case of Dirac neutrino scenarios, there is another symmetry $U(1)_{\chi}$, which in an appropriate linear combination with $U(1)_{Y}$ corresponds to $U(1)_{B-L}$ \cite{BHSV, EPOINT}, much as in conventional four-dimensional $SO(10)$ GUT models. In both cases, these $U(1)$'s originate from the Cartan subalgebra of $SU(5)_{\bot}$. In the case where $U(1)_{B-L}$ is explicitly broken due to the choice of tilting in the high scale theory, we can generically expect there to be both $(B+L)$ and $(B-L)$ violating processes.

Focussing on theories which preserve $U(1)_{\chi}$, we note that some of the $3-7_{hid}$ strings will be charged under this factor as well. Let us note that the interactions of (\ref{SMcoupling}) are allowed insofar as they are invariant under $SU(5)_{GUT} \times SU(5)_\bot$ and descend from representations of $E_8$. In such cases, it is immediate that there will be interactions that violate $(B+L)$ while preserving a generalized notion of $(B-L)$. For instance, in cubic interactions involving two Standard Model fields, there will be interactions such as
\begin{equation}
(5, 10) \otimes (10, \overline{5}) \otimes (10, \overline{5}) + (\overline{5}, \overline{10}) \otimes (\overline{5}, \overline{10}) \otimes (10, \overline{5}).
\end{equation}
As far as the $SU(5)_{GUT}$ part is concerned, these are precisely the $(B+L)$ violating interactions. Moreover, there will also be quartic or higher-point interactions which still involve two Standard Model fields and violate $(B+L)$ number. For instance, quartic interactions of the type
\begin{equation}
(5, 10) \otimes (10, \overline{5}) \otimes (10, \overline{5}) \otimes (1, 75) + (\overline{5}, \overline{10}) \otimes (\overline{5}, \overline{10}) \otimes (10, \overline{5}) \otimes (1, 75)
\end{equation}
involve precisely the same field content as the cubic interactions and one additional state transforming in the $(1, 75)$ of $SU(5)_{\bot}$. Again, let us stress that although these interactions involve more than three fields, because we are at strong coupling, they are just as relevant.

We can also see that in a given decay, there is typically a non-zero amount of $U(1)_{\chi}$ induced in the visible sector, and a compensating amount generated in the hidden sector. To illustrate this, it is helpful to return to the case of the ``Higgs triplet decay''. This mode is uncharged under $U(1)_{\chi}$ as it is a $3-7_{vis}$ string. Note, however, that $Q^{SM}$ will have non-zero charge under $U(1)_{\chi}$. The excess charge is carried by the $3-7_{hid}$ strings. Thus, there is a net conservation of $U(1)_{\chi}$ between the two sectors. Note, however, that by charge conservation \textit{each} sector has non-zero charge density under $U(1)_{\chi}$. Similar considerations also hold in Majorana neutrino scenarios which conserve an overall $U(1)_{PQ}$. In other words, though there is no net $(B-L)$ violation once the visible and hidden sectors are taken into account, there is a violation in a given sector, when treated separate from the rest of the system.

\subsubsection*{Comments on Proton Decay}

Adding in an additional source of vector-like states to the Standard Model introduces another potential source of modes which can induce proton decay. For example, under $SU(5)_{GUT}$, there will be states of the probe sector with the same quantum numbers as Higgs triplets. It is known that Higgs triplet exchange in minimal four-dimensional $SU(5)$ GUTs is quite problematic, so it is of interest to study whether similar concerns would be present here as well.

First let us note that dimension five operators such as:
\begin{equation}
\lambda_{ijkl}\int d^{2} \theta \frac{Q^{i}Q^{j}Q^{k}L^{l}}{M_{GUT}}
\end{equation}
can be forbidden in F-theory GUT models by including an additional $U(1) \subset E_8$ symmetry under which the Standard Model fields are charged \cite{BHVII, HVGMSB}. In principle, this symmetry may be broken at some lower scale, though this is a more model dependent issue. However, to give the broadest possible application to our scenario, let us now consider the contribution to this dimension five operator from the probe sector in a fictitious model where there is no protection from an effective $U(1)$. The most stringent constraint comes from the limit on $\lambda_{112l} \lesssim 10^{-10}$ (see \cite{Salati:1982bw,Ellis:1983qm,Ibanez:1991pr}).

In this fictitious scenario, we observe that the coupling between the Standard Model to hidden sector will involve an intermediate channel of probe sector states, coupled to four external Standard Model states. Since these heavy states are expected to have mass on the order of $M_{\cancel{CFT}}$ (as they are charged under $SU(5)_{GUT}$), we obtain our estimate for the dimension five operator:
\begin{equation}
\lambda_{3333}\int d^{2} \theta \frac{Q^{3}Q^{3}Q^{3}L^{3}}{M_{GUT}} \sim
\alpha_{GUT} \cdot \epsilon^{4} \int d^{2} \theta \frac{Q^3Q^3Q^3L^3}{M_{\cancel{CFT}}} \sim \alpha_{GUT} \cdot \left(\frac{M_{\cancel{CFT}}}{M_{GUT}} \right)^{4 \nu - 1} \int d^{2} \theta \frac{Q^3Q^3Q^3L^3}{M_{GUT}}.
\end{equation}
Here we have included an additional suppression factor for each Standard Model mode of order $\alpha_{GUT}^{1/4}$ coming from wrapping matter fields on GUT scale curves.

To obtain our estimate of $\lambda_{112l}$, we should next consider the overlap of two first generation Standard Model fields with a second generation field, and a third generation field. This really depends on the details of the monodromy scenario, which should then be folded in with a discussion of wave function profiles. However, to get a rough sense, we use the same estimates for wave function profiles and associated Yukawas obtained for example in \cite{HVCKM}. In this case, Yukawas involving the third, second and first generation are respectively multiplied by powers of $\alpha_{GUT}^{0}$, $\alpha_{GUT}^{1}$, $\alpha_{GUT}^{2}$. In this fictitious scenario, this yields the net estimate for $\lambda_{112l}$ of order:
\begin{equation}
\lambda_{112l} \sim \alpha_{GUT}^{6} \cdot \left(\frac{M_{\cancel{CFT}}}{M_{GUT}} \right)^{4 \nu - 1}.
\end{equation}
Anticipating the numerical values of $M_{\cancel{CFT}}$ and $\nu$ to be found later in section \ref{sec:COSMO} and taking $\alpha_{GUT} \sim 1/25$, we can for example take a high value of $M_{\cancel{CFT}} \sim 10^{13}$ GeV and $\nu \sim 0.6$ associated with a ``high dilution scenario'', obtaining $\lambda_{112l} \sim 10^{-13}$, which is below the present bound. As another example, we can also consider $M_{\cancel{CFT}} \sim 10^{13}$ GeV and $\nu \sim 1.2$, associated with a ``low dilution scenario'' in which case $\lambda_{112l} \sim 10^{-21}$. Again, we stress that additional suppression is expected from approximate $U(1)$ symmetries. We therefore conclude that even in the most conservative case, there is little danger of violating present constraints from proton decay.

\subsection{C and CP Violation at Strong Coupling}
We now turn to sources of C and CP violation in the coupling between the probe and visible sectors. Recall that the D3-bane probe theory descends from an $\cN=2$ supersymmetric theory, so in the UV, its spectrum is vector-like. Therefore, we expect that coupling constants that enter $H^{(0)}_{eff}$ do not source any C and CP violation.\footnote{In principle, however, the non-zero value of the strongly coupled theta angle for the $U(1)_{D3}$ gauge theory
introduces another source of CP violation.} We also expect this continues to be so even after we deform the $\cN=2$ theory to $\cN=1$ configurations. Further deformation by coupling to the Standard Model is different, since the chiral matter fields of the Standard Model are involved. So, we expect the coupling constants $\epsilon$'s in (\ref{SMcoupling}) must be complex valued, breaking C and CP explicitly. Note that in general this has the structure of a large mixing matrix, which is also different from the weakly coupled case.

The overall decay rate for $3-7_{vis}$ strings is dominated by decays to a single Standard Model state. Indeed, as explained earlier using general scaling properties of the approximate conformal theory, we have:
\begin{equation}
\Gamma_{D} \equiv \Gamma\left( Q_{med} \rightarrow  \Psi^{\dag}_{SM} \otimes any^{\dag}_{hid} \right) \sim \vert \epsilon \vert^{2} M_{\cancel{CFT}}.
\end{equation}
we observe that this decay rate is actually suppressed below the SCFT breaking scale by $|\epsilon|^2$. An important feature of the strongly coupled probe theory is that it is now more difficult to give a weakly coupled particle interpretation to the final $3-7_{hid}$ strings and in particular, may involve many soft $3-7_{hid}$ states. This is analogous to QCD, where the hadronization of a jet will typically contain, in addition to a few leptons, many soft pions. This also means that it is difficult to relate in precise terms the net number of hidden sector states which are created simultaneously with the visible sector states.

The decays involving a single Standard Model final state do not induce a net baryon asymmetry. This is basically because the $3-7_{vis}$ string and the Standard Model state can always form a vector-like pair with respect to $SU(5)_{GUT}$. Indeed, summing over all hidden sector final states, we see that no asymmetry can be generated from such processes. Generating an asymmetry requires us to include higher order decay processes involving at least two Standard Model fields. Let us now estimate the amount of CP violation expected from such processes.

\begin{figure}
[ptb]
\begin{center}
\includegraphics[
trim=0.000000in 9.102140in 0.000000in 0.714558in,
height=1.6207in,
width=6.8978in
]%
{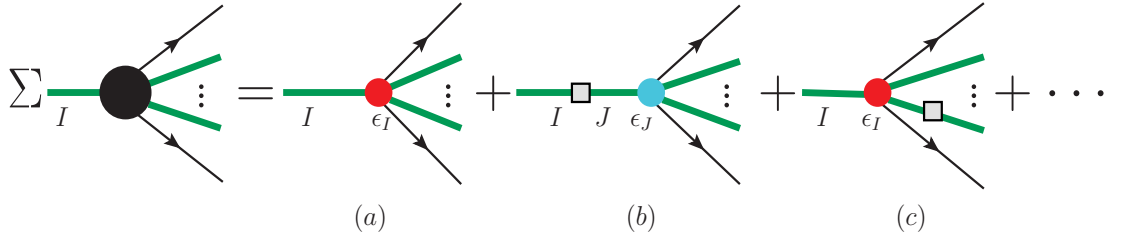}%
\caption{Decay amplitude of $3-7_{vis}$ state at strong coupling. The decay amplitude consists of (a) diagonal interaction proportional to $\epsilon^{SM,SM}_I$, (b) transitional interaction proportional to $\epsilon^{SM,SM}_J$ times self-energy of initial state and (c) transitional interaction proportional to $\epsilon^{SM,SM}_I$ times self-energy of the final state.}%
\label{strong-decay}%
\end{center}
\end{figure}

The first step involves expanding the decay process according to the number of SM fields. We argued above that $(B+L)$-violating, C and CP violating interactions involve {\sl at least} two SM fields. Since the mixing causes a suppression factor $O(\epsilon)$ per each SM field, the leading contribution to the transition amplitude comes from the interactions involving two SM fields. On the other hand, the number of ${\cal O}_{hid}$ operators involved in the transition is not suppressed because of strong coupling of D3-brane dynamics. Consequently, for the estimate, it is sufficient to consider the semi-inclusive process
\begin{equation}
Q^I_{med} \rightarrow \Psi_{SM}^\dagger \Psi_{SM}^\dagger  \otimes \sum_{n=1}^\infty Q_{hid}^{\overline{J}_1 \dagger} \cdots Q_{hid}^{\overline{J}_n \dagger} .
\end{equation}
We depict the process in figure \ref{strong-decay}. Note that each individual process is $(B+L), C$ and $CP$-violating.

The decay amplitude $\mathcal{A}$ involving $Q^I_{med} \rightarrow \Psi_{SM}^\dagger \Psi_{SM}^\dagger \otimes \sum_{n=1}^\infty Q_{hid}^{\overline{J}_1 \dagger } \cdots Q_{hid}^{\overline{J}_n \dagger }$ is:
\begin{equation}
\mathcal{A} \simeq \epsilon^{SM,SM}_{I} + \sum_J \left( {1 \over \Sigma + i \Upsilon} \right)_{IJ} \epsilon^{SM,SM}_J + \epsilon^{SM,SM}_I \sum_{hid}
\left( {1 \over \widetilde{\Sigma} + i \widetilde{\Upsilon}} \right)_{hid} + \cdots
\end{equation}
The three terms are contributions of the processes (a), (b) and (c) in figure \ref{strong-decay}. The mixing coefficients $\epsilon^{SM,SM}_I, \epsilon^{SM,SM}_J, \cdots$ are of order $O(\epsilon^2)$. Effects of strong D3-brane dynamics to the initial SM states and of the final flavor states are summarized in the processes (b) and (c), respectively. Of these, diagonal contribution of (b) (viz. $I=J$ term) and all of (c) are proportional to $\epsilon^{SM,SM}_I$ and hence to (a). On the other hand, off-diagonal contribution of (b) (viz. $I \ne J$ terms) are proportional to $\epsilon^{SM,SM}_J$ and hence are out-of-phase to (a). Quite importantly, unitarity dictates that propagators on the mass-shell $[\Sigma + i \Upsilon]^{-1}$ and $[\widetilde{\Sigma} + i \widetilde{\Upsilon}]^{-1}$ have non-vanishing imaginary part, which takes the Breit-Wigner form with decay width $\Upsilon, \widetilde{\Upsilon}$ and which is $C$ and $CP$ invariant. Comparing the decay rate for $Q^I_{med}$ and $Q^{I \dagger}_{med}$, we estimate the amount of CP violation to be
\begin{align}
\varepsilon_{CP}^{I}  & ={\frac{\Gamma(Q_{med}^{I}\rightarrow\Psi
_{SM}^{\dagger}\Psi_{SM}^{\dagger}\otimes any_{hid}^{\dagger})-\Gamma
(Q_{med}^{I\dagger}\rightarrow\Psi_{SM}\Psi_{SM}\otimes any_{hid})}%
{\Gamma(Q_{med}^{I}\rightarrow any^{\dagger})+\Gamma(Q_{med}^{I\dagger
}\rightarrow any)}}\nonumber\\
& \simeq\sum_{J}{\frac{\Upsilon_{IJ}}{M_{CFT}}}{\frac{\operatorname{Im}%
(\epsilon_{I}\epsilon_{J}^{\ast})}{|\epsilon_{I}|^{2}}}\\
& \sim O(\epsilon^{2}).
\end{align}
Let us note that in the above estimate, we have used the fact that above the CFT breaking scale, the spectrum is scale invariant. In particular, this means that there will be $3-7_{vis}$ strings which can decay to other $3-7_{vis}$ strings. This accounts for the various $I-J$ mixing terms appearing in figure \ref{strong-decay}. Such imaginary parts lead to a relative mismatch between the leading order contribution depicted in figure (a) and the mixing terms of figures (b) and (c), which are the analogues at strong coupling of loop order interference terms.

\section{Cosmological Timeline\label{sec:COSMO}}

\begin{figure}
[ptb]
\begin{center}
\includegraphics[
trim=2.066752in 6.884788in 2.067577in 0.693508in,
height=3.9211in,
width=3.9237in
]%
{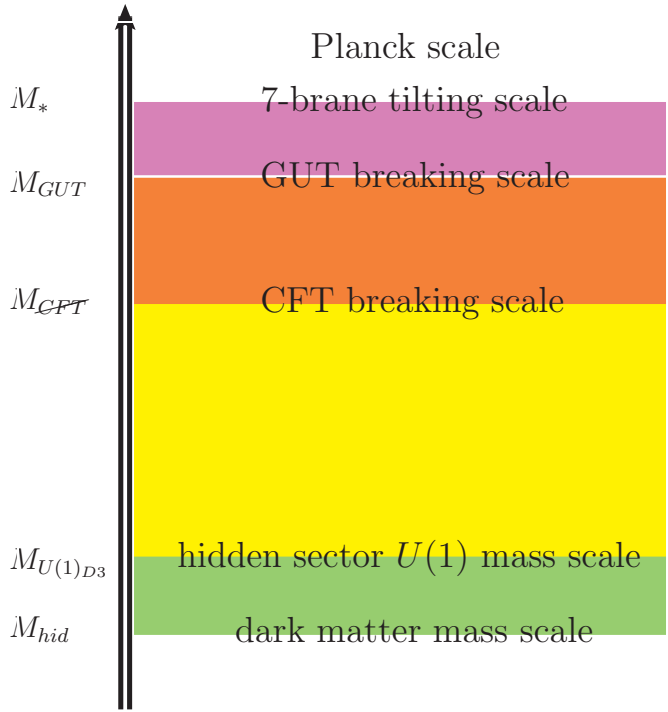}%
\caption{Mass scales of the probe sector. At a scale $M_{\ast}$, the visible
sector seven-brane configuration is specified. GUT\ breaking takes place at a
lower scale $M_{GUT}$. The characteristic mass of the mediator states or
$3-7_{vis}$ strings is $\sim M_{\cancel{CFT}}$. The hidden sector contains a dark
$U(1)_{D3}$ gauge boson, and dark matter given by $3-7_{hid}$ strings. In the above we
depict a typical scenario in which the mass of this gauge boson is higher than
that of the dark matter.}%
\label{mass-scale}%
\end{center}
\end{figure}

So far, we have focused on the particle physics content of a probe D3-brane near the Standard Model.
The main features of the model are summarized in figure \ref{mass-scale}. There are mediator $3-7_{vis}$ states
which connect the probe to the SM. These heavy states then decay, simultaneously
generating a matter asymmetry in the visible and hidden sectors. The light $3-7_{hid}$ strings then correspond to the dark matter of the model.

One of the important features of this scenario is that the dynamics of the probe determine both the \textit{mass} of the dark matter, as well as
the overall matter asymmetry. Phenomenological considerations then require that if the dark matter yield is similar to the baryon asymmetry yield, the mass of the dark matter and the yield are:
\begin{equation}
m_{DM} \sim 10 \,\,\text{GeV} \, \qquad \mbox{and} \qquad Y_{DM} \sim 10^{-10}.
\end{equation}
In this section we show that remarkably, \textit{both} of these conditions as well as several other cosmological constraints can
be satisfied for a range of parameters natural for such probe sectors.

\begin{figure}
[ptb]
\begin{center}
\includegraphics[
trim=2.066752in 6.884788in 2.067577in 0.693508in,
height=3.9211in,
width=3.9237in
]%
{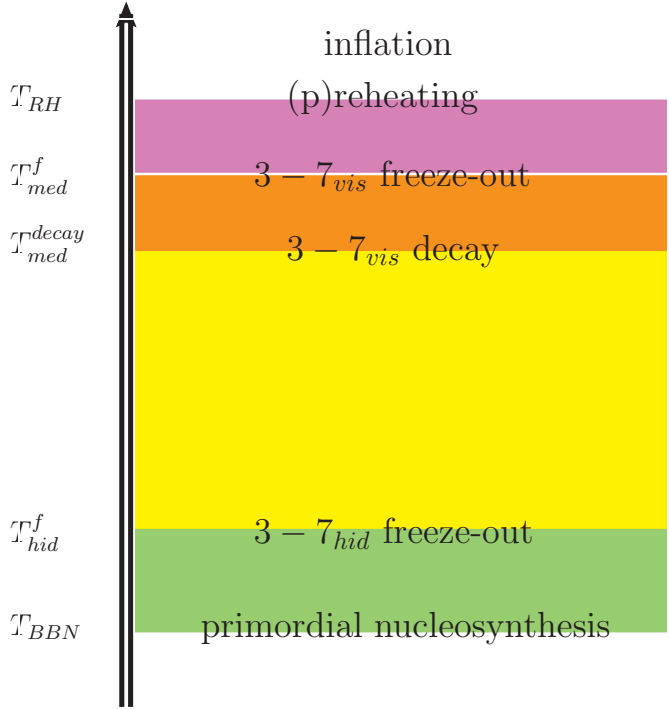}%
\caption{Cosmological timeline for visible and hidden sector matter production.
After inflation ends, the visible and hidden sectors remain in thermal contact due to $3-7_{vis}$ string
mediator states. These states freeze out at a temperature $T^{f}_{med}$ which is within an order of magnitude of
$M_{\cancel{CFT}}$. These modes decay at a a lower temperature $T^{decay}_{med}$. This decay generates a correlated baryon and dark
matter asymmetry. The visible and hidden sectors remain in thermal contact due
to kinetic mixing between the visible sector $U(1)_{vis}$ and hidden sector
$U(1)_{D3}$. At lower temperatures $T^{f}_{hid}$ near $\sim 10$ GeV, the dark matter $3-7_{hid}$
strings freeze out and the symmetric component of dark matter annihilates via
kinetic mixing effects.}%
\label{timeline}%
\end{center}
\end{figure}

We begin by discussing the cosmological timeline for this scenario. See figure \ref{timeline} for
a summary of the thermal history of this scenario.
Our analysis begins after the Universe has exited a period of inflation at a
temperature $T_{RH}$, given by the \textquotedblleft reheating
temperature\textquotedblright\ of the inflaton. We shall assume that the inflaton decays predominantly to the Standard Model/probe sector. After inflation ends, the Universe enters an era of radiation domination. In what follows we assume $T_{RH} > M_{\cancel{CFT}}$, in which case the probe theory and the Standard Model are initially in thermal contact after inflation ends. This is achieved through the exchange of the $3-7_{vis}$ strings which
are charged under both the visible and hidden sector gauge groups. Let us note that even if we lower $T_{RH}$, resonant preheating processes can still provide a mechanism for efficiently populating an initial distribution of $3-7_{vis}$ strings. Such issues require specifying much more about the details of the inflationary phase, however, and so for now we shall defer such questions to future work.

For cosmological considerations, it is important to have a rough
estimate for the number of relativistic degrees of freedom associated with the
probe theory. Though the E-type probe theories do not appear to admit a weakly
coupled Lagrangian description, we can estimate the number of effective ``free field''
degrees of freedom in terms of the central charge $a_{IR}$ associated with any conformal field
theory in four dimensions. In the examples studied in \cite{HVW}, it was found
that the value of $a_{IR}$ for realistic T-brane configurations is $a_{IR} \sim 3 \pm 1$. In
our normalization a chiral superfield has
$a\left(  \text{chiral superfield}\right)  = 1/48$. On the other hand, the
number of relativistic degrees of freedom associated with a free chiral superfield
is:%
\begin{equation}
g_{\ast}(\text{chiral superfield})=2\times\left(  1+\frac{7}{8}\right)
=\frac{15}{4},%
\end{equation}
where $1$ is the contribution from the scalar components, and $7/8$ is the
contribution from the fermionic components. We therefore conclude that the number of
effective free field degrees of freedom associated with the probe D3-brane theory is roughly:%
\begin{equation}
g_{\ast}(\text{D3-brane})=\frac{15/4}{1/48}\times a_{IR}\sim 550 \pm 200.
\end{equation}
This should be compared with the number of relativistic species in the MSSM,
which is:%
\begin{equation}
g_{\ast}(\text{MSSM})=228.75.
\end{equation}
In other words, there are a significant number of effective ``free field'' degrees of freedom
in the probe sector. Our expectation is that most of these degrees of freedom are associated with the
$3-7_{vis}$ strings. This is because the initial tilting of the seven-brane configuration takes place within the $SU(5)_{\bot}$ factor,
and so mainly leads to heavy masses for the $3-7_{hid}$ strings.\footnote{Of course, this also produces a seesaw effect which makes some of the
$3-7_{hid}$ much lighter than the $3-7_{vis}$ strings.} This means that after the $3-7_{vis}$ strings freeze out, most of the free field relativistic degrees of freedom of the probe sector will have been depleted. Let us note, however, that the running of the beta function leads to an effect on the order of two to four $5 \oplus \overline{5}$'s, which is a much smaller number.

The main source of communication between the probe D3-brane sector and the Standard
Model sector is via the $3-7_{vis}$ strings, which are charged under both gauge groups.
Kinetic mixing between the visible and hidden sector $U(1)$'s provides another source of communication.

At the temperature $T\sim T^{f}_{med}$, the $3-7_{vis}$ strings and potentially other heavy states of the probe decouple from the thermal bath and freeze out. In most examples of freeze out, this temperature is comparable to that of the mass of the species in question, though there can be distinctions by factors of order $10$ so that $m/T^{f} \sim 10$. To compute the relic abundance generated from the decay of $3-7_{vis}$ states we should in principle integrate the Boltzmann equations, taking account of $3-7_{vis}$ decays as well as the effects of the inverse processes. This latter effect leads to thermal washout, the effect of which is included through a multiplicative factor $\eta_{washout}$. Washout is expected when the decay rate is fast compared to the size of the Hubble patch.

The out-of-equilibrium decay of these heavy $3-7_{vis}$ strings violates the matter number of the Standard Model and, as we have already seen also violates C
and CP. Thus, the Sakharov conditions are satisfied, and leads to a net
amount of baryon asymmetry. Note that this decay also includes light $3-7_{hid}$ strings which are the dark matter of this scenario. The relic abundances between the two sectors are thus correlated.\footnote{In principle, we should also consider the decay of the other heavy GUT singlets of the probe sector, though such modes dominantly decay just to probe sector states, and so do not obviously generate an asymmetry. The amount of entropy generated by such decays is comparable to that of the $3-7_{vis}$ decays, so at this level of analysis we can neglect this effect.} Electroweak sphaleron processes
equilibrate the amount of baryon and lepton number in the visible sector, but as we explain later, do not affect the matter asymmetry in the hidden sector.

The decay of the $3-7_{vis}$ strings is expected to create a large number of $3-7_{hid}$ strings. Even so, the overall energy density is expected to be roughly comparable between the two sectors. The overall number density of such hidden sector states consists of an asymmetric component, which is correlated with the visible sector matter asymmetry, and a large symmetric component, which is part of the hidden sector thermal bath. Due to the strongly coupled nature of the $U(1)$ gauge theory, this symmetric component can efficiently annihilate to dark $U(1)$ gauge bosons which then kinetically mix to visible sector radiation. Upon freezing out, the $3-7_{hid}$ strings continue to annihilate, depleting the symmetric component of the dark matter. See for example \cite{Davoudiasl:2010am} for related discussions of kinetic mixing between visible and hidden sectors as a mechanism for depopulating the symmetric component of the dark matter.

The matter asymmetry in the visible and hidden sectors is
conveniently expressed in terms of the yield of the species, which is the ratio of the
number density to the entropy density. This is a convenient
quantity to compute because it remains constant as the Universe expands. In our case, we have:
\begin{equation}
Y_{\Delta B}\sim Y_{DM}\sim \varepsilon_{CP} \cdot \frac{g_{eff,med}}{g_{\ast}(M_{\cancel{CFT}})}\frac{n_{med}}{n_{rad}}
\sim \varepsilon_{CP} \cdot \frac{g_{eff,med}}{g_{\ast}(M_{\cancel{CFT}})} \eta_{washout}%
\end{equation}
where in the above, $g_{eff,med} \sim 1 - 10$ is the number of ``free field'' degrees of freedom associated with the mediator states, counted in terms of GUT multiplets. The size of the beta function contribution to $SU(5)_{GUT}$ to the probe sector is of order two to four $5 \oplus \overline{5}$'s \cite{HVW}, so we expect $g_{eff,med} \sim 1 - 10$. As usual, $g_{\ast}(M_{\cancel{CFT}}) \sim 10^{3}$ is the number of relativistic degrees of freedom at a temperature $T \sim M_{\cancel{CFT}}$. Finally, $0 \leq \eta_{washout} < 1$ is a factor associated with thermal washout effects. In what follows, we shall typically take $g_{eff,med} / g_{\ast}(M_{\cancel{CFT}}) \sim 5 \times 10^{-3}$. The final equality is obtained
by equating the number density of mediator $3-7_{vis}$ states $n_{med}$ with that of the radiation as is the case for thermal baryogenesis, with a washout factor given by $\eta_{washout}$. As we explain later, the decay of the $3-7_{med}$ states is slow enough that there is not much washout.

The overall relic abundance stored in each species is then given as:%
\begin{align}
\Omega_{\Delta B}h^{2}  &  =\frac{\rho_{B}^{0}h^{2}}{\rho_{c}^{0}}=\frac{s^{0}h^{2}%
}{\rho_{c}^{0}}m_{B}Y_{\Delta B}\\
\Omega_{DM}h^{2}  &  =\frac{\rho_{DM}^{0}h^{2}}{\rho_{c}^{0}}=\frac{s^{0}%
h^{2}}{\rho_{c}^{0}}m_{DM}Y_{DM}.
\end{align}
where the superscripts denote present day values for the energy density ($\rho^{0}$) and entropy density ($s^{0}$).
To achieve an appropriate relic abundance, we require $m_{DM}\sim10$ GeV as usual in asymmetric dark
matter scenarios  (see for example \cite{Kaplan:1991ah, Thomas:1995ze,
Kaplan:2009ag, Cohen:2010kn}). Let us note, however, that this value should
be viewed only as a rough estimate, since we do not know the exact number density
in the hidden sector. This subtlety can be potentially important
in other scenarios which aim to correlate the visible and hidden sector
relic abundance, see for example \cite{Buckley:2010ui,Falkowski:2011xh}.

In the remainder of this section, we discuss some additional details of this
scenario. First, we discuss how the \textit{symmetric} component of the
dark matter can efficiently annihilate away to visible sector radiation,
thus leading to the expected relation between the visible and hidden sector dark matter \textit{asymmetries}. Next, we discuss the effects of sphaleron
interactions, and explain why even when $U(1)_{B-L} \subset E_8$ is a symmetry of the seven-brane configuration,
sphaleron processes do not eliminate the visible sector matter asymmetry.
An additional subtlety in generating an appropriate relic abundance is the potential
effects of other issues in cosmology such as the over-production of
gravitinos (especially in high scale gauge mediation models common to some F-theory GUT scenarios \cite{HVGMSB,MarsanoGMSB,HKSV,Heckman:2010xz}), and the
potential effects of late decaying scalars. We explain how to account for these effects in our scenario. After taking into account these subtleties, we discuss whether the probe scenario is capable of satisfying
various cosmological constraints. Quite remarkably, three a priori different requirements
-- a correlated relic abundance, correct mass for the dark matter, and the absence of significant thermal
washout effects -- are all satisfied for SCFT breaking scales in the range of $M_{\cancel{CFT}} \sim 10^{9} - 10^{13}$ GeV.
This is a remarkably \textit{predictive} feature of this scenario. After this, we briefly comment on prospects for detecting signatures of
this scenario.

\subsection{Depleting the Symmetric Component of the Dark Matter}

In this subsection we discuss in greater detail how the visible and hidden sector maintain thermal contact after the freeze out of the $3-7_{vis}$ strings due to kinetic mixing. After this, we explain how kinetic mixing leads to annihilation of a significant component of the symmetric component of the dark matter.

Let us begin by discussing in more detail the thermal history of the visible and
hidden sectors after the freeze out of the $3-7_{vis}$ strings. Even though the
$3-7_{vis}$ strings freeze out and decay, the visible and hidden sectors remain in contact due to kinetic mixing effects. There are various annihilation modes between the visible and hidden sector. In terms of a weakly coupled Lagrangian description, the dominant annihilation mode is expected to be s-channel annihilation $Q^{\dag}_{hid}Q_{hid} \rightarrow \Psi_{SM}^{\dag}\Psi_{SM}$ which proceeds via an intermediate $U(1)_{D3}$ which kinetically mixes with the visible sector photon. It is convenient to characterize such annihilations in terms of a hidden sector ``Fermi-constant'':
\begin{equation}
G_{hid} \sim \frac{\kappa_{mix} \cdot g_{hid} g_{vis}}{M^{2}_{U(1)_{D3}}}
\end{equation}
where $\kappa_{mix}$ is the amount of kinetic mixing. The expected annihilation
cross section at high and low temperatures is then given by:
\begin{equation}
\sigma_{ann}\sim\left\{
\begin{array}
[c]{c}%
G_{hid}^{2}M_{U(1)_{D3}}^{4}T^{-2}\text{ for }T\gtrsim M_{U(1)_{D3}}\\
G_{hid}^{2}M_{hid}^{2} \ \ \ \ \ \ \ \ \ \ \text{for }T\lesssim M_{U(1)_{D3}}%
\end{array}
\right\}
\end{equation}
The presence of such an annihilation mode allows the hidden sector to remain in thermal contact with the visible sector. Since we expect the mass of the $U(1)_{D3}$ to be somewhat similar to the mass of the dark matter, which is in turn close to the weak scale, this is similar to visible sector annihilations which are mediated by $Z^{0}$-bosons. In other words, thermal contact is maintained between the visible and hidden sectors until quite late.

Let us now discuss the fate of the symmetric component of the dark matter. By convention,
the number $n_{DM} - n_{\overline{DM}} > 0$ is referred to as the number density for the asymmetric component, and $n_{\overline{DM}}$ refers to the
number density of the symmetric component. To compute the overall relic abundance between the visible and hidden sector, recall that we can track the matter \textit{asymmetry} via the conserved $U(1)_{7_{hid}} \subset E_8$ of the $7_{hid}$-brane. In the visible sector, most of the matter and anti-matter will annihilate away efficiently to massless radiation, leaving only a net matter asymmetry. In the hidden sector, the situation is somewhat different, because the annihilation involves a gauge boson which has a mass. It is therefore important to study how much of the symmetric component of the dark matter candidate is able to annihilate back to visible sector radiation.

If no annihilation occurs, there is a potentially serious issue with the overall relic abundance. Indeed, in such a world, the yield for the symmetric component would be order one, and would therefore be significantly higher than that of the asymmetric component. this is a generic issue in ``asymmetric dark matter'' scenarios, discussed for example in \cite{Kaplan:2009ag}. The basic idea is that there must be some way to deplete the symmetric component of dark matter.
The presence of kinetic mixing is one way to accomplish this transfer discussed for example in \cite{Davoudiasl:2010am, Shelton:2010ta}.

There are two different thermal histories for the symmetric component in our probe sector scenario, which depend on the relative masses of the $3-7_{hid}$ strings and the dark $U(1)_{D3}$ gauge boson. As we now explain, in both cases we find that most of the symmetric component is eliminated by kinetic mixing to the visible sector. Consider first the case $M_{U(1)_{D3}} < M_{hid}$. Here, the symmetric component annihilates away into hidden sector radiation, given by dark $U(1)_{D3}$ gauge bosons. These dark photons then decay to visible sector states, with decay rate $M_{U(1)_{D3}} \times \kappa_{mix}^{2}$. For typical values of $\kappa_{mix} \sim 10^{-3}$, we see that this decay occurs on timescales much faster than the onset of big bang nucleosynthesis.

Next consider the case $M_{hid} < M_{U(1)_{D3}}$, which is in some sense the more generic situation to expect if the hidden sector develops weak scale vevs induced by supersymmetry breaking effects. Here, the dark photons instead decay back into the symmetric component of dark matter. After the $3-7_{hid}$ strings freeze out from the thermal bath they can still efficiently annihilate to Standard Model particles. Let us now discuss the net yield expected from the symmetric component.

In the special case where there is no matter asymmetry for the $3-7_{hid}$ strings, we can apply the standard freeze out computation of the yield for the symmetric component (see for example \cite{KT}). This yields a very conservative estimate for the overall yield of the symmetric component:
\begin{equation}
Y_{symm} \lesssim Y_{cnsrv} \sim \frac{\sqrt{g_{\ast}(T^{f}_{hid})}}{g_{\ast S}(T^{f}_{hid})}\frac{x^{f}_{hid}}{M_{hid} M_{pl} \sigma_{ann}} \sim \frac{G_{hid}^{-2}}{M^{3}_{hid} M_{PL}}
\end{equation}
where here, $M_{pl} \sim 10^{19}$ GeV is the Planck mass and $x^{f}_{hid} = M_{hid} / T^{f}_{hid}$, with $T^{f}_{hid}$ the freeze out temperature for the $3-7_{hid}$ strings. In other situations, $x^{f} \sim 1- 10$, and we shall use a similar estimate here. Hence, we see that the symmetric component of dark matter is of order $10^{-10}$ or less provided $G^{-1/2}_{hid} < 10^{5}$ GeV. Insofar as we expect $M_{U(1)_{D3}}$ to be close to the mass of the dark matter, e.g. $10$ GeV, and $\kappa_{mix} \sim \sqrt{\alpha_{D3} \cdot \alpha_{Y}} / 4 \pi \sim 10^{-3} - 10^{-2}$, we see that the yield for the symmetric component will be smaller than that of the visible sector. Hence, we see that in a scenario where a dark matter asymmetry is also created, there is little danger of overproducing the symmetric component. Let us comment that in scenarios with additional dilution effects, the yield $Y_{cnsrv}$ can be even bigger than $10^{-10}$ because the asymmetric yield $Y_{asymm}$ must also be increased. This enables smaller values of $\kappa_{mix}$, and larger values of $M_{U(1)_{D3}}$. In this way, we can accommodate larger masses for the $U(1)_{D3}$ gauge boson, up to around $10^{3}$ GeV.

As we have already mentioned, this is actually a quite conservative overestimate for the yield of the symmetric component of dark matter. The reason is that there is a further suppression from the presence of an overall dark matter asymmetry. Indeed, annihilations of the $\overline{DM}$'s can proceed for longer, because they see an effectively large number of $DM$ states. This can also be seen directly by integrating the Boltzmann equations, where the existence of an asymmetric part changes the boundary conditions for these differential equations. This leads to an effective exponential suppression of the form $\exp{(-A)}$ in $n_{\overline{DM}}$ \cite{Griest:1986yu}. Here, $A \sim Y_{cnsrv} / Y_{asymm}$. This point has been stressed in the specific context of related scenarios with kinetic mixing, such as \cite{Davoudiasl:2010am}.

For all of these reasons, we see that generically, the symmetric component is expected to be depleted for generic values of the kinetic mixing parameter $\kappa_{mix}$ and ``dark photon'' mass $M_{U(1)_{D3}}$. Let us note that such considerations are also in accord with various model independent bounds on the mass of the $U(1)_{D3}$ and the overall amount of kinetic mixing, studied for example in \cite{Hook:2010tw}. For example, taking $M_{U(1)_{D3}} \sim 20$ GeV, $\kappa_{mix} \sim 10^{-3}$ is compatible with present bounds.

\subsection{$B-L$ and Sphalerons}

Since we are assuming $M_{\cancel{CFT}}$ is above the scale
of the electroweak phase transition, sphaleron interactions are expected to enforce the condition $B+L=0$
after equilibration \cite{Kuzmin:1985mm}. Sphaleron processes are a non-perturbative phenomenon associated with the
fact that $B$ and $L$ are anomalous in the Standard Model. In the Standard
Model, the corresponding currents satisfy the relation:%
\begin{equation}
\partial_{\mu}j_{B}^{\mu}=\partial_{\mu}j_{L}^{\mu}=N_{\text{gen}}%
\times\left(  \frac{g^{2}}{32\pi^{2}}W_{\mu\nu}\widetilde{W}^{\mu\nu}%
-\frac{g^{\prime2}}{32\pi^{2}}F_{\mu\nu}\widetilde{F}^{\mu\nu}\right)
\end{equation}
where $W_{\mu \nu}$ is the $SU(2)_{L}$ field strength, $\widetilde{W}_{\mu \nu}$ is its dual, and
$F_{\mu \nu}$ is the hypercharge field strength, with $\widetilde{F}_{\mu \nu}$ its dual. The
conversion of the baryon number current to the lepton number current can be
viewed as mediated by the combination of two triangle diagrams joined by a
non-perturbative sphaleron process. In each triangle diagram, chiral matter fields
of the Standard Model run in the corresponding loop. The importance of the
sphaleron process is that it enforces $B+L=0$ in the end.

In the simplest GUT baryogenesis scenarios with conserved
$(B-L)$, this can be problematic, because simultaneously
setting $B+L = 0$ and $B-L = 0$ also eliminates all baryon asymmetry. On the
other hand, sphaleron processes are actually beneficial in leptogenesis scenarios, as they
provide a way to convert an initial lepton asymmetry to a baryon asymmetry \cite{Fukugita:1986hr}.

As we now explain, the specific $(B-L)$ charge assignment for the $3-7_{vis}$
strings leads to a net violation of $(B-L)$ in the visible sector
which cannot be removed by sphaleron processes. Let us first note that in
F-theory GUT scenarios with Majorana neutrinos, $U(1)_{B-L}$ is
already explicitly broken at a very high scale. In the probe D3-brane this
translates to a specific choice of tilting parameter $\Phi(Z_1,Z_2)$. Since
$U(1)_{B-L}$ is not a symmetry of the probe D3-brane, there is no reason to
expect any $B-L$ conservation. In such a situation we generically expect
sphaleron processes to erase some, but not all of the initial asymmetry.

Focusing now on scenarios where $(B-L)$ is preserved as can happen in the
Dirac neutrino scenarios of \cite{BHSV,EPOINT}, $U(1)_{B-L}$ is given as a linear combination of
$U(1)_{Y}$ and an additional $U(1)_{\chi} \subset SU(5)_{\bot} \subset E_{8}$. As noted
in subsection \ref{BVIOL}, if one just tracks the $U(1)_{\chi}$ charge in the visible sector,
the decay of $3-7_{vis}$ strings appears to violate $(B-L)$. What has happened is that the
$3-7_{hid}$ strings carry net $U(1)_{\chi}$ charge. From the
form of the decay processes we have been considering, we see that the excess $(B-L)$ charge has been
\textquotedblleft hidden away\textquotedblright in the dark sector! This is similar to the asymmetric
dark matter scenarios of \cite{Davoudiasl:2010am,Shelton:2010ta}. As a brief aside, let us observe that symmetry considerations 
alone strongly constrain the way the $(B-L)$ is generated in the visible and hidden sector. For example, due to the interactions amongst the 
$3-7_{vis}$ strings, there is no way to consistently assign a non-zero $U(1)_{\chi}$ charge to the messenger $3-7_{vis}$ strings.

This excess $(B-L)$ charge in the hidden sector cannot be removed by electroweak
sphaleron processes. Indeed, since the D3-brane probe sector is a vector-like theory, its contribution to all triangle
diagrams vanishes. In other words, there is no way for a sphaleron process to
convert the excess of $U(1)_{\chi}$ charge present in the hidden sector back
to the visible sector. Since $B_{vis} - L_{vis} \neq 0 $ in the visible sector, such sphalerons will simply impose
a relation between the net baryon and lepton asymmetry via the equilibration condition $B_{vis} + L_{vis} = 0$.
In the context of weakly coupled scenarios such as leptogenesis, one can compute
the exact conversion rate from leptons to baryons, which is an order one number somewhat less than one.
In the present context, we cannot deduce this exact value, because it depends
on the excess amount of $10$'s versus $\overline{5}$'s created. Nevertheless, this will again be an order one number.

\subsection{Gravitinos and Late Entropy Production}

Another issue which is quite common in supersymmetric theories, especially, in high-scale gauge mediation
scenarios is the potential over-production of gravitinos \cite{Pagels:1981ke,Weinberg:1982zq}.
The production rate was carefully studied in \cite{Ellis:1995mr}. As reviewed for
example in \cite{FGUTSCosmo}, for a $10-100$ MeV mass gravitino, the resulting relic abundance
turns out to be larger than the dark matter relic abundance by a factor of
$\sim10^{4}$. One way to evade this over-closure is to simply lower the reheating
temperature of inflation so that the relic abundance is sufficiently low.
Another way to evade this constraint in the context of F-theory GUT\ scenarios
is to consider the dynamics of a late decaying scalar such as the saxion. As
found in \cite{FGUTSCosmo} (see also \cite{Ibe:2006rc,KitanoIbeSweetSpot}), the oscillations of this scalar can come
to dominate the energy density of the Universe. Its subsequent decays can then
lead to an effective dilution of order $\mathcal{D}_{decay} \sim 10^{-4}$. Depending
on the exact choice of initial amplitude for this scalar, this can either dilute away all gravitinos to
a negligible value, or in certain cases allow such gravitinos to make up a
significant component of the dark matter relic abundance.

Though we do not wish to commit to a particular scenario for gravitino
cosmology or late decaying scalars in this paper, let us note that
our previous considerations can accommodate such dilution effects. The main change is to the
overall yield, which must be multiplied by a \textquotedblleft dilution
factor\textquotedblright\ $\mathcal{D}$, which is the ratio of the entropy
densities before and after the decay of a mode which has come to dominate the energy density of the universe:
\begin{equation}
\mathcal{D}_{decay} = \frac{s^{before}}{s^{after}}.
\end{equation}
Recall that the yield is given by the ratio of the number density to the
present entropy density. Hence, the yields of all species before and after decay are related by:
\begin{equation}
Y^{after} = \mathcal{D}_{decay} \cdot Y^{before}.
\end{equation}
Note, however, that this relation preserves the basic relation connecting the yields of visible and hidden sector abundances:
\begin{equation}
\frac{Y^{after}_{\Delta B}}{Y^{after}_{DM}} \sim \frac{Y^{before}_{\Delta B}}{Y^{before}_{DM}}.
\end{equation}
Thus, even in the presence of dilution the visible and hidden sector
relic abundances generated by the probe theory will still be
correlated, still requiring the dark matter mass to be on the order of $10$ GeV.

\subsection{A Confluence of Parameters}

Up to now, we have written all estimates in terms of the parameters of the SCFT
breaking scale. This sets up a potentially significant source of tension for the probe D3-brane sector, because
this single scale and various order one parameters of the SCFT must satisfy three a priori
\textit{independent} requirements. First, we have required that the relic abundances of dark matter and
visible matter be correlated. This imposes a condition on the mass of the dark matter, which in
turn fixes the SCFT breaking scale. Further, we have demanded that the actual relic abundance comes out correctly.
Finally, to generate the right relic abundance, there must not be significant thermal washout effects. In this subsection
we show that all of these requirements are satisfied when the CFT breaking scale is on the order of $\sim 10^{9} - 10^{13}$ GeV, where
the specific value depends on the choice of intersecting seven-brane configuration for the visible sector.

The first point we should address is the correlation between the visible
sector matter asymmetry, and the hidden sector asymmetry. As mentioned
previously, in a given decay process of a $3-7_{vis}$ string mediator state into Standard
Model states and some number of $3-7_{hid}$ states, we should not expect a simple
match between the visible and hidden sector degrees of freedom. However, as we
have already noted, there is an effective $7_{hid}$ charge deposited to the
visible sector and the hidden sector. Since the initial relic abundance is
neutral under the local charge from the $7_{hid}$-brane, we conclude that the magnitude of the charge densities
in each sector is equal. This is quite important, because although we do not have an exact count of the number of degrees of freedom created through an individual decay process, most of these states rejoin a strongly coupled thermal bath anyway. In this sense, we can still correlate the effective matter asymmetries in the visible and hidden sectors:%
\begin{equation}
n_{\Delta B}\sim n_{DM}%
\end{equation}
which in turn leads to a correlation between the yields of the two species:%
\begin{equation}
Y_{\Delta B}\sim Y_{DM}.
\end{equation}
Since the energy density stored in a non-relativistic species is proportional
to its mass, we then obtain the phenomenological requirement that to explain
the appropriate relic abundance of dark matter from this hidden sector,
\begin{equation}
\frac{m_{DM}}{m_{B}}\sim 10.
\end{equation}
This leads to the requirement that the dark matter is on the order of $10$ GeV
in mass. This is the same condition common to most asymmetric dark matter scenarios (see
for example \cite{Kaplan:1991ah,Thomas:1995ze,Kaplan:2009ag}). Of course, this
precise value can be affected by various order one
quantities which can push the required mass of the $3-7_{hid}$ masses either
up or down.

Let us now turn to the actual relic abundance computation. We expect there to be only a small
amount of washout from inverse decay processes which convert $3-7_{hid}$ and Standard Model states back
to $3-7_{vis}$ strings. The reason is that in order for inverse processes to proceed efficiently, these
decay products must be produced before they can separate by more than a Hubble patch.
Washout effects are expected mainly when the decay rate is faster than the scale set by one Hubble patch:
\begin{equation}
H|_{T = M_{\cancel{CFT}}} < \Gamma_{D}
\end{equation}
where $H|_{T = M_{\cancel{CFT}}}$ is the value of the Hubble parameter at the temperature of SCFT breaking, and
$\Gamma_{D}$ is the decay rate of the $3 - 7_{vis}$ states. Plugging in our
estimate of $\Gamma_{D}$ obtained in section \ref{sec:DECAY}, fast washout occurs when:
\begin{equation}
1.66 \sqrt{g_{\ast}(M_{\cancel{CFT}})} \frac{M_{\cancel{CFT}}^{2}}{M_{pl}} < M_{\cancel{CFT}} \cdot \left( \frac{M_{\cancel{CFT}}}{M_{GUT}} \right)^{2 \nu}
\end{equation}
where here $\nu \sim 1$.

In our case, the decay rate for the $3-7_{vis}$ strings is comparatively slow. For example, when $M_{\cancel{CFT}} \sim 10^{13}$ GeV and $\nu = 1$, we obtain $\Gamma_{D} \sim 10^{6}$ GeV, while $H|_{T = M_{\cancel{CFT}}} \sim 4 \times 10^{8}$ GeV. In other words, the decay of the $3-7_{vis}$ states is slow enough that we do not expect significant thermal washout.

In some scenarios, especially those where over-production of the matter asymmetry is required (as when $\mathcal{D}_{decay} \ll 1$), some washout may occur. This is because increasing the production also requires increasing the decay rate $\Gamma_{D}$, which in turn can lead to some washout since $\Gamma_D \gtrsim H|_{T = M_{\cancel{CFT}}}$.  This leads to a Boltzmann suppression factor in the overall yield, which amounts to a value of $\eta_{washout} \sim H|_{T=M_{\cancel{CFT}}} / \Gamma_{D}$. It would be interesting to perform a more precise analysis of washout in such cases, though this would require computation in strongly coupled probe sector. In such cases we shall simply absorb such washout effects into a net dilution factor:
\begin{equation}
\mathcal{D}_{net} \equiv \mathcal{D}_{decay} \cdot \eta_{washout}.
\end{equation}

Taking into account various dilution effects, the baryon asymmetry yield is:
\begin{equation}
Y_{\Delta B} \sim \mathcal{D}_{net} \frac{g_{eff,med}}{g_{\ast}(M_{\cancel{CFT}})} \cdot \left(\frac{M_{\cancel{CFT}}}{M_{GUT}} \right)^{2 \nu}
\end{equation}
where $\nu$ is an order one number which is set by the scaling
dimensions of operators in the CFT, and is expected to be less than one.
On the other hand, obtaining the correct dark matter relic abundance
leads to a \textit{different} relation:
\begin{equation}
\frac{\Omega_{DM}}{\Omega_{\Delta B}} \sim \frac{M_{\cancel{CFT}}}{m_{B}} \cdot \left( \frac{M_{\cancel{CFT}}}{M_{\ast}}\right)^{\alpha}
\end{equation}
for $\alpha \sim O(1)$ depending on the details of the seven-brane monodromy scenario. In weakly coupled models, $\alpha = 1 ,2 ,3 $, though there can be corrections to this based on D-term contributions. As explained earlier, we expect such effects to be small.

Varying over $\alpha \sim O(1)$, we can solve for $M_{\cancel{CFT}}$ and $\nu$ subject to the conditions
imposed on the rest of the scenario. For concreteness, we take $\alpha = n = 1,2,3$ an integer to get a sense of the various allowed values. 
The ratio of the CFT breaking scale to $M_{\ast}$ must then simultaneously satisfy the constraints:
\begin{equation}
10^{-10} \sim \frac{\mathcal{D}_{net}}{g_{\ast}(M_{\cancel{CFT}})} \cdot \left(\frac{M_{\cancel{CFT}}}{M_{GUT}} \right)^{2 \nu},
\,\,\,\, \frac{\Omega_{DM}}{\Omega_{\Delta B}} \sim \frac{M_{\cancel{CFT}}}{m_{B}} \cdot \left( \frac{M_{\cancel{CFT}}}{M_{\ast}}\right)^{n}.
\end{equation}
Here we display the numerical values of the resulting parameters, in the cases with and without dilution.
For the purposes of illustration, we take $g_{\ast}(M_{\cancel{CFT}}) \sim 10^{3}$,
$g_{eff,med} / g_{\ast}(M_{\cancel{CFT}}) \sim 5 \times 10^{-3}$,
$M_{GUT} \sim 2 \times 10^{16}$ GeV and $M_{\ast} \sim 10^{17}$ GeV.

\begin{equation}%
\begin{tabular}
[c]{|c|c|c|c|c|}\hline
$\mathcal{D}_{net}\sim1$ & $\nu$ & $M_{\cancel{CFT}}$ (GeV) & $\Gamma_{D}$ (GeV) & $H|_{T=M_{\cancel{CFT}}}$
(GeV)\\\hline
$n=1$ & $0.5$ & $1\times10^{9}$ & $5\times10^{1}$ & $4\times10^{0}$\\\hline
$n=2$ & $0.8$ & $5\times10^{11}$ & $2\times10^{4}$ & $1\times10^{6}$\\\hline
$n=3$ & $1.1$ & $1\times10^{13}$ & $5\times10^{5}$ & $4\times10^{8}$\\\hline \hline
$\mathcal{D}_{net}\sim10^{-2}$ & $\nu$ & $M_{\cancel{CFT}}$ (GeV) & $\Gamma_{D}$ (GeV) & $H|_{T=M_{\cancel{CFT}}%
}$ (GeV)\\\hline
$n=1$ & $0.4$ & $1\times10^{9}$ & $5\times10^{3}$ & $4\times10^{0}$\\\hline
$n=2$ & $0.6$ & $5\times10^{11}$ & $1\times10^{6}$ & $2\times10^{6}$\\\hline
$n=3$ & $0.8$ & $1\times10^{13}$ & $5\times10^{7}$ & $4\times10^{8}$\\\hline \hline
$\mathcal{D}_{net}\sim10^{-4}$ & $\nu$ & $M_{\cancel{CFT}}$ (GeV) & $\Gamma_{D}$\ (GeV) &
$H|_{T=M_{\cancel{CFT}}}$ (GeV)\\\hline
$n=1$ & $0.2$ & $1\times10^{9}$ & $5\times10^{5}$ & $4\times10^{0}$\\\hline
$n=2$ & $0.4$ & $5\times10^{11}$ & $2\times10^{8}$ & $1\times10^{6}$\\\hline
$n=3$ & $0.5$ & $1\times10^{13}$ & $5\times10^{9}$ & $4\times10^{8}$\\\hline
\end{tabular}
\end{equation}%
In the scenarios where $\Gamma_{D} > H|_{T=M_{\cancel{CFT}}}$, some amount of thermal washout is expected.
In the above table, this has been absorbed into the value of $\mathcal{D}_{net}$.

Let us comment on the numerical values found in this table. In all cases, the CFT breaking scale is in the rather narrow range of $10^{9} - 10^{13}$ GeV. Further, we see that the exponent $\nu$ associated with details of scaling dimensions in the CFT is on the order of $\nu \sim 1$, though it is typically less than one. In the ``chiral approximation'' to scaling dimensions found in \cite{HVW}, one can estimate $\nu$ to be on the order of $\sim 0.2$ in many examples. Of course, as we have already mentioned, the actual decay amplitude involves various non-chiral data, and further, the precise value of $\nu$ can be tuned by including contributions from $W_{flux}$. By inspection of the table, we find that in most cases, the amount of washout is compatible with achieving the correct yield. We also see that in scenarios where dilution from $\mathcal{D}_{decay}$ is significant, the more favorable scenarios are those with larger values of $n$.

Quite remarkably, even without detailed information about the strongly coupled sector, we have landed on a
consistent scenario of stringy matter genesis which simultaneously correlates the mass \textit{and}
the relative relic abundances of visible and dark matter!

\subsection{Prospects for Detection}

In the previous sections we have seen that in a range of parameters natural
for probe D3-branes, the mass of the dark matter, and relic abundances are
compatible with observation. Though a full treatment is beyond
the scope of the present paper, in this subsection we briefly comment on the prospects
for detecting this type of dark matter.

A novelty of the scenario considered here is that it involves a strongly coupled $U(1)$
gauge theory. Moreover, as explained in section \ref{sec:MASS}, we expect there to generically
be both light electric and magnetic states charged under the hidden $U(1)_{D3}$.
These states can in turn interact with the visible sector through
electric and magnetic kinetic mixing terms. As far as we are aware, the
phenomenological signatures associated with magnetic mixing are not well
explored, though it would clearly be of interest to study such effects. See
\cite{Chang:2010en} for some discussion on the effects of dark magnetic dipoles as a potential explanation of
various anomalies in some dark matter experiments.

Putting aside such concerns, let us now turn to constraints from various
direct detection experiments. At some level, such signatures are unavoidable,
because in order for the present model to avoid over-producing a symmetric
component of dark matter, it is necessary for this component to annihilate to
visible sector radiation. This sort of cross section may also show up in
the scattering of dark matter off of nuclei.\footnote{The dark matter mass is
lighter than $\sim 100$ GeV, so we do not expect a signal to be
generated for the PAMELA and FERMI experiments.} To illustrate the basic idea, we
focus on scenarios where the mass of the $U(1)$ is heavier than that of the
dark matter. The estimate for the annihilation cross-section is roughly:%
\begin{equation}
\mathcal{\sigma}_{ann}\sim G_{hid}^{2}\cdot M_{hid}^{2}.
\end{equation}
Typical values for this cross section can range from $10^{-38}$ cm$^{2}$ in
the case of $M_{hid}\sim M_{U(1)_{D3}}\sim10$ GeV to $10^{-50}$ cm$^{2}$ in
the case of $M_{DM}\sim10$ GeV and $M_{U(1)_{D3}}\sim10^{3}$ GeV. Viewing this
as a very rough estimate for the characteristic size of a t-channel dark
matter/nucleon cross-section, we see that nucleon scattering with dark matter
may be in an accessible range in such models. We again must caution that at
this point, various theoretical uncertainties, as well as the competition between electric and magnetic mixing and how
this interacts with nuclei all figure into the discussion. For example, note
that holding all other parameters fixed, increasing $M_{U(1)_{D3}}$ from $10$
GeV up to $10^{3}$ GeV, will cause the cross section to decrease by a factor
of $10^{-12}$. One expects this to be correlated with the value of the conservative estimate on the symmetric component of the yield,
$Y_{cnsrv}$, though the precise relation involves various theoretical uncertainties.

Suspending such concerns, let us at least see whether this scenario could be compatible with present
experimental bounds. As an example, modulo various astrophysics effects, figure 5 of \cite{Aprile:2010um} indicates a bound on
the spin-independent elastic WIMP/nucleon cross section of order $10^{-41}-10^{-42}$ cm$^{2}$ for a $10$ GeV WIMP. Note, however, that
even changing the mass of the WIMP to $5$ GeV already significantly
weakens the bound to $10^{-40}$ cm$^{2}$. Given all of the theoretical
uncertainties in the present model, we see that our present considerations are
close to this range. Considering how tightly the other elements of the model
hang together, it would clearly be of interest to investigate further the
expected prospects for direct detection. Additional novel signatures may be
present due to the presence of both electric and magnetic mixing
with the visible sector. We leave this as an exciting direction for future investigation.

\section{Conclusions\label{sec:CONCLUDE}}

D3-branes probes of Yukawa points are well-motivated extensions of the Standard Model in F-theory GUTs.
In this paper we have seen that such probe theories lead to a rich additional sector which can extend the
Standard Model at higher energy scales. Breaking the CFT leads to a hierarchy of mass scales in
which matter charged under $SU(5)_{GUT}$ is naturally heavier than some of the $3-7_{hid}$ strings neutral under the GUT group.
We have also seen that in a cosmological scenario, the freeze out and subsequent decay of such $3-7_{vis}$ strings induce
a baryon asymmetry, with a correlated relic abundance in the hidden sector. For suitable values of the CFT breaking scale,
both the correct relic abundances, and baryon asymmetry will be achieved. In the remainder of this section
we discuss potential future directions of investigation.

Let us note that although we have phrased our discussion in
terms of a D3-brane probe in F-theory, similar considerations also hold for
model building in heterotic M-theory, where via dualities, the role of
the D3-brane is replaced by an M5-brane. This widens the applicability of this
class of models, providing a natural mechanism for generating a baryon asymmetry
and correlated dark matter relic abundance in many string based models.

A novel feature of these probe sectors is that both electric and magnetic kinetic mixing are generically present. This sector
features a strongly coupled $U(1)$ sector, and rather exotic dark matter states which may lead to novel
phenomenological signatures, especially in various direct detection experiments.
It would clearly be interesting to understand such effects better.

In this paper we have focused on the effects of such a probe D3-brane after the Universe has already inflated. We have also seen that
the D3-brane contains various higher dimension operators, and in particular, that such scaling effects can lead to flattened out effective
potentials for the motion of the D3-brane. It would be interesting to see
whether the motion of the D3-brane would lead to a phenomenologically viable inflationary phase, given that in general, a
strongly coupled inflaton sector is in conflict with observations \cite{Rey:1986zk}.
Note that the D3-brane could naturally couple to the visible sector matter, and the inflaton decay (if made of $3-7_{vis}$ strings) would, as in this paper, simultaneously generate a visible and hidden sector relic abundance. This would be a non-thermal production mechanism, which could nevertheless retain the
main features of the scenario presented here.

We have also seen that depending on the details of the T-brane configuration, various hierarchical mass scales can be realized in the
probe sector. Indeed, one of the main points of this paper has been that the $3-7_{vis}$ strings charged under the Standard Model gauge group
are heavier than those of the $3-7_{hid}$ sector. Though in this paper we have focused on scenarios where the SCFT breaking scale
is in the range of $\sim 10^{9} - 10^{13}$ GeV, it is also interesting to contemplate the effects of far lower, TeV scale SCFT breaking, perhaps correlated with the scale of visible sector superpartner masses. This would provide a concrete stringy realization of a hidden
valley scenario \cite{Strassler:2006im}, which could also naturally accommodate some GUT-like structures. It is tempting to also speculate that due to their very low mass, there may be a sense in which the dynamics of the $3-7_{hid}$ sector is still approximately conformal, even below the scale of CFT breaking. This may realize an unparticle scenario \cite{Georgi:2007ek}, with a possibly colored unparticle sector \cite{Cacciapaglia:2007jq}. It would be interesting to study the associated collider signatures for such a scenario.

\section*{Acknowledgements}

We thank N. Arkani-Hamed, N. Craig, D. Gaiotto, D. Green, P. Langacker, M. Lisanti, C. Vafa,
B. Wecht and N. Weiner for many helpful discussions. We also thank D. Morrissey, C. Vafa, S. Watson
and B. Wecht for helpful comments on an earlier draft of this paper.
The work of JJH is supported by NSF grant PHY-0969448. The work
of SJR was supported in part by the National Research Foundation
of Korea grants 2005-009-3843, 2009-008-0372,
2010-220-C00003 and by the DOE grant DE-FG02-90ER40542.


\appendix
\section{CFT Scaling Dimensions}
In this Appendix we review the scaling dimensions of operators in the $\mathcal{N} = 1$ $\Phi$-deformation of the
$\mathcal{N} = 2$ Minahan-Nemeschansky theory with $E_8$ flavor symmetry studied in \cite{FCFT}.

In the $\mathcal{N} = 2$ Minahan-Nemeschansky theory with $E_8$ flavor symmetry, the Coulomb branch is parameterized by a dimension six
field $Z$, and the Higgs branch is parameterized by a dimension two operator $\mathcal{O}$ in the adjoint of $E_8$. The probe D3-brane
also contains a decoupled hypermultiplet $Z_1 \oplus Z_2$.
The dimensions of the operators of the $\cN=1$ $\Phi$-deformed theory are organized according to their $SU(5)_{\bot}$
representation content. For example, the infrared R-symmetry of the corresponding $\cN=1$ theory is given by:%
\begin{equation}\label{RIRt}
R_{IR}=R_{UV}+\left
(  \frac{t}{2}-\frac{1}{3}\right)  J_{\mathcal{N}=2}%
-tT_{3}+u_{1}U_{1}+u_{2}U_{2}. %
\end{equation}
Here, $U_{1}$ and $U_{2}$ are $U(1)$ generators associated with phase rotation of the
fields $Z_1$ and $Z_2$, $J_{\cN=2}$ is a $U(1)$ such that the $\mathcal{O}$'s
parameterizing the Higgs branch of the $\mathcal{N} = 2$ theory
have charge $-2$, $Z$ has charge $12$ and $Z_1$ and $Z_2$ each have charge $-1$.
$T_{3}$ is a Cartan generator in $SU(5)_{\bot}$ specified by the Jordan block structure of
the constant part of $\Phi$. In addition, the coefficients $u_{1}$ and $u_{2}$
are:%
\begin{align}
u_{1}  &  =\left(  S_{1}+\frac{3}{2}\right)  t-1\\
u_{2}  &  =\left(  S_{2}+\frac{3}{2}\right)  t-1
\end{align}
where $S_{1}$ and denotes the $T_3$ charge of the corresponding operator
$\mathcal{O}_{S_1}$ with lowest $T_3$ charge which multiplies
$Z_1$ in $Tr_{E_8}(\Phi(Z_1 , Z_2) \cdot \mathcal{O})$. Similar considerations
hold for $S_2$. Here, $t$ is a coefficient which is fixed by $a$-maximization \cite{Intriligator:2003jj}.
We parameterize all operator dimensions in terms of the ratio of the IR to UV scaling
dimension of the operator $Z$ via \cite{FCFT}:%
\begin{equation}
\rho=\frac{\Delta_{IR}(Z)}{\Delta_{UV}(Z)}=\frac{3}{2}t.
\end{equation}
In terms of $\rho$, the scaling dimensions of the chiral operators $Z_1$, $Z_2$,
$Z$ and ${\cal O}_s$ are:%
\begin{align}
\Delta\left(  Z_1\right)   &  =\left(  S_{1} + 1\right)  \rho\\
\Delta\left(  Z_2\right)   &  =\left(  S_{2} + 1\right)  \rho\\
\Delta\left(  Z\right)   &  =\Delta_{UV}(Z)\rho\\
\Delta\left(  \mathcal{O}_{s}\right)   &  =3-(s+1)\rho
\end{align}
where $\mathcal{O}_{s}$ is an operator with $T_3$ charge $s$.

Denote by $w_{i}$ the spin content of the five components of the
fundamental of $SU(5)_{\bot}$ under the $T_{3}$ generator. There are various
representations of $SU(5)_{\bot}$ to keep track of; the singlet, given by
$s=0$, the adjoint representation $s=w_{i}-w_{j}$, the $10$ given by
$s=w_{i}+w_{j}$ for $i\neq j$, the $5$ given by $s=w_{i}$, and the various
conjugate representations. We can now see that the scaling dimensions of the
various $\mathcal{O}$ operators are:
\begin{align}
\Delta\left(  \mathcal{O}_{(24,1)}\right)   &  =3-\rho\\
\Delta\left(  \mathcal{O}_{(1,24)}\right)   &  =3-(w_{i}-w_{j}+1)\rho\\
\Delta\left(  \mathcal{O}_{(10,5)}\right)   &  =3-(w_{i}+1)\rho\\
\Delta\left(  \mathcal{O}_{(\overline{5},10)}\right)   &  =3-(w_{i}+w_{j}+1)\rho\\
\Delta\left(  \mathcal{O}_{(\overline{10},\overline{5})}\right)   &  =3+(w_{i}-1)\rho\\
\Delta\left(  \mathcal{O}_{(5,\overline{10})}\right)   &  =3+(w_{i}+w_{j}-1)\rho.
\end{align}
By inspection the operator $\mathcal{O}$ with the lowest scaling
dimension descends from the $(1,24)$ of $SU(5)_{GUT}\times SU(5)_{\bot}$. In
other words, the $\mathcal{O}$ operators which are not charged under $SU(5)_{GUT}$ have
lower scaling dimension.

\section{Fixed Points with $W_{flux}$}
In this Appendix, we study whether adding terms quadratic in the $Z_{i}$ fields
will induce a flow to a new IR fixed point of the probe theory. The case
of linear deformations was studied in \cite{FCFT}, where it was found that
deformations linear in $Z_{1}$ and~$Z_{2}$ do not induce a flow to a new
interacting fixed point. By contrast, in most viable models based on a point
of $E_{8}$, $Z$ is an irrelevant operator, and does not affect the IR
behavior of the theory.

Our starting point is the $E_{8}$ Minahan-Nemeschansky theory, which we
imagine deforming by terms of the form:%
\begin{equation}
\delta W_{tilt} + \delta W_{flux} =Tr_{E_{8}}\left(  \Phi\left(  Z_{1},Z_{2}\right)  \cdot
\mathcal{O}\right)  +m_{ab}Z_{a}Z_{b}+O(Z_{a}^{3})
\end{equation}
we do not consider terms linear in the $Z_{a}$ since flavor physics
considerations already require a critical point at the origin. If
necessary, we can forbid all of $\delta W_{flux}$ because it breaks
additional flavor symmetries of the probe sector.

To analyze whether we flow to a new interacting SCFT, we demand that some of
these deformation are marginal in the infrared. Marginality
of some of the terms in $Tr_{E_8}(\Phi\left(  Z_{1},Z_{2}\right)  \cdot\mathcal{O})$
has been studied in \cite{FCFT} and fixes the form of the IR R-symmetry
up to one undetermined parameter $t$. If we now also require some of the terms
quadratic in the $Z_{a}$'s to be marginal in the IR, we can now solve
for $t$. First, we observe that not all of these terms can be simultaneously
marginal in the infrared. Following the conventions of \cite{FCFT} that
$Z_{1}$ multiplies a component of ${\cal O}$ with the lowest scaling dimension, it
follows that the dimension of $Z_{1}$ is always greater than or equal to the
dimension of $Z_{2}$. It is therefore enough to focus on demanding
$(Z_{2})^{2}$ is marginal in the IR.

If we demand that $(Z_{2})^{2}$ is marginal in the IR, we obtain the
condition:%
\begin{equation}
2=R_{IR}((Z_{2})^{2})=R_{UV}((Z_{2})^{2})+\left(  \frac{t}{2}-\frac{1}%
{3}\right)  J_{\mathcal{N}=2}((Z_{2})^{2})+u_{1}U_{1}((Z_{2})^{2})
\end{equation}
with conventions as in Appendix A. The $J_{\mathcal{N}=2}$ charge of $Z_{2}$
is $-1$. Solving for $t$ yields:
\begin{equation}
t=\frac{2}{2\mu_{2}-1}=\frac{1}{S_{2}+1}. \label{tcomp}%
\end{equation}
Note that due to the additional constraints, we do not need to invoke
a-maximization to determine the IR R-symmetry. Let us
compute the scaling dimensions of the operators at this new candidate fixed point. Using
the summary of operator dimensions given in Appendix A, we find:%
\begin{align}
\Delta\left(  Z_{1}\right)   &  =\frac{3}{2}\frac{S_{1}+1}{S_{2}+1}\\
\Delta\left(  Z_{2}\right)   &  =\frac{3}{2}\\
\Delta\left(  Z\right)   &  =\frac{3}{2}\frac{\Delta_{UV}(Z)}{S_{2}+1}\\
\Delta\left(  \mathcal{O}_{s}\right)   &  =3-\frac{3}{2}\frac{s+1}{S_{2}+1}.
\end{align}
In our conventions, $S_{1}\geq S_{2}$, which is consistent with $Z_{2}$ having
the lowest scaling dimension of all the $Z_{i}$. Let us quickly check that
there are no violations of the unitarity bound in this example. This is
immediate for all fields except $\mathcal{O}_{s}$. In that case, we observe
that in the most generic examples of T-brane configurations, the value of $s$
will be bounded above by $S_{1}$. In such situations, the dimension of
$\Delta\left(  \mathcal{O}_{s}\right)  $ is bounded below by:%
\begin{equation}
\Delta\left(  \mathcal{O}_{s}\right)  \geq 3-\frac{3}{2}\frac{S_{1}+1}{S_{2}+1}.
\end{equation}
In most realistic models, $S_{2}$ is typically $1$ or $2$ and $S_1$ is typically $2$ or $3$, so we see that
there are no obvious violations of the unitarity bound.

It is also of interest to compute the further effects of deforming by
Standard Model fields through the couplings $\Psi^{SM}_{R} \cdot \mathcal{O}_{\overline{R}}$ \cite{HVW}.
Here, $\Psi^{SM}_{R}$ refers to a third generation chiral matter field, or either of the Higgs fields. This is because these modes
have maximal overlap with the probe D3-brane. In \cite{HVW} a more complete analysis of such fixed points is
performed, where it is shown that with the contribution from $W_{flux}$
switched off, the perturbation by the Standard Model fields leads to only a
small shift in the overall scaling dimension of the Standard Model fields.
Here, we consider the effects of quadratic terms in the $Z_{i}$ as well as
further perturbations by operators such as $\Psi_{R}^{SM}\cdot \mathcal{O}_{\overline
{R}}$. If the mode $\Psi_{R}^{SM}$ does not couple to other Standard\ Model
fields, it will come equipped with an additional $U(1)$ symmetry which acts
only on this field. Denote this by $U(1)_{\Psi}$. This rephasing symmetry will
be broken by the presence of non-trivial Yukawa couplings in the visible sector.

There is a subtlety here, because the visible sector Yukawa couplings are generated by the
same fluxes which induce a potential for the D3-brane \cite{FGUTSNC}! Thus,
once we have switched on the fluxes to generate the D3-brane potential and the
Yukawa interactions, there are no additional abelian symmetries available in
the UV which the IR R-symmetry can mix with. Of course, in the IR,
it is still possible that there could be some emergent symmetries. This is
possible, for example, if some of the Standard Model fields develop an
anomalous dimension. Then, the Yukawa couplings will become irrelevant in the
IR as they involve three fields. In this limit, it is as if the Yukawa
couplings have been switched off. It is therefore consistent to consider
switching off the Yukawa couplings, but keeping terms quadratic in the $Z_{i}%
$'s as possible deformations. In this limit, there is an additional set of
$U(1)_{\Psi}$ symmetries which also mix with the infrared R-symmetry
\cite{HVW}:%
\begin{equation}
R_{IR}=R_{UV}+\left(  \frac{t}{2}-\frac{1}{3}\right)  J_{\mathcal{N}=2}%
-tT_{3}+u_{1}U_{1}+u_{2}U_{2}+u_{\Psi}U_{\Psi}%
\end{equation}
We now demand that the coupling $\Psi_{R}^{SM}\cdot \mathcal{O}_{\overline{R}}$ is
marginal in the infrared. This composite operator is neutral under $T_{3}$,
and $\Psi_{R}^{SM}$ is uncharged under $J_{\mathcal{N}=2}$. Hence, we obtain
the condition:%
\begin{equation}
2=R_{IR}(\Psi_{R}^{SM}\cdot \mathcal{O}_{\overline{R}})=2-2\left(  \frac{t}{2}-\frac
{1}{3}\right)  +u_{\Psi}%
\end{equation}
or:%
\begin{equation}
u_{\Psi}=t-\frac{2}{3}.
\end{equation}
In the infrared, the dimension of the Standard Model field is therefore:%
\begin{equation}
\Delta_{IR}(\Psi_{R}^{SM})=\frac{3}{2}t(1-T_{3}(\Psi_{R}^{SM})) = \frac{3}{2 S_{2} + 2}(1-T_{3}(\Psi_{R}^{SM}))
\end{equation}
The exact value of $S_{2}$ for the
other modes is somewhat model dependent, and this choice
correlates with the value of the $T_{3}$ charge.
As computed in \cite{HVW}, the $T_{3}$ charge of $H_{u}$ is typically zero or positive, so that
$H_{u} \mathcal{O}_{H_{u}}$ is irrelevant, and $H_{u}$ actually remains
dimension one in the infrared.

Let us now turn to some examples. Consider first a $%
\mathbb{Z}
_{2}$ monodromy scenario with tilting parameter:%
\begin{equation}
\Phi=\left[
\begin{array}
[c]{ccccc}
& 1 &  &  & \\
Z_{2} &  &  &  & \\
&  & \ast &  & \\
&  &  & \ast & \\
&  &  &  & \ast
\end{array}
\right]
\end{equation}
so that $S_{2}=1$. In one choice of matter curve assignments, the $T_{3}$
charges of the visible sector and Standard Model fields are $T_{3}%
(\overline{5}_{M})=T_{3}(10_{M})=-1/2$, while $T_{3}(\overline{5}_{H})=T_{3}(5_{H})= 0$ \cite{HVW}.
This means that the Higgs fields retain their free field scaling dimensions in
the infrared, and the $\overline{5}_{M}$ and $10_{M}$ attain scaling
dimensions:%
\begin{equation}
\Delta_{IR}(\overline{5}_{M})=\Delta_{IR}(10_{M})=\frac{9}{8}=1.125
\end{equation}
This is a small shift to the scaling dimensions.

Not all deformations by the Standard Model fields lead to such small
deformations. Consider a $%
\mathbb{Z}
_{2}\times%
\mathbb{Z}
_{2}$ monodromy scenario with tilting parameter:%
\begin{equation}
\Phi=\left[
\begin{array}
[c]{ccccc}
& 1 &  &  & \\
Z_{1} & \ast &  &  & \\
&  &  & 1 & \\
&  & Z_{2} & \ast & \\
&  &  &  & \ast
\end{array}
\right]  .
\end{equation}
This is an interesting case to consider, because now, terms quadratic in both
$Z_{1}$ and~$Z_{2}$ will be marginal in the infrared. In this case, $S_{2}=1$
as before, and the $T_{3}$ charge assignments for the Standard Model fields are $T_{3}%
(\overline{5}_{M})=T_{3}(10_{M})=-1/2$ and $T_{3}(\overline{5}_{H})=-1$ \cite{HVW}. This
leads to scaling dimensions:
\begin{align}
\Delta_{IR}(\overline{5}_{M})  & =\Delta_{IR}(10_{M})=\frac{9}{8}=1.125\\
\Delta_{IR}(\overline{5}_{H})  & =\frac{3}{2}=1.5.
\end{align}
In this case, $H_{d}$ mixes more strongly with the CFT. $H_u$ remains dimension one in this example.

The situation becomes more pronounced in a $Dih_{4}$ monodromy scenario,
where the tilting parameter is given as:%
\begin{equation}
\Phi=\left[
\begin{array}
[c]{ccccc}
& 1 &  &  & \\
&  & 1 &  & \\
&  &  & 1 & \\
Z_{1} &  & Z_{2} & \ast & \\
&  &  &  & \ast
\end{array}
\right]  .
\end{equation}
In this case, we have $S_{2}=1$, $T_{3}(5_{H}) = -2$, $T_{3}(10_{M})= T_{3}(\overline{5}_{H})=-3/2$, and
$T_{3}(\overline{5}_{M}) = 0$. Hence, $\overline{5}_{M}$ remains dimension one, and the other fields develop scaling
dimensions:
\begin{align}
\Delta_{IR}(5_{H}) & =\frac{9}{4}=2.25\\
\Delta_{IR}(10_{M})  & =\Delta_{IR}(\overline{5}_{H})=\frac{15}{8}=1.875.
\end{align}
which are significant deviations from the free field value.

Let us stress, however, that we can treat the quadratic mass terms in the $Z_i$ as a small
perturbation of the probe sector. This is a justified approximation because in the limit in which we take all flavor
seven-branes to be non-compact, so that $1/M_{GUT} \rightarrow \infty$, the flux induced superpotential is also
diluted, and we can effectively ignore such deformations. Also,
such deformations break additional symmetries of the probe sector,
so it is technically natural for such perturbations to be small. Turning
the issue around, we can use $W_{flux}$ as a way to tune the value of the
parameter $\nu$ which controls the amount of baryon
asymmetry generated by the model.

\section{Kinetic Mixing at Strong Coupling}
In this Appendix we discuss additional details of kinetic mixing with the hidden sector $U(1)_{D3}$ of the probe D3-brane. In contrast to many other
scenarios considered in the literature, here, the $U(1)_{D3}$ gauge theory on the Coulomb branch of the D3-brane is strongly coupled. As we have mentioned, this is due to the presence of electric and magnetic states of comparable mass. Our discussion follows and extends that given in \cite{Funparticles}. For additional discussion of kinetic mixing, see for example \cite{Holdom:1985ag,Babu:1996vt,Dienes:1996zr,Babu:1997st,Lust:2003ky,Abel:2003ue,Abel:2004rp,
Abel:2006qt,Feldman:2006wd,Feldman:2007wj,Abel:2008ai,Brummer:2009cs,Bruemmer:2009ky,Benakli:2009mk}.

For a single $U(1)$ gauge theory, we write:
\begin{equation}
L_{eff} = {1 \over 8 \pi} \text{Im} \int d^{2} \theta \,\, \tau \mathcal{W}^{\alpha} \mathcal{W}_{\alpha}
\end{equation}
where in our conventions, the gauge coupling and theta angle are packaged together as:
\begin{equation}
\tau = \frac{4 \pi i}{g^{2}} + \frac{\theta}{2 \pi}.
\end{equation}
Next consider the case of mixing in a $U(1)^{n}$ theory. In
terms of superfields, this type of mixing is conveniently expressed in
terms of an $n$-component vector $\mathcal{W}^{i}_{\alpha}$ which couple to
each other through an $n \times n$ matrix of couplings $\tau_{ij}$:
\begin{eqnarray}
L_{mix} = {1 \over 8 \pi} \text{Im} \int d^2 \theta \tau_{ij} \mathcal{W}^{\alpha}_{i} \mathcal{W}_{\alpha j}
\end{eqnarray}
where $i,j = 1,...,n$ index the various $U(1)$ factors of the theory. The superfield gauge field strengths $\mathcal{W}^{\alpha}_{i}$ are chosen with respect to a particular convention for what is an electric state and what is a magnetic state, which is a somewhat basis dependent statement. Fixing a convention, we can expand out into component fields, yielding the couplings between the field strengths:
\begin{equation}
L_{mix} \supset -\frac{1}{4 g_{ij}^{2}} F^{\mu \nu}_{i} F_{j \mu \nu} + \frac{\theta_{ij}}{32 \pi^{2}} F^{\mu \nu}_{i} \widetilde{F}_{j \mu \nu}
\end{equation}
in the obvious notation. In general, we can move to another duality frame by applying a $Sp(2n,\mathbb{Z})$ transformation. This acts on the $2n$-component vector of field strengths $(F_{1},\widetilde{F}_{1},...,F_{n},\widetilde{F}_{n})$. Our conventions are that $Sp(2,\mathbb{Z}) \simeq SL(2,\mathbb{Z})$ is the duality group for a single $U(1)$. The actual duality group may be a modular subgroup of $Sp(2n ,\mathbb{Z})$, dictated by the particular details of the theory. Let us note that the Dirac-Zwanziger-Schwinger quantization condition basically amounts to the condition that for $n$ $U(1)$ gauge fields, there is a $2n$ dimensional lattice of electric and magnetic charges with half-integer valued symplectic pairing. This can lead to interesting apparent violations of ``charge quantization'' if one artificially restricts to a single $U(1)$ factor of the theory (see for example \cite{Brummer:2009cs,Bruemmer:2009ky} for related discussion). Let us note that here we are considering the case of a gapless spectrum. After the spectrum develops a mass gap, the remaining gapless modes can be treated by similar means.

It is well-known that in the weakly coupled setting, such kinetic mixing is expected from one loop corrections associated with matter fields which are charged under different gauge group factors. It is also generic in many $\mathcal{N} = 2$ theories at strong coupling. This is basically because the $\mathcal{N} =2$ Seiberg-Witten curve is typically a curve of higher genus, and the modular parameters of this curve determine the matrix of couplings $\tau_{ij}$.\footnote{It is not, however, guaranteed that such mixing will occur. For example, it is known that the $\mathcal{N} = 2$ theory of $N$ D3-branes probing a $D_4$ F-theory singularity has $\tau_{ij}$ given by a diagonal matrix \cite{Douglas:1996js}.}

Let us now turn to the specific probe/SM kinetic mixing terms of this paper. We begin by discussing mixing in the UV theory. A significant complication for analyzing such theories is that the D3-brane theory in the UV does not have a known Lagrangian description. In the IR, however, we can work in terms of the low energy theory on the Coulomb branch. For example, in the $\mathcal{N} = 2$ supersymmetric theory, we can still discuss $\mathcal{W}_{\alpha}$ as associated with a descendant of the Coulomb branch parameter $Z$. Since $Z$ has high scaling dimension in the theories we are considering, the kinetic mixing is expected to be small at the CFT breaking scale.

In the weakly coupled $D_{4}$ probe theory, this is basically the statement that the $U(1)_{D3}$ is embedded inside $SU(2)_{D3}$, and so the corresponding field strength must multiply a factor of the adjoint valued $\varphi$ in order to generate a non-zero $SU(2)_{D3}$ invariant. There is an additional complication in GUT theories, which is that in the UV, $\mathcal{W}_{vis}$ embeds inside of $SU(5)_{GUT}$. This means that GUT breaking effects must also be included in the UV to generate such an operator. This sets a UV boundary condition that at the scale of CFT breaking, we expect the kinetic mixing between the two sectors to be nearly zero.

Below the CFT breaking scale we can still expect non-trivial kinetic mixing to be generated. No symmetry principle forbids such mixing terms at low energies. For example, there are various C and CP violating interactions such as $H_u \cdot \mathcal{O}_{H_u}$ and $H_d \cdot \mathcal{O}_{H_d}$ which explicitly break $SU(5)_{GUT}$ and induce radiative corrections. One can check in weakly coupled toy models that such effects generically lead to mixing between the different $U(1)$ factors. Basically, the renormalization group evolution of the mixing term has UV boundary condition set by the CFT breaking scale, and IR boundary condition set by the mass scale for the various threshold particles running in the loop. Some of these particles running in the loop are expected to be quite light, such as the Standard Model and $3-7_{hid}$ strings, while some such as the $3-7_{vis}$ are much closer to the CFT breaking scale. Here, the precise mass of such states and the dependence on K\"ahler data becomes important. In general, we expect a non-zero value for such mixing terms. Let us note that at strong coupling, there is little penalty in having many internal $U(1)_{D3}$ gauge bosons running in a loop diagram. It would be interesting to perform an exact computation of such mixing terms.

In the absence of such a computation, we must instead rely on genericity arguments. In
terms of canonically normalized gauge field strengths, we expect:
\begin{equation}
L_{mix} \supset -\frac{1}{4} F_{vis}^{2} -\frac{1}{4} F_{hid}^{2} + \frac{\theta_{hid} \alpha_{hid}}{8 \pi} F_{hid} \widetilde{F}_{hid} +   {\kappa_{elec} \over 2} F_{vis} F_{hid} + {\kappa_{mag} \over 2} F_{vis} \widetilde{F}_{hid}
\end{equation}
where the $\kappa$ mixing terms are expected to be of order:
\begin{equation}
\kappa_{mix} \sim \frac{\sqrt{\alpha_{vis} \alpha_{hid}}}{4 \pi} \sim 10^{-3} - 10^{-2}.
\end{equation}
In practice we allow this parameter to be fit by various phenomenological
considerations. It would clearly be interesting to study this type of effect further.

\newpage
\bibliographystyle{utphys}
\bibliography{D3gen}

\end{document}